\documentclass[10pt]{article}

\usepackage{etoolbox}

\usepackage{graphicx}
\usepackage{multicol,multirow}
\usepackage{amsmath,amssymb, amscd, amsthm, amsfonts, latexsym}
\usepackage{mathrsfs}
\usepackage{booktabs}
\usepackage{appendix}
\usepackage{enumitem}
\usepackage[scr=boondoxo,scrscaled=1.05]{mathalpha}
\usepackage{bbm}
\usepackage[french,english]{babel}
\usepackage{caption,subcaption}
\usepackage{adjustbox}
\usepackage{siunitx} 
\usepackage{rotating, delarray}
\usepackage{lipsum} 
\usepackage[colorlinks,allcolors=blue]{hyperref}
\usepackage{thmtools}

\newtheorem{theorem}{Theorem}[section]

\newtheorem{corollary}[theorem]{Corollary}
\newtheorem{proposition}[theorem]{Proposition}
\theoremstyle{definition}
\newtheorem{remark}[theorem]{Remark}
\newtheorem{example}[theorem]{Example}
\numberwithin{equation}{section}
\newtheorem{definition}[theorem]{Definition}
\usepackage{natbib}
\usepackage{authblk}

\usepackage{bbm}

\makeatletter
\renewcommand\@biblabel[1]{\textbullet}
\makeatother

\newcommand{\alphav}[1]{\alpha_{(\mathrm{pa}(#1), #1)}}
\newcommand{\pa}[1]{\mathrm{pa}(#1)}
\newcommand{\alphavj}[2]{\alpha_{(#1, #2)}}

\usepackage{pgfplotstable}

\usepackage[linesnumbered,ruled,vlined]{algorithm2e}
\SetKwInput{KwData}{Input} 
\SetKwInput{KwResult}{Output} 
\usepackage{multicol}
\usepackage{tikz, pgfplots}
\usetikzlibrary{trees,arrows.meta,shapes,positioning,shadows,trees, decorations.pathreplacing,calligraphy,fit,shapes.multipart,
arrows.meta,bending, calc, positioning, patterns}
\usepackage{float}
\usepackage{wrapfig}
\pgfplotsset{compat=1.17}
\usepackage{threeparttable}

\usepackage[onehalfspacing]{setspace}

\pgfplotsset{
    colormap={batlow}{%
rgb255(0cm)=(255,254,203)
rgb255(1cm)=(255,252,196)
rgb255(2cm)=(254,249,189)
rgb255(3cm)=(253,247,182)
rgb255(4cm)=(252,244,175)
rgb255(5cm)=(253,242,168)
rgb255(6cm)=(252,239,161)
rgb255(7cm)=(251,236,153)
rgb255(8cm)=(251,233,145)
rgb255(9cm)=(249,230,138)
rgb255(10cm)=(248,226,131)
rgb255(11cm)=(248,222,123)
rgb255(12cm)=(246,218,117)
rgb255(13cm)=(245,214,110)
rgb255(14cm)=(244,211,105)
rgb255(15cm)=(243,207,100)
rgb255(16cm)=(243,203,96)
rgb255(17cm)=(241,199,92)
rgb255(18cm)=(241,196,90)
rgb255(19cm)=(241,192,88)
rgb255(20cm)=(240,189,87)
rgb255(21cm)=(238,185,85)
rgb255(22cm)=(238,181,84)
rgb255(23cm)=(238,179,85)
rgb255(24cm)=(237,175,84)
rgb255(25cm)=(236,172,84)
rgb255(26cm)=(236,169,83)
rgb255(27cm)=(234,166,83)
rgb255(28cm)=(235,163,82)
rgb255(29cm)=(233,161,82)
rgb255(30cm)=(233,157,83)
rgb255(31cm)=(233,154,82)
rgb255(32cm)=(231,151,82)
rgb255(33cm)=(231,148,82)
rgb255(34cm)=(230,146,82)
rgb255(35cm)=(229,142,81)
rgb255(36cm)=(229,140,81)
rgb255(37cm)=(228,136,80)
rgb255(38cm)=(229,134,80)
rgb255(39cm)=(228,130,81)
rgb255(40cm)=(227,127,80)
rgb255(41cm)=(227,124,80)
rgb255(42cm)=(226,121,79)
rgb255(43cm)=(224,118,79)
rgb255(44cm)=(224,115,78)
rgb255(45cm)=(224,112,78)
rgb255(46cm)=(222,108,79)
rgb255(47cm)=(221,105,79)
rgb255(48cm)=(219,102,78)
rgb255(49cm)=(219,99,78)
rgb255(50cm)=(216,95,77)
rgb255(51cm)=(214,91,76)
rgb255(52cm)=(212,89,77)
rgb255(53cm)=(208,85,76)
rgb255(54cm)=(205,83,76)
rgb255(55cm)=(201,80,75)
rgb255(56cm)=(197,79,74)
rgb255(57cm)=(193,76,73)
rgb255(58cm)=(188,76,73)
rgb255(59cm)=(184,74,72)
rgb255(60cm)=(178,72,70)
rgb255(61cm)=(173,71,70)
rgb255(62cm)=(167,69,68)
rgb255(63cm)=(162,69,67)
rgb255(64cm)=(157,67,65)
rgb255(65cm)=(152,66,64)
rgb255(66cm)=(147,64,61)
rgb255(67cm)=(141,64,60)
rgb255(68cm)=(136,62,57)
rgb255(69cm)=(131,61,56)
rgb255(70cm)=(125,59,53)
rgb255(71cm)=(120,57,50)
rgb255(72cm)=(116,56,48)
rgb255(73cm)=(111,54,45)
rgb255(74cm)=(106,53,44)
rgb255(75cm)=(102,51,41)
rgb255(76cm)=(97,50,39)
rgb255(77cm)=(93,48,36)
rgb255(78cm)=(89,47,34)
rgb255(79cm)=(85,45,32)
rgb255(80cm)=(81,44,30)
rgb255(81cm)=(77,43,28)
rgb255(82cm)=(73,41,25)
rgb255(83cm)=(69,41,24)
rgb255(84cm)=(66,39,22)
rgb255(85cm)=(63,38,21)
rgb255(86cm)=(59,37,19)
rgb255(87cm)=(57,36,18)
rgb255(88cm)=(53,35,16)
rgb255(89cm)=(51,34,15)
rgb255(90cm)=(48,33,13)
rgb255(91cm)=(45,32,11)
rgb255(92cm)=(43,32,10)
rgb255(93cm)=(40,31,8)
rgb255(94cm)=(38,29,8)
rgb255(95cm)=(35,28,6)
rgb255(96cm)=(34,28,5)
rgb255(97cm)=(31,28,3)
rgb255(98cm)=(30,26,2)
rgb255(99cm)=(27,25,0)
rgb255(100cm)=(25,25,0)
}
}
\begin{document}

\sloppy
	\title{On a risk model with tree-structured Poisson Markov random field frequency, with application to rainfall events}
	\author{Hélène Cossette$^*$, Benjamin Côté$^\dagger$, Alexandre Dubeau$^*$, Etienne Marceau$^*$ \and 
		\textit{$^*$École d'actuariat, Université Laval, Québec, Canada}\\
    \textit{$^\dagger$ University of Waterloo, Ontario, Canada}}
        \date{November 30, 2024 \\
        This version: \today}
	\maketitle
	
\begin{abstract}
In many insurance contexts, dependence between risks of a portfolio may arise from their frequencies. We investigate a dependent risk model in which we assume the vector of count variables to be a tree-structured Markov random field with Poisson marginals. The tree structure translates into a wide variety of dependence schemes. We study the global risk of the portfolio and the risk allocation to all its constituents. We provide asymptotic results for portfolios defined on infinitely growing trees. To illustrate its flexibility and computational scalability to higher dimensions, we calibrate the risk model on real-world extreme rainfall data and perform a risk analysis.
\end{abstract}


\section{Introduction}
\label{sect:Introduction} 

Compound Poisson distributions serve as a basis for building risk models with applications in property and casualty insurance, such as risk management, pricing, and reserves. In these contexts, the loss $X_v$ for a given risk $v$ is often assumed to follow a compound Poisson distribution, and the portfolio's aggregate loss is thereby defined as
\begin{equation}
 S = X_1 + X_2 + \cdots + X_d,  \text{ with } X_v = \sum_{j=1}^{N_{{v}}} B_{v,j}, \; \text{for every }  v\in\mathcal{V}=\{1,\ldots,d\}, \, d\in\mathbb{N}_1=\mathbb{N}\backslash\{0\},   
 \label{eq:model}
\end{equation}
where $N_v \sim \mathrm{Poisson}(\lambda_v)$ is referred to as the risk's frequency and $\{B_{v, j}, j \in \mathbb{N}_1\}$ as its sequence of severities, with the convention $\sum_{j=1}^0 B_{v,j} = 0$.

Dependence between risks of a portfolio may arise from their frequencies. The model in \eqref{eq:model} has the advantages, as discussed in \cite{cummins1983estimating},  of allowing for proper accommodation of this dependence while explicitly accounting for events of different sources, which may have distinct claim amount distributions. Mathematically, this translates into components of the frequency random vector $\boldsymbol{N}= (N_v,\, v\in\mathcal{V})$ being dependent in \eqref{eq:model}. In this paper, the sequences of claim amounts $\{B_{1,j},\, j\in\mathbb{N}_{1}\},\ldots,\{B_{d,j},\, j\in\mathbb{N}_1\}$ are assumed to be mutually independent and independent of $\boldsymbol{N}$. Let the random variables within each sequence be identically distributed, and thus we may refer to claim amounts
 with stand-in random variables $B_1,\ldots, B_d$ for convenience. The portfolio $\boldsymbol{X} = (X_v,\, v\in\mathcal{V})$ thus follows a multivariate compound distribution of Type 2 according to the terminology in \cite{sundt2009recursions}; see Chapters 19--20. 
 In a risk modelling setting, multivariate compound Poisson distributions of Type 2 have been studied notably in \cite{cossette2012tvarbased} and \cite{kim2019capital}.

One may rely on three approaches to conceive a multivariate Poisson distribution for the random vector $\boldsymbol{N}$: copulas, common shocks and binomial thinning; see for instance \cite{inouye2017review} and \cite{liu2024maximum}.  The copula approach allows separate modelling of marginals and dependence but faces theoretical and computational challenges in a discrete setting \citep{genest2007primer,henn2022limitations}. The common shock approach, dating back to \cite{m1925applications} and later extended to higher dimensions \citep{krishnamoorthy1951multivariate,teicher1954multivariate}, offers a clear stochastic interpretation but quickly becomes intractable due to the exponential growth in parameters \citep{karlis2003algorithm}. The family of common-shock-based models, while widely referred to as \emph{multivariate Poisson}, does not encompass all Poisson-marginal distributions; see \cite{ccekyay2023computing} for a historical recap. 

The third approach is to rely on stochastic representations employing binomial thinning. Binomial thinning was introduced in \cite{steutel1983integer} and first employed to incorporate dependence between Poisson random variables in \cite{mckenzie1985some} and \cite{mckenzie1988some}. Such an approach has been used for risk modelling in \cite{yuen2002comparing}, \cite{lindskog2003common} and \cite{wang2005correlated}. In \cite{cote2025tree}, the authors encapsulate binomial thinning operations within a tree structure to provide a much wider variety of dependence schemes under this approach. 
The resulting tree-structured Markov random field (MRF) has explicit probability mass function (pmf) and probability generating function (pgf) expressions, enabling efficient computation in high dimensions. 

In a risk modelling context, scalability of computations and estimation methods to high dimensions is of the most important consideration. In particular, to introduce dependence between Poisson risk frequencies, the current methods based on common shocks are either little flexible or computationally intensive. This work draws inspiration from the MRF in \cite{cote2025tree} to reconcile flexibility and scalability without recourse to copulas. In this paper, the authors' main focus is the MRF's distributional properties. Here, we put forth its relevance for risk modelling. To do so, we begin by allowing more flexibility for the modelling of marginals by letting the MRF's means be heterogeneous, despite a slight tradeoff in the disconnect between marginal and dependence. We also show the MRF's connection to the multivariate Poisson and discuss methods for the model's estimation.

We put forth the MRF's relevance for risk modelling through two objectives. One of our objectives is to highlight the computational methods' practicality and their applicability to actuarial science. We will discuss this through two tasks. First, we aim to evaluate the aggregate risk of the portfolio by studying the distribution of $S$ in \eqref{eq:model} and developing efficient methods to evaluate its pmf without resorting to approximations. Second, we aim to assess the contribution of every component of the portfolio $\boldsymbol{X}$ to the aggregate claim amount, and we perform this risk allocation twofold. For an allocation \textit{ex-ante}, we resort to the contribution to the TVaR under Euler's principle, see \cite{tasche2007capital}; for an allocation \textit{ex-post}, we turn to conditional-mean risk-sharing, see \cite{denuit2012convex} and subsequent work. Algorithms for its exact computation are developed, inspired from the methods put forth in \cite{blier2025efficient}. We furthermore provide results allowing for a better understanding of the portfolio's asymptotic behavior by defining the model on infinite-dimensional trees. 

Another objective is to illustrate the practical relevance and effectiveness of the proposed risk model in real-world applications. Notably, from Theorem~7.1 of \cite{coles2001classical}, counts of extreme events follow Poisson distributions. This insight provides a natural avenue for applying our model. We perform a detailed risk management study on extreme rainfall events data, in which we evaluate tail risk measures (e.g., TVaR) and assess the allocation of extreme losses across different station locations. Our study showcases the risk model's ability to capture dependence structures in multivariate extreme events while maintaining interpretability. This paper contributes to the growing use of graphical models in actuarial science, including \cite{oberoi2020graphical}, \cite{denuit2022conditional} and \cite{boucher2024modeling}.

The structure of the paper is as follows. In Section~\ref{sect:extension}, we present the tree-structured MRF with Poisson marginal distributions and discuss its connection with the distributions obtained through the common shock approach. In Section~\ref{sect:aggregate}, we perform our first risk management task, evaluating the risk associated to $S$. In Section~\ref{sect:allocation}, we perform our second risk management task, allocating that risk to the components of $\boldsymbol{X}$; we also provide results for asymptotic cases of infinitely large portfolios. In Section~5, we discuss the MRF's parameters' estimation. Section~\ref{sect:data} comprises the application to extreme rainfall data. All proofs are relegated to Appendix~\ref{sect:proofs}. Section~7 provides concluding remarks.

\section{Tree-structured MRFs with Poisson marginal distributions}
\label{sect:extension}
In this section, we present the tree-structured MRF with Poisson marginal distributions, which will be the center of consideration in the following sections. Distributions of this family describe tree-structured MRFs. We recall below the definition of a MRF \citep[Chapter 4.2]{cressie2015statistics} and some elementary notions pertaining to trees.  Let $\mathcal{V} = \{1, 2, \dots, d\}$, with $d \in \mathbb{N}_1$, represent a set of vertices, and $\mathcal{E}\subseteq \mathcal{V} \times \mathcal{V}$ be a set of edges.    
\begin{definition}[MRF] \label{def:mrf}
A vector of random variables $ \boldsymbol{X} = (X_v, v \in \mathcal{V}) $ is a MRF if it satisfies the local Markov property with respect to a graph $\mathcal{G}=(\mathcal{V},\mathcal{E})$; that is, for any two of its components, say $ X_u $ and $ X_w $, such that $(u,w) \notin \mathcal{E} $,
\begin{equation}\label{eq:local-mrf}
X_u \perp\!\!\!\perp X_w \mid \{X_j,\, (u, j) \in \mathcal{E}\}, \; u, w \in \mathcal{V},
\end{equation}
where $ \perp\!\!\!\perp $ denotes conditional independence. A MRF is tree-structured if its underlying graph is a tree.
\end{definition}

A tree, denoted by $\mathcal{T}$, is a simple and connected undirected graph such that no path from a vertex to itself exists. A path from vertex $u$ to vertex $v$, written $\mathrm{path}(u,v)$, is the set of edges $e \in \mathcal{E}$ such that $u$ and $v$ participate in an edge once and any other involved vertices twice. All graphs considered in this paper are trees. Labeling a specific vertex $r \in \mathcal{V}$ as the root of a tree, we define $\mathcal{T}_{\!\!r}$ as the $r$-rooted version of $\mathcal{T}$. A root for the tree defines a parent $\mathrm{pa}(v)$ (except for the root), children $\mathrm{ch}(v)$ and descendants $\mathrm{dsc}(v)$ for each $v \in \mathcal{V}$. An example of this notation is provided in Figure~\ref{fig:tree-notation}, where vertex 1 acts as the tree's root. We refer to Section~3.3 of \cite{saoub2021graph} for further insight on the terminology surrounding rooted trees.
 
\begin{figure}[t]
\centering
\resizebox{0.125\textwidth}{!}{\begin{tikzpicture}[every node/.style={circle, draw, inner sep=2pt}, scale=0.8, transform shape]

\node[fill=black!20] (n1) at (0,0) {1};
\node[below left = 0.8 cm of n1,xshift=0.3cm] (n2) {2};
\node[below right  = 0.8 cm of n1,xshift=-0.3cm] (n3) {3};
\node[below left = 0.8cm of n3, xshift=0.3cm] (n4) {4};
\node[below right = 0.8cm of n3,xshift=-0.3cm] (n5) {5};
\node[below left = 0.8cm of n4, xshift=0.3cm] (n6) {6};
\node[below right = 0.8cm of n4, xshift=-0.3cm] (n7) {7};
\draw (n1) -- (n2);
\draw (n1) -- (n3);
\draw (n3) -- (n4);
\draw (n3) -- (n5);
\draw (n4) -- (n6);
\draw (n4) -- (n7);

\path (n3) -- (n5) node[midway, sloped, draw=none, font=\footnotesize, yshift=0.25cm] {};
\path (n1) -- (n3) node[midway, sloped, draw=none, font=\footnotesize, yshift=0.25cm] {};

\begin{scope}[overlay]
\node[draw = none, below = 0.7cm of n5] (n8) {};
\draw[decorate, decoration={brace, raise=0.5cm, amplitude=5pt}, thick] (n5.north)  -- (n8.south) node[midway, draw=none, right=0.8cm, scale=1] {Descendants of 3};
\draw[decorate, decoration={brace, mirror, raise=0.8cm, amplitude=5pt}, thick] (n4.north) -- (n4.south) node[midway, draw=none, left=1.3cm, scale=1] {Children of 3};
\draw[decorate, decoration={brace, raise=1.5cm, amplitude=5pt}, thick] (n1.north) -- (n1.south) node[midway, draw=none, right=2.1cm, scale=1] {Parent of 3};
\end{scope}

\end{tikzpicture}}
\caption{Filial relations in a rooted tree.}
\label{fig:tree-notation}
\end{figure}

To encrypt a MRF on a tree, the random vector's dependence scheme inherits the structural properties of the latter. For instance, the correlation between two of its components decays multiplicatively along the path between them. Although constraining feasible dependence schemes, this comes with the benefit of a greater interpretability: this is the idea underlying probabilistic graphical models. 

The construction of tree-structured MRFs with Poisson marginal distributions relies on the binomial thinning operator, denoted by $\circ$, defined in terms of a random variable $Y$ taking values in $\mathbb{N}$ as $\theta \circ Y := \sum_{j=1}^{Y} I_{j}^{(\theta)}, \; \theta \in [0, 1],$ where $ \{I_{j}^{(\theta)}, j \in \mathbb{N}_1 \} $ is a sequence of independent Bernoulli random variables with a probability of success $\theta$; see \cite{steutel1983integer}. This operation can be interpreted as sampling among $Y$ elements, each with probability $\theta$ of being selected independently.
We refer the interested reader to \cite{weiss2008thinning} 
for further insight on the binomial thinning operator.

\subsection{Main characteristics of tree-structured MRFs with Poisson marginal distributions}
To construct a risk model with heterogeneous marginal behaviors in Section~\ref{sect:aggregate}, we define tree-structured MRFs with Poisson marginals where each component has its own mean parameter $\lambda_v$, $v \in \mathcal{V}$.
\begin{theorem}\label{th:flexible}
Consider a tree $ \mathcal{T} = (\mathcal{V}, \mathcal{E})$, and let $\mathcal{T}_{\!\!r}$ be its rooted version, for some $r \in \mathcal{V}$. Given a vector of mean parameters $\boldsymbol{\lambda}=(\lambda_v,\,v\in\mathcal{V})$ where $\lambda_v>0$ for every $v\in\mathcal{V}$ and a vector of dependence parameters $\boldsymbol{\alpha} = (\alpha_e,\, e\in \mathcal{E})$ where \mbox{$\alphav{v} \in (0, \min(\sqrt{{\lambda_{v}}/{\lambda_{\pa{v}}}}, \sqrt{{\lambda_{\pa{v}}}/{\lambda_{v}}})]$} for every $(\pa{v}, v)\in \mathcal{E}$. Let $\boldsymbol{L}=(L_v,\,v\in\mathcal{V})$ be a vector of independent random variables such that \mbox{$L_v\sim \text{Poisson}(\lambda_v ( 1- \alphav{v}\sqrt{\lambda_{\pa{v}}/\lambda_v}))$} for every $v\in\mathcal{V}$, with $\alphav{r}=0$ since the root has no parent. Define $\boldsymbol{N} = (N_v,\, v \in \mathcal{V})$ as a vector of random variables such that 
\begin{equation}
N_v = 
    \begin{cases}
    L_r, & \text{if } v=r \\ 
    \left(\alphav{v}\sqrt{\frac{\lambda_v}{\lambda_{\mathrm{pa}(v)}}} \right)\circ N_{\mathrm{pa}(v)}+ L_v,& \text{if } v\in\mathrm{dsc}(r)
    \end{cases},\quad \text{for every } v\in\mathcal{V}.
\label{eq:StoConstruct}
\end{equation} 
Then, $\boldsymbol{N}$ is a MRF with a unique joint distribution whichever the chosen root of $\mathcal{T}$, where the random variable $N_v$ follows a Poisson distribution of parameter $\lambda_v$, for all $v\in \mathcal{V}$.

Henceforth, we write \mbox{$\boldsymbol{N}\sim \text{MPMRF}(\boldsymbol{\lambda}, \boldsymbol{\alpha}, \mathcal{T})$} to signify $\boldsymbol{N}$ admits the stochastic representation in \eqref{eq:StoConstruct}, and we let ${\boldsymbol{\Lambda}}$ denote the set of admissible parameters $(\boldsymbol{\lambda}, \boldsymbol{\alpha})$. We write $\mathbb{MPMRF}$ the family of all such distributions for $\boldsymbol{N}$. 
\end{theorem}

The MRF studied in \cite{cote2025tree} is a special case of the MRF constructed in Theorem~\ref{th:flexible}, where the mean parameters are homogeneous. In \eqref{eq:StoConstruct}, the components of $\boldsymbol{N}$, except the root's, are defined as the sum of two independent random variables. We interpret them as the \textit{propagation} and the \textit{innovation} random variables, respectively. The propagation random variable $(\alphav{v}\sqrt{{\lambda_v}/{\lambda_{\mathrm{pa}(v)}}} )\circ N_{\mathrm{pa}(v)}$ expresses the number of events that have duplicated by binomial thinning from $N_{\mathrm{pa}(v)}$ to $N_v$. The innovation random variable  $L_v$ expresses the number of events occurring at vertex $v$ that have not propagated from vertex $\mathrm{pa}(v)$.  The thinning parameter \mbox{$\alphav{v}\sqrt{{\lambda_v}/{\lambda_{\mathrm{pa}(v)}}}$} for the propagation random variable is an adjustment of the dependence parameter $\alphav{v}$ taking into account the heterogeneous means, so that the correlation between $N_{\mathrm{pa}(v)}$ and $N_v$ is  $\alphav{v}$.  
The upper bound  $\alpha_{(\mathrm{pa}(v),v)} <
\min(\sqrt{\lambda_v/\lambda_{\mathrm{pa}(v)}},\sqrt{\lambda_{\mathrm{pa}(v)}/\lambda_v})$ stems from the fact that dependence is characterized by this thinning mechanism. The fraction of events propagating from one random variable to another cannot exceed the root of the ratio of the marginal rates, otherwise the innovation component would require more inherited events than the number of events possible on the parent. The admissible parameter space therefore reflects a structural feature of event-sharing mechanisms in count data. 
 Note that any vertex can be chosen as the root of the tree; accordingly, the constraint on the dependence parameters  takes into account the reversibility of parent-child relationships 
that may occur, thus preserving the root-free formulation of the distribution; see Remark~\ref{rem:root}. 
For illustration, if $(\lambda_1,\lambda_2) = (5,5)$, the upper bound is $1$, maximal correlation. If $(\lambda_1,\lambda_2) = (3,5)$, the bound is $\sqrt{0.6} \approx 0.77$, compared to the Fréchet upper bound of $\approx 0.97$.

\begin{remark}\label{rem:root}
As put forth in the proof of Theorem~\ref{th:flexible}, the symmetry of pairwise stochastic dynamics, combined with the local Markov property, yields a root-free distribution of the MRF. The directionality seemingly implied by the necessity to choose a root is illusory: to define a parent-child dynamic is necessary to have a binomial-thinning representation, 
but it has no impact on the joint distribution of $\boldsymbol{N}$. 
The model is thus truly undirected; we will provide in Section~\ref{sect:shocks} a root-free stochastic construction. The artificial directionality, required for the binomial thinning represetation, is what allows
an iterative way of proceeding along trees, resulting in algorithmic efficiency.
A similar situation occurs for tree-structured Ising models; see \cite{cote2025ising}.    
\end{remark}

\begin{remark}
    Compared to the homogeneous-mean model in \cite{cote2025tree}, $\boldsymbol{\alpha}$ is not a true dependence parameter, as it is constrained by $\boldsymbol{\lambda}$. To circumvent this, one could let $\theta_v = \alphav{v}\sqrt{{\lambda_v}/{\lambda_{\mathrm{pa}(v)}}} $, and $\boldsymbol{\theta}$ would be a true dependence parameter, ranging from 0 to 1. The values of $\boldsymbol{\theta}$ would however vary with the root, so this parameter change would come at the cost of the undirectionality of the model. Also note that, because the constraints on $\boldsymbol{\alpha}$ come directly from the binomial thinning operations, they should not be limiting if the stochastic dynamics underlying the data are effectively of propagation and innovation.
\end{remark}

The following sequence of joint pgfs \mbox{\{$\eta_{v}^{\mathcal{T}_{\!\!r}},\,v\in\mathcal{V}\}$} will prove useful throughout the paper.  

\begin{definition}
\label{def:eta}
Consider a tree $ \mathcal{T} = (\mathcal{V}, \mathcal{E})$, and let $\mathcal{T}_{\!\!r}$ be its rooted version, $r \in \mathcal{V}$. For any vector $\boldsymbol{\theta}$  of thinning parameters, let $\boldsymbol{\theta}_{\mathrm{dsc}(v)} = (\theta_{j },\, j \in \mathrm{dsc}(v))$ for all $v\in\mathcal{V}$. We define \mbox{$\{\eta_{v}^{\mathcal{T}_{\!\!r}},\,v\in\mathcal{V}\}$} as a sequence of joint pgfs through the recursive relation 
\begin{equation}
\eta_{v}^{\mathcal{T}_{\!\!r}}(\boldsymbol{t}_{v\mathrm{dsc}(v)};\boldsymbol{\theta}_{\mathrm{dsc}(v)}) := t_v \prod_{j\in\mathrm{ch}(v)} \left(1 - \theta_j 
+ \theta_j \eta_{j}^{\mathcal{T}_{\!\!r}}(\boldsymbol{t}_{j\mathrm{dsc}(j)};\boldsymbol{\theta}_{\mathrm{dsc}(j)}) \right),\quad\quad\boldsymbol{t}\in[-1,1]^d,
\label{eq:eta}
\end{equation}
where $\boldsymbol{t}_{v\mathrm{dsc}(v)}$ is a short-hand notation for the vector $(t_j, \,  j \in \{v\} \cup \mathrm{dsc}(v))$, and with the convention $\eta_{j}^{\mathcal{T}_{\!\!r}}(\boldsymbol{t}_{j\mathrm{dsc}(j)};\boldsymbol{\theta}_{\mathrm{dsc}(j)}) = t_j$ for vertices $j$ that have no children according to the rooting in $r$.
\end{definition}

In the following proposition, we present the joint pmf and joint pgf of $\boldsymbol{N}$ as in \eqref{eq:StoConstruct}.

\begin{proposition}\label{th:propN} Let $\boldsymbol{N} \sim \text{MPMRF}(\boldsymbol{\lambda}, \boldsymbol{\alpha}, \mathcal{T})$, where $(\boldsymbol{\lambda},\boldsymbol{\alpha})\in\boldsymbol{\Lambda}$. For a chosen root $r \in \mathcal{V}$, let $\mathcal{T}_{\!\!r}$ be the rooted version of $\mathcal{T}$ and $\zeta_{L_v} = \lambda_v ( 1- \alphav{v}\sqrt{\lambda_{\pa{v}}/\lambda_v})$ for $v \in \mathcal{V}\backslash\{r\}$. Then, 
\begin{enumerate}[label = (\roman*), ref=(\roman*)]
\item \label{item:pmfN} the joint pmf of $\boldsymbol{N}$ is given by
\begin{equation}
p_{\boldsymbol{N}}(\boldsymbol{x}) = \frac{\mathrm{e}^{-\lambda_r}\lambda_r^{x_r}}{x_r!} \prod_{v\in\mathrm{dsc}(r)}
 \sum_{k=0}^{\min(x_{\mathrm{pa}(v)},x_v)} \frac{{ \mathrm{e}}^{-\zeta_{L_v}} (\zeta_{L_v})^{x_v-k}}{(x_v-k)!} 
 \binom{x_{\mathrm{pa}(v)}}{k} \left( \theta_v \right)^{k} \left(1- \theta_v\right)^{x_{\mathrm{pa}(v)}-k}, 
\label{eq:jointpmf-vecteurN}
\end{equation}
for $\boldsymbol{x} \in \mathbb{N}^d$, where $\theta_v = \alpha_{(\mathrm{pa}(v),v)}\sqrt{{\lambda_v}/{\lambda_{\mathrm{pa}(v)}}}$
for all $v\in \mathrm{dsc}(r)$;
\item \label{item:pgfN} the joint pgf of $\boldsymbol{N}$ is given by 
\begin{equation}
\mathcal{P}_{\boldsymbol{N}}(\boldsymbol{t}) = \prod_{v\in\mathcal{V}} \mathrm{e}^{\zeta_{L_v}(\eta_{v}^{\mathcal{T}_{\!\!r}}(\boldsymbol{t}_{v\mathrm{dsc}(v)};\,\boldsymbol{\theta}_{\mathrm{dsc}(v)}) - 1)},\quad\quad \boldsymbol{t}\in[-1,1]^d,
\label{eq:jointpgf}
\end{equation}
where $\boldsymbol{\theta}_{\mathrm{dsc}(v)}= (\alphav{k}\sqrt{{\lambda_k}/{\lambda_{\mathrm{pa}(k)}}}, \, k \in \mathrm{dsc}(v))$ is the vector of thinning parameters for the propagation random variables according to a rooting in $r$.
\end{enumerate}
\end{proposition}

The joint pmf in \eqref{eq:jointpmf-vecteurN} factorizes along the tree, underscoring that $\boldsymbol{N}$ is a MRF. For further discussions on Gibbs distributions and pmfs factorization, we refer to Appendix~\ref{sect:factorization}, as well as Chapter~4.2 of \cite{cressie2015statistics}, and \cite{cote2025tree}. The analytical form of the joint pmf in~\eqref{eq:jointpmf-vecteurN} enables efficient numerical evaluation of the likelihood, making it particularly well-suited for parameter estimation procedures, as shown in Section~\ref{sec:estimation} and Section~\ref{sect:data}. 

In the upcoming subsection, we show that every distribution of $\mathbb{MPMRF}$ may be reparameterized such that $\boldsymbol{N}$ admits an alternative stochastic representation based on common shocks. Whereas models based on the common-shock approach, whose family of distributions we write $\mathbb{MPCS}$, may become intractable in high dimensions due to the exponential growth in the number of possible shock configurations, the $\mathbb{MPMRF}$ family scales conveniently to high dimensions.

\subsection{A subfamily of the multivariate Poisson distribution based on common shocks}
\label{sect:shocks}

Let ${\mathcal{V}}  = \{1, 2, \dots, d\}$  be a set of indices and let $\mathscr{P}(\mathcal{V})$ be the power set of ${\mathcal{V}}$, that is the set of all subsets of $\mathcal{V}$, including the empty set and $\mathcal{V}$ itself.  For every $v\in\mathcal{V}$, let $\mathscr{P}(\mathcal{V};v) = \{\mathcal{W} \in \mathscr{P}(\mathcal{V}): v\in \mathcal{W} \}$, that is, $\mathscr{P}(\mathcal{V};v)$ comprises the elements of $\mathscr{P}(\mathcal{V})$ in which $v$ participates. Hence, $\bigcup_{v\in\mathcal{V}} \mathscr{P}(\mathcal{V};v) = \mathscr{P}(\mathcal{V})$. We define \mbox{$\boldsymbol{Y} = (Y_{\mathcal{W}}, \mathcal{W} \in\mathscr{P}(\mathcal{V}))$} as a vector of independent Poisson distributed random variables with a corresponding mean-parameter vector \mbox{$\boldsymbol{\gamma}= (\gamma_{\mathcal{W}}, \, \mathcal{W}\in\mathscr{P}(\mathcal{V}))$},  with $\gamma_{\mathcal{W}}\geq 0$ for every $\mathcal{W} \in \mathscr{P}(\mathcal{V})$. We use the convention $Y_{\mathcal{W}} = 0$ whenever $\gamma_{\mathcal{W}}  = 0$. Letting $\boldsymbol{D} = (D_v,\,v \in \mathcal{V}) \sim \text{MPCS}(\boldsymbol{\lambda})$, we have $D_v = \sum_{\mathcal{W} \in \mathscr{P}(\mathcal{V};v)} Y_{\mathcal{W}}, \;  v \in \mathcal{V},$ where, from the closure on convolution of the Poisson distribution, each component of $\boldsymbol{D}$ is Poisson distributed with parameter $\lambda_v = \sum_{\mathcal{W} \in \mathscr{P}(\mathcal{V};v)} \gamma_{\mathcal{W}}$. The joint pgf of $\boldsymbol{D}$ is given by
\begin{equation}
    \mathcal{P}_{\boldsymbol{D}}(\boldsymbol{t}) = \mathrm{e}^{\left({\gamma_0 + \sum_{\mathcal{W} \in \mathscr{P}(\mathcal{V})} \gamma_{\mathcal{W}} \prod_{v \in \mathcal{W}} t_{v}}\right)}, \quad \boldsymbol{t}\in[-1,1]^d,
    \label{eq:PGFmultivariatepoisson}
\end{equation}
with $\gamma_0 = - \sum_{\mathcal{W}\in \mathscr{P}(\mathcal{V})} \gamma_{\mathcal{W}}$.
We recall that the parameters vector $\boldsymbol{\gamma}$ is of length $|\mathscr{P}(\mathcal{V})| = 2^d$. This may make computations regarding the multivariate Poisson distribution cumbersome, as discussed earlier.

The following proposition provides an alternative parameterization and stochastic representation of $\boldsymbol{N}$ in terms of common shocks.
\begin{proposition} 
\label{th:StoRepCommonShocks}
Consider a tree $\mathcal{T}= (\mathcal{V},\mathcal{E})$ and, for every $v\in\mathcal{V}$, let $\Theta_v$ be the set of all subtrees of $\mathcal{T}$ in which $v$ participates, meaning $\Theta_v  = \{\mathcal{W} \in \mathscr{P}(\mathcal{V};v)$: for every $i, j \in \!\mathcal{W}, \,\,  k,l \in\! \mathcal{W}$ for every $(k, l) \in\! \mathrm{path}(i,j)\}$.  If \mbox{$\boldsymbol{N}\sim\text{MPMRF}(\boldsymbol{\lambda}, \boldsymbol{\alpha}, \mathcal{T})$}, with $(\boldsymbol{\lambda},\boldsymbol{\alpha})\in{\boldsymbol{\Lambda}}$, then $\boldsymbol{N}$ admits the following alternative stochastic representation:
\begin{equation}
    N_v = \sum_{\mathcal{W}\in\Theta_v} Y_{\mathcal{W}},\quad   v\in\mathcal{V},
    \label{eq:commonshockMPMRF}
\end{equation}
    where $\{Y_{\mathcal{W}},\, \mathcal{W}\in\bigcup_{v\in\mathcal{V}}\Theta_v\}$ are independent Poisson random variables of respective parameters
     \begin{equation}
     \gamma_{\mathcal{W}} = \left( \prod_{w\in \mathcal{W}} \lambda_w \right)\left(\prod_{(i,j)\in\mathcal{E}_{\mathcal{W}}} \frac{\alpha_{(i,j)}}{\sqrt{\lambda_i\lambda_j}}\right)\left(\prod_{(i,j)\in\mathcal{E}^{\dagger}_{\mathcal{W}}} \left(1-\alpha_{(i,j)}\sqrt{\frac{\lambda_j}{\lambda_i}}\right)\right), \quad \mathcal{W}\in \bigcup_{v\in\mathcal{V}}\Theta_v,
     \label{eq:gammaparamMPMRF}
     \end{equation}
     with $\mathcal{E}_{\mathcal{W}} = \{(i,j)\in\mathcal{E} : i,j\in \mathcal{W}\}$ and $\mathcal{E}_{\mathcal{W}}^{\dagger} = \{(i,j)\in\mathcal{E} : i\in \mathcal{W}, j\not\in \mathcal{W}\}$.
\end{proposition}

The upper limit for $\alpha_e$, $e\in\mathcal{E}$, for $(\boldsymbol{\lambda},\boldsymbol{\alpha})$ to be in ${\boldsymbol{\Lambda}}$, ensures $\gamma_{\mathcal{W}}\geq0$ for every $\mathcal{W}\in\bigcup_{v\in\mathcal{V}}\Theta_v$. 

Given Proposition~\ref{th:StoRepCommonShocks}, one easily sees that $\boldsymbol{N}$ follows a multivariate Poisson with vector of parameters $\boldsymbol{\gamma}= (\gamma_V, \, V\in\mathscr{P}(\mathcal{V}))$ such that 
\begin{equation*}
    \gamma_{V} = 
\begin{cases}
\gamma_{\mathcal{W}} 
,&\text{if }{\mathcal{W}\in{\bigcup_{v \in \mathcal{V}}\Theta_v}}\\
0,&\text{else} 
\end{cases},
\quad V \in \mathscr{P}(\mathcal{V}).
\end{equation*}
Hence, Proposition~\ref{th:StoRepCommonShocks} shows $\mathbb{MPMRF}\subseteq\mathbb{MPCS}$. For a further discussion on the connection between the thinning and the common-shock approaches for Poisson random variables, see \cite{liu2024maximum} and their Remark~2.3 in particular.  Although the number of non-zero parameters in the common shock representation of $\mathbb{MPMRF}$ is lower than $2^d$ (as for $\mathbb{MPCS}$),  the reduction is not substantial enough to overcome computation challenges. Moreover, the parameterization in terms of $\boldsymbol{\gamma}$ intertwines the dependencies and the marginals, thereby removing their intended parametric disconnection. Theorem~\ref{th:flexible} remains a simpler representation, as put forth in Example~\ref{ex:MPCS} below. The family $\mathbb{MPMRF}$ being a subset of $\mathbb{MPCS}$, it cannot grow out of the latter's limitations in terms of achievable dependence structures. The advantage of $\mathbb{MPMRF}$ over generic elements of $\mathbb{MPCS}$ comes from its binomial-thinning stochastic representation \eqref{eq:StoConstruct} and the computational efficiency ensuing.

\begin{example}\label{ex:MPCS}
A 5-variate distribution in $\mathbb{MPCS}$ generally requires $2^5 = 32$ parameters. Consider \mbox{$\boldsymbol{N} \sim \text{MPMRF}(\boldsymbol{\lambda}, \boldsymbol{\alpha}, \mathcal{T})$} where $\mathcal{T}$ is structured as in Figure~\ref{fig:five-nodes}. Using \eqref{eq:commonshockMPMRF}, we develop $\boldsymbol{N}$ into its common shock representation in Figure~\ref{fig:five-nodes}. One notices that constructing $\boldsymbol{\gamma}$ demands $|\bigcup_{v \in \mathcal{V}}\Theta_v| = 17$ non-zero parameters, which is a meaningful diminution, but still much higher than the 9 parameters required by the representation in Theorem~\ref{th:flexible}. A comparison of $N_1$ and $N_2$ in Figure~\ref{fig:five-nodes} reveals that a change in $\gamma_{\{1,2\}}$ affects both mean parameters of the random variables $N_1$ and $N_2$. The parameters $\gamma_{\mathcal{W}}$ associated to each $Y_{\mathcal{W}}$ in Figure~\ref{fig:five-nodes} are given in Table~\ref{tab:gamma-values}. We verify easily that $N_v\sim\text{Poisson}(\lambda_v)$ for every $v\in\{1,\ldots,5\}$. In Table~1 of Supplement~\ref{sup:equivalence}, a numerical instance illustrates the equivalence between $\mathbb{MRMRF}$ and $\mathbb{MPCS}$.

\begin{figure}[t]
    \centering
\begin{subfigure}[t]{0.1\textwidth}
\centering
\resizebox{!}{1.15\textwidth}{{\begin{tikzpicture}[every node/.style={circle, draw, inner sep=2pt, minimum size=6mm},  node distance = 1.5mm, scale=0.6, transform shape, thick]
\node (1a) {$1$};
\node [below= 0.4cm of 1a] (2a) {$2$};
\node [below= 0.4cm of 2a] (3a) {$3$};
\node [below left= 0.4cm of 3a] (4a) {$4$};
\node [below right= 0.4cm of 3a] (5a) {$5$};

\draw (1a) --  (2a);
\draw (2a) -- (3a);
\draw (3a) -- (4a);
\draw (3a) -- (5a);
\begin{scope}[overlay]
\end{scope}
\end{tikzpicture}}}
\end{subfigure}
\hfill
\begin{subfigure}[b]{0.85\textwidth}
    \centering
    \input{Figures/table-commonshock}
\end{subfigure}   
\caption{Tree $\mathcal{T}$ of Example~\ref{ex:MPCS} and $\boldsymbol{N}$ components' common shock representations.}
    \label{fig:five-nodes}
\end{figure}
\begin{table}[t]
\centering
\small 
\begin{tabular}{ll}
\toprule
Set $\mathcal{W}$ & Parameter $\gamma_{\mathcal{W}}$ \\ 
\midrule
$\{1\}$   & $\lambda_1(1 - \alpha_{(1,2)} \sqrt{\lambda_2/\lambda_1})$ \\[1ex]
$\{2\}$   & $\lambda_2(1 - \alpha_{(1,2)} \sqrt{\lambda_1/\lambda_2})(1 - \alpha_{(2,3)} \sqrt{\lambda_3/\lambda_2})$ \\[1ex]
$\{3\}$   & $\lambda_3(1 - \alpha_{(2,3)} \sqrt{\lambda_2/\lambda_3})(1 - \alpha_{(3,4)} \sqrt{\lambda_4/\lambda_3})(1 - \alpha_{(3,5)} \sqrt{\lambda_5/\lambda_3})$ \\[1ex]
$\{4\}$   & $\lambda_4(1 - \alpha_{(3,4)} \sqrt{\lambda_3/\lambda_4})$ \\[1ex]
$\{5\}$   & $\lambda_5(1 - \alpha_{(3,5)} \sqrt{\lambda_3/\lambda_5})$ \\[1ex]
$\{1,2\}$ & $\sqrt{\lambda_1 \lambda_2} \alpha_{(1,2)} (1 - \alpha_{(2,3)} \sqrt{\lambda_3/\lambda_2})$ \\[1ex]
$\{2,3\}$ & $\sqrt{\lambda_2 \lambda_3} \alpha_{(2,3)} (1 - \alpha_{(1,2)} \sqrt{\lambda_1/\lambda_2})(1 - \alpha_{(3,4)} \sqrt{\lambda_4/\lambda_3})(1 - \alpha_{(3,5)} \sqrt{\lambda_5/\lambda_3})$ \\[1ex]
$\{3,4\}$ & $\sqrt{\lambda_3 \lambda_4} \alpha_{(3,4)} (1 - \alpha_{(2,3)} \sqrt{\lambda_2/\lambda_3})(1 - \alpha_{(3,5)} \sqrt{\lambda_5/\lambda_3})$ \\[1ex]
$\{3,5\}$ & $\sqrt{\lambda_3 \lambda_5} \alpha_{(3,5)} (1 - \alpha_{(2,3)} \sqrt{\lambda_2/\lambda_3})(1 - \alpha_{(3,4)} \sqrt{\lambda_4/\lambda_3})$ \\
$\{1,2,3\}$   & $\sqrt{\lambda_1 \lambda_3} \alpha_{(1,2)} \alpha_{(2,3)} (1 - \alpha_{(3,4)} \sqrt{\lambda_4/\lambda_3})(1 - \alpha_{(3,5)} \sqrt{\lambda_5/\lambda_3})$ \\[1ex]
$\{2,3,4\}$   & $\sqrt{\lambda_2 \lambda_4} \alpha_{(2,3)} \alpha_{(3,4)} (1 - \alpha_{(1,2)} \sqrt{\lambda_1/\lambda_2})(1 - \alpha_{(3,5)} \sqrt{\lambda_5/\lambda_3})$ \\[1ex]
$\{2,3,5\}$   & $\sqrt{\lambda_2 \lambda_5} \alpha_{(2,3)} \alpha_{(3,5)} (1 - \alpha_{(1,2)} \sqrt{\lambda_1/\lambda_2})(1 - \alpha_{(3,4)} \sqrt{\lambda_4/\lambda_3})$ \\[1ex]
$\{3,4,5\}$   & $\sqrt{\lambda_4 \lambda_5} \alpha_{(3,4)} \alpha_{(3,5)} (1 - \alpha_{(2,3)} \sqrt{\lambda_2/\lambda_3})$ \\[1ex]
$\{1,2,3,4\}$ & $\sqrt{\lambda_1 \lambda_4} \alpha_{(1,2)} \alpha_{(2,3)} \alpha_{(3,4)} (1 - \alpha_{(3,5)} \sqrt{\lambda_5/\lambda_3})$ \\[1ex]
$\{1,2,3,5\}$ & $\sqrt{\lambda_1 \lambda_5} \alpha_{(1,2)} \alpha_{(2,3)} \alpha_{(3,5)} (1 - \alpha_{(3,4)} \sqrt{\lambda_4/\lambda_3})$ \\[1ex]
$\{2,3,4,5\}$ & $\sqrt{\lambda_2 \lambda_4 \lambda_5/\lambda_3} \alpha_{(2,3)} \alpha_{(3,4)} \alpha_{(3,5)} (1 - \alpha_{(1,2)} \sqrt{\lambda_1/\lambda_2})$ \\[1ex]
$\{1,2,3,4,5\}$ & $\sqrt{\lambda_1 \lambda_4 \lambda_5/\lambda_3} \alpha_{(1,2)} \alpha_{(2,3)} \alpha_{(3,4)} \alpha_{(3,5)}$ \\
\bottomrule
\end{tabular}
\caption{Parameters $\gamma_{\mathcal{W}}$ for each set $\mathcal{W}$ of vertices in Figure~\ref{fig:five-nodes}}
\label{tab:gamma-values}
\end{table}
\end{example}

Proposition~\ref{th:StoRepCommonShocks} makes clear the difference between $\mathbb{MPMRF}$ and the tree-structured multivariate Poisson distribution examined in \cite{kizildemir2017supermodular}. In the latter, there are only random variables $Y_{\mathcal{W}}$ from the representation in (\ref{eq:commonshockMPMRF}) for $\mathcal{W}\in\mathscr{P}(\mathcal{V};v)$ comprising two elements, given by the set of edges $\mathcal{E}$ of the graph. There are no shock random variables $Y_{\mathcal{W}}$ for $|\mathcal{W}|\geq 3$. As a consequence, the multivariate distribution does not exhibit the conditional independence relations from Definition~\ref{def:mrf} to render a MRF. 

While previous work has extended the multivariate Poisson distribution based on common shocks to higher dimensions, no method combines minimal parameters with the wide variety of dependence structures achievable by $\mathbb{MPMRF}$. For instance, \cite{schulz2021multivariate} generalize the bivariate Poisson model from \cite{genest2018new} to higher dimensions, requiring only $d+1$ parameters, but this approach imposes limitations on the correlation structure by restricting dependence to a single parameter. \cite{murphy2025multivariate} address this limitation with the multivariate Poisson distribution based on triangular comonotonic shocks, but requires  $d+d(d-1)/2 = \mathcal{O}(d^2)$ parameters, still computationally intensive in high-dimensional settings. The $\mathbb{MPMRF}$ family, by comparison, achieves complex dependence structures with only $2d-1$ parameters, scaling more efficiently at $\mathcal{O}(d)$.  This allows for convenient estimation in higher dimensions; further discussion is provided in Section~\ref{sect:data}.  The comonotonic shock approach in the above papers accommodates a broader range of correlations, but it cannot be expressed as a thinning operation. One could modify $\mathbb{MPMRF}$ by letting the construction in \eqref{eq:commonshockMPMRF} be in terms of comonotonic shocks, but would therefore be unable to reexpress it in terms of an iterative stochastic construction as in \eqref{eq:StoConstruct}. The latter is what enables efficient computational methods, as we discuss in the upcoming sections.

\section{Aggregate analysis of the portfolio}\label{sect:aggregate}
A risk model $\boldsymbol{X} = (X_v  =\sum_{j=1}^{N_v} B_{v,j}, \; v\in\mathcal{V})$ as defined in \eqref{eq:model} with $\boldsymbol{N}$ from Theorem~\ref{th:flexible}  benefits from analytical and computable expressions, even if the dimension $d =|\mathcal{V}|$ is high. The flexibility in choosing parameters $(\boldsymbol{\lambda},\boldsymbol{\alpha})\in\boldsymbol{\Lambda}$ and the underlying tree $\mathcal{T}$ provides a diversity of dependence structures.

The joint Laplace-Stieltjes transform (LST) of $\boldsymbol{X}$, denoted $\mathcal{L}_{\boldsymbol{X}}$, used to obtain the distribution of the aggregate claim amount for the portfolio, is given by
\begin{equation}
\mathcal{L}_{\boldsymbol{X}}(\boldsymbol{t}) = 
\mathbb{E}\left[\prod_{v \in \mathcal{V}} \mathrm{e}^{-t_v X_v} \right] =
\mathcal{P}_{\boldsymbol{N}}(\mathcal{L}_{B_1}(t_1), \dots, \mathcal{L}_{B_d}(t_d)) = \prod_{v\in\mathcal{V}} \mathrm{e}^{\zeta_{L_v}\left(\eta_{v}^{\mathcal{T}_{\!\!r}}(\pmb{\mathcal{L}}_{B_v}(\boldsymbol{t}_{v\mathrm{dsc}(v)}); \, \boldsymbol{\theta}_{\mathrm{dsc}(v)}) - 1\right)}, 
\label{eq:jointtls-vecteurX}
\end{equation}
for \mbox{$\boldsymbol{t} \in \mathbb{R}_+^d$,} with the sequence of joint pgfs \mbox{$\{\eta_{v}^{\mathcal{T}_{\!\!r}},\,v\in\mathcal{V}\}$} defined by the recursive relation in \eqref{eq:eta}, and with the vectors {$\pmb{\mathcal{L}}_{B_v}(\boldsymbol{t}_{v\mathrm{dsc}(v)}) = (\mathcal{L}_{B_{j}}(t_j), j \in \{v\} \cup \mathrm{dsc}(v))$} and {$\boldsymbol{\theta}_{\mathrm{dsc}(v)}= (\alphav{k}\sqrt{{\lambda_k}/{\lambda_{\mathrm{pa}(k)}}}, k \in \mathrm{dsc}(v))$}  for every $v\in\mathcal{V}$.

Given $\mathcal{L}_{S}(t) = \mathcal{L}_{\boldsymbol{X}}(t,\ldots,t) = \mathcal{P}_{\boldsymbol{N}}(\mathcal{L}_{B_{1}}(t), \ldots, \mathcal{L}_{B_{d}}(t))$, $t \geq 0$  (Theorem~1 of \citealp{wang1998aggregation}),  the joint LST in \eqref{eq:jointtls-vecteurX} leads to the following LST of the aggregate claim $S$:
\begin{equation}\label{eq:lst-S-Poisson}
\mathcal{L}_S(t) 
= \mathrm{e}^{ \sum_{v \in \mathcal{V}} \zeta_{L_v} \left( \sum_{v \in \mathcal{V}} \frac{\zeta_{L_v}}{ \sum_{v \in \mathcal{V}} \zeta_{L_v}} \eta_v^{\mathcal{T}_{\!\!r}} (\pmb{\mathcal{L}}_{B_v}(t \, \mathbf{1}_{|\{v\} \cup \mathrm{dsc}(v)|}); \, \boldsymbol{\theta}_{\mathrm{dsc}(v)}) - 1 \right)} = \mathrm{e}^{\lambda_S (\mathcal{L}_{C_S}(t) - 1)}, \quad  t \geq 0,
\end{equation}
implying that $S$ follows a compound Poisson distribution with primary mean parameter $\lambda_S = \sum_{v \in \mathcal{V}} \zeta_{L_v}$, and secondary LST given by \mbox{$\mathcal{L}_{C_S}(t) = \sum_{v \in \mathcal{V}}\left({\zeta_{L_v}}/{\lambda_S}\right)  \eta_v^{\mathcal{T}_{\!\!r}} (\pmb{\mathcal{L}}_{B_v}(t \,\mathbf{1}_{|\{v\} \cup \mathrm{dsc}(v)|}); \,\boldsymbol{\theta}_{\mathrm{dsc}(v)})$}, $t\geq 0$.

Generating realizations of $\boldsymbol{X}$ is straightforward, given that, the stochastic representation of $\boldsymbol{N}$ allows for an easily scalable sampling method. One generates realizations for each component of $\boldsymbol{N}$ successively; this is well-suited for high-dimensional contexts. In this vein, by adapting Algorithm~2 from \cite{cote2025tree} to accommodate flexible mean parameters, one can efficiently produce a realization of $\boldsymbol{X}$ by independently producing a realization of $\boldsymbol{N}$ and of the claim amounts.

For discrete claim amount random variables, values of the pmf of $S$ can directly be computed using the the fast Fourier transform (FFT) algorithm or Panjer's recursion. Algorithm~1 in Supplement~\ref{sup:algo-s} illustrates a procedure for our context using FFT. Discretization  methods or mixed Erlang approximations may be used for continuous claim amounts.
Mixed Erlang distributions are known to approximate any continuous positive distribution effectively; see for instance, \cite{tijms1994stochastic}.

\begin{remark}\label{rem:tls-mixederlang}
Let $\boldsymbol{X}$ be a multivariate compound Poisson with $\boldsymbol{N} \sim \text{MPMRF}(\boldsymbol{\lambda}, \boldsymbol{\alpha}, \mathcal{T})$.
We assume each ${B_v},\,v \in \mathcal{V}$,  follows a mixed Erlang distribution with parameters $(\boldsymbol{\pi}_v, \beta_v)$ where $\boldsymbol{\pi}_v = (\pi_{v,k}, k \in \mathbb{N}_1)$  is a vector of non-negative weight parameters, $\sum_{k=1}^n \pi_{v,k} = 1$, and $\beta_v > 0$. The LST of $S$ in  \eqref{eq:lst-S-Poisson} becomes
\begin{equation}\label{eq:tls-CS} 
\mathcal{L}_{S}(t) 
=  \exp\left\{\lambda_S \left( \sum_{v \in \mathcal{V}} \frac{\zeta_{L_v}}{ \lambda_S} \mathcal{P}_{\boldsymbol{G}^{\mathcal{T}_{\!\!r}}_v}\left\{ \left(\mathcal{P}_{\widetilde{K}_j}(\mathcal{L}_{B_{\max}}(t)),\, j\in \{v\}\cup\mathrm{dsc}(v) \right) \right\} \right) \right\} = \mathcal{P}_W(\mathcal{L}_{B_{\max}}(t)),
\end{equation}
for \mbox{$t \geq 0$}, where
$\widetilde{K}_j$ is as defined in Appendix~\ref{app:mixederlang}, $B_{\max}\sim$ Exp$(\max_{v\in\mathcal{V}} \beta_{v})$ and $\boldsymbol{G}^{\mathcal{T}_{\!\!r}}_v = (G^{\mathcal{T}_{\!\!r}}_{v,j}, j \in \{v\} \cup \mathrm{dsc}(v))$ is a vector of discrete random variables whose joint pgf is given by $\eta^{\mathcal{T}_{\!\!r}}_v(\boldsymbol{t}_{v\mathrm{dsc}(v)};\boldsymbol{\theta}_{\mathrm{dsc}(v)})$, $\boldsymbol{t}\in[-1,1]^d$,
as in Definition \ref{def:eta}. 
We recognize in \eqref{eq:tls-CS} the LST of a mixed Erlang distribution. 
\end{remark}

Hence, to perform computations regarding $S$, one must simply compute the pmf of $W$, relying on (\ref{eq:tls-CS}) and Algorithm~1. This is at the core of Algorithm~2 in Supplement~\ref{sup:algo-s}, which computes the cumulative distribution function (cdf) of $S$ under mixed Erlang claim amounts. With the distribution of $S$, one may compute the portfolio's required capital through different risk measures. This allows to complete our first risk management task regarding the quantification of the portfolio's risk. 

\section{Risk sharing}
\label{sect:allocation}
A subsequent risk management task involves the proper allocation of the portfolio's required capital to each component. This allocation can be performed ex ante; the allocation rule divides the overall portfolio's risk,  quantified by a risk measure, into shares for each component of $\boldsymbol{X}$ based on their respective levels of risk. When dealing with positively homogeneous risk measures, Euler's principle can be utilized to determine the value of these shares. A well-known example of such risk measures is the Tail Value-at-Risk (TVaR). For a random variable $Z$, the TVaR at confidence level $\kappa\in[0,1)$ is given by $    \mathrm{TVaR}_{\kappa}(Z) = \frac{1}{1 - \kappa} \int_{\kappa}^1 \text{VaR}_u(Z) \, \mathrm{d}u$, where $\mathrm{VaR}_u(Z) = \inf\{x\in\mathbb{R} : F_Z(x)\geq u\}$, and $u\in[0,1)$. Let us recall that mixed-Erlang distributions may approximate any continuous claim amount distributions; we showed in Section~\ref{sect:aggregate} that the pmf of $S$ can be exactly computed in this case. The results from \cite{cossette2012tvarbased} are thus readily applicable for computing the exact contribution to the TVaR based on Euler's rule. If claim amount distributions are discrete, additional manipulations are required to allocate risk. In such a case, the contribution of $X_v$, $v\in\mathcal{V}$, to the TVaR of $S$ under Euler's principle is given by
\begin{align}
 \mathcal{C}^{\mathrm{TVaR}}_{\kappa}&(X_v;\, S) = \frac{1}{1-\kappa}\left( \mathbb{E}[X_v\mathbbm{1}_{\{S>\mathrm{VaR}_{\kappa}(S)\}}] + \mathbb{E}[X_v|S=\mathrm{VaR}_{\kappa}(S)](F_S(\mathrm{VaR}_{\kappa}(S))-\kappa)\right) \notag\\
  &= \frac{1}{1-\kappa}\left( \mathbb{E}[X_v] -  \sum_{k=0}^{\mathrm{VaR}_{\kappa}(S)}\mathbb{E}[X_v\mathbbm{1}_{\{S=k\}}] + \frac{F_S(\mathrm{VaR}_{\kappa}(S))-\kappa}{p_S(\mathrm{VaR}_{\kappa}(S))} \mathbb{E}[X_v\mathbbm{1}_{\{S=\mathrm{VaR}_{\kappa}(S)\}}]\right),\label{eq:contribTVaR}
\end{align}
 for $\kappa \in [0,1)$; see, for instance, Section~2 in \cite{mausser2018long}. 

A risk modeler may prefer the covariance-based allocation rule instead of $C^{\mathrm{TVaR}}_\kappa(X_v, S)$ for $v \in \mathcal{V}$. The contribution amount of risk $X_v$, denoted by $C^{\mathrm{Cov}}_\kappa(X_v, S)$, is given by
\begin{equation*}
C^{\mathrm{Cov}}_\kappa(X_v, S) = \mathbb{E}[X_v] + \frac{\mathrm{Cov}(X_v, S)}{\mathrm{Var}(S)} \left( \mathrm{TVaR}_\kappa(S) - \mathbb{E}[S] \right), \quad v \in \mathcal{V}.
\end{equation*}

Both allocation rules ensure that the sum of the contributions
equals $\mathrm{TVaR}_\kappa(S)$, the required capital for the tail risk of the portfolio. Moreover, both rules satisfy Euler's principle. For detailed discussions on these allocation principles, we refer the reader to \cite{tasche1999risk} and \cite{mcneil2015quantitative}, and to \cite{hesselager2002risk} for further information on the covariance-based allocation rule.

To allocate the aggregate risk \textit{ex-post}, one may choose a fair risk-sharing rule. A \emph{risk-sharing rule} is a mapping that assigns to each participant a contribution $h_{v,d}(S)$ such that $\sum_{v=1}^d h_{v,d}(S) = S$. A rule is said to be \emph{fair} if it also satisfies $\mathbb{E}[h_{v,d}(S)] = \mathbb{E}[X_v] = \mu_v$ for all $v$, ensuring participants pay their expected loss on average. In the context of peer-to-peer insurance, for instance, risk-sharing rules serve to determine each participant's contribution to the pool \citep{denuit2022risk}.
 
Linear fair rules take the form $h_{v,d}^{\text{lin}}(S) = \mu_v + a_{v,d}(S - \mathbb{E}[S])$, where the coefficients $a_{v,d}$ satisfy $\sum_{v \in \mathcal{V}} a_{v,d} = 1$. Two notable examples include
the {proportional rule}, where $a_{v,d} = \mu_v / \mathbb{E}[S]$, allocating risk in proportion to expected losses; the {linear regression rule}, where $a_{v,d} = \mathrm{Cov}(X_v,S)/\mathrm{Var}(S)$, allocating deviations according to volatility. This rule minimizes the mean squared error $\mathbb{E}[(X_v - h_{v,d}(S))^2]$ among all linear fair rules.

The (nonlinear) {conditional mean risk sharing rule} \citep{denuit2012convex},  defined by 
$h_{v,d}^{\star}(S) = \mathbb{E}[X_v|S]$, minimizes $\mathbb{E}[(X_v - h_{v,d}(S))^2]$ over all measurable functions $h_{v,d}$ with finite variance. This rule is Pareto-optimal under risk aversion and does not rely on individual preference inclusion, making it particularly suitable for peer-to-peer insurance frameworks \citep{denuit2012convex}. For discrete distributions, we have $\mathbb{E}[X_v|S=k] = {\mathbb{E}[X_v\mathbbm{1}_{\{S=k\}}]}/{p_{S}(k)},$ $v\in\mathcal{V}, \; k\in\mathrm{supp}(S)$. 

\subsection{Computation of risk allocations}
A crucial component for calculating both $\mathcal{C}^{\mathrm{TVaR}}_{\kappa}(X_v;S)$ and $\mathbb{E}[X_v|S=k]$ is the expected allocation: $\mathbb{E}[X_v\mathbbm{1}_{\{S=k\}}]$, for $k\in\mathbb{N}$. The significance of expected allocations in the context of capital allocation is thoroughly discussed in \cite{blier2025efficient}. The authors introduce an ordinary generating function for expected allocations, which is defined as follows.

\begin{definition}[OGFEA]
    Consider a vector of discrete random variables $\boldsymbol{Z}=(Z_1,\ldots, Z_d)$ taking values in $\mathbb{N}^{d}$. The ordinary generating function of expected allocations (OGFEA) of $Z_v$, $v\in\{1,\ldots,d\}$, to the sum of components $\sum_{v=1}^d Z_v$ is given by $\mathcal{P}^{[v]}_{\sum_{v=1}^d Z_v}(t) = \sum_{k=0}^{\infty} \mathbb{E}[Z_v\mathbbm{1}_{\{\sum_{v=1}^d Z_v = k\}}] t^k,$ $t\in[-1,1].$
\end{definition}
The convenience of OGFEAs lies in the fact that information on expected allocations for all total outcomes is encapsulated within a single power series. We present the OGFEA for our model in the 
following theorem.

\begin{theorem}
\label{th:OGFEAX}
Consider the risk model in \eqref{eq:model}, where $\boldsymbol{N} = (N_v,\,v \in \mathcal{V}) \sim \text{MPMRF}(\boldsymbol{\lambda}, \boldsymbol{\alpha}, \mathcal{T})$, 
for $(\boldsymbol{\lambda},\boldsymbol{\alpha})\in\boldsymbol{\Lambda}$, and a tree $\mathcal{T}=(\mathcal{V},\mathcal{E})$. The OGFEA for $X_v$ to $S$ is given by
\begin{equation}
\mathcal{P}_S^{[v]}(t) = \lambda_{v} \, \mathbb{E}[B_v] \, \eta_{v}^{\mathcal{T}_{\!\!v}}\left(\boldsymbol{s}(t); \boldsymbol{\theta}\right) \,  \mathcal{P}_S(t), \quad t\in[-1,1],
\label{eq:OGFEA}
\end{equation}		
{where $\boldsymbol{s}(t) = (s_j(t),\, j\in\mathcal{V})$ is a vector with $s_v(t) = t\tfrac{\mathrm{d}}{\mathrm{dt}}\mathcal{P}_{B_v}(t)/\mathbb{E}[B_v]$ for $j = v$, and $s_j(t) = \mathcal{P}_{B_j}(t)$ for $j\in\mathcal{V}\backslash\{v\}$, $t\in[-1,1]$, and \mbox{$\boldsymbol{\theta}_{\mathrm{dsc}(v)} = (\alphav{k}\sqrt{{\lambda_k}/{\lambda_{\mathrm{pa}(k)}}}, k \in \mathrm{dsc}(v))$}.
}\end{theorem}

Let us highlight that $\eta_v^{\mathcal{T}_{\!\!v}}$ and $\theta_v^{\mathcal{T}_{\!\!v}}$ in \eqref{eq:OGFEA} are predicated on the tree rooted at vertex $v$ specifically. 
In addition to $\lambda_v\,\mathbb{E}[B_v]$, the other two factors in \eqref{eq:OGFEA} are pgfs. Their product can be interpreted as the sum of two independent random variables. Therefore, the coefficients of the OGFEA can be expressed in terms of the pmf of that sum. This is illustrated in the following corollary, which offers a stochastic interpretation of expected allocations.
\begin{corollary}
\label{cor:OGFEAstorep}
Consider the risk model in \eqref{eq:model}, where $\boldsymbol{N} = (N_v,\,v \in \mathcal{V}) \sim \text{MPMRF}(\boldsymbol{\lambda}, \boldsymbol{\alpha}, \mathcal{T})$, 
for $(\boldsymbol{\lambda},\boldsymbol{\alpha})\in\boldsymbol{\Lambda}$ and a tree $\mathcal{T}=(\mathcal{V},\mathcal{E})$.
Define $\boldsymbol{G}^{\mathcal{T}_{\!\!v}}=(G_w^{\mathcal{T}_{\!\!v}},\,w\in\mathcal{V})$ as a vector of random variables with joint pgf given by $\eta_{w}^{\mathcal{T}_{\!\!v}}(\boldsymbol{t}_{w\mathrm{dsc}(w)}
;\,\boldsymbol{\theta}_{\mathrm{dsc}(w)})$ as in~\eqref{eq:eta}, $\boldsymbol{t}\in[-1,1]^d$. Consider the random variable 
\begin{equation}
K^{(v)} = \sum_{i=1}^{G_v^{\mathcal{T}_{\!\!v}}} B^{*}_{v,i} + \sum_{j\in\mathrm{dsc}(v)}\sum_{i=1}^{G_{j}^{\mathcal{T}_{\!\!v}}} B_{j,i}, \label{eq:Kv-expalloc}
\end{equation}
where $B^*_v$ is the size-biased transform of $B_v$, that is $p_{B^*_v}(x) = \tfrac{x}{\mathbb{E}[B_v]} p_{B_v}(x)$, for $x\in\mathbb{R}$. 
The expected allocation of $X_v$ to $S$ for a total outcome $k\in\mathbb{N}$ is 
 \begin{equation}
     \mathbb{E}\left[X_v\mathbbm{1}_{\{S=k\}}\right] = \lambda_v \, \mathbb{E}[B_v] \, p_{K^{(v)}+S}(k),
     \label{eq:expealloc}
 \end{equation}
 with $K^{(v)}$ and $S$ mutually independent. 
\end{corollary}

{Since $\sum_{k = 0}^{\infty} p_{K^{(v)} + S}(k) = 1$, the summation of $\mathbb{E}[X_v\mathbbm{1}_{\{S =k\}}]$ over $k \in \mathbb{N}$ is equal to $\lambda_v \mathbb{E}[B_v]$, as expected.} Note that, in the case of independent compound Poisson, that is, $\boldsymbol{\alpha} = {0} \mathbf{1}_{|\mathcal{E}|}$, we have $K^{(v)}=B^{*}_{v}$ and thus recover results of Section~3.2 of \cite{denuit2019size}. The result in Corollary~\ref{cor:OGFEAstorep} allows for an explicit expression of contributions to the TVaR under Euler's rule. 
\begin{corollary}\label{cor:sizebiasedcontribution}
Consider the risk model in \eqref{eq:model}, where $\boldsymbol{N} = (N_v,\,v \in \mathcal{V}) \sim \text{MPMRF}(\boldsymbol{\lambda}, \boldsymbol{\alpha}, \mathcal{T})$, 
for $(\boldsymbol{\lambda},\boldsymbol{\alpha})\in\boldsymbol{\Lambda}$, and a tree $\mathcal{T}=(\mathcal{V},\mathcal{E})$. For $v\in\mathcal{V}$, the contribution of ${X_v}$ to the TVaR of $S$ under Euler's rule at confidence level $\kappa\in[0,1)$ is 
\begin{equation*}
    \mathcal{C}^{\mathrm{TVaR}}_{\kappa}(X_v;S) = \frac{\lambda_v \mathbb{E}[B_v]}{1-\kappa} \left(1- {F}_{K^{(v)}+S}(\mathrm{VaR}_{\kappa}(S)) + \frac{F_S(\mathrm{VaR}_{\kappa}(S)) - \kappa}{p_S(\mathrm{VaR}_{\kappa}(S))} p_{K^{(v)}+S}(\mathrm{VaR}_{\kappa}(S)) \right),
\end{equation*}
where the random variable $K^{(v)}$ admits the stochastic representation given in \eqref{eq:Kv-expalloc}.
\end{corollary}

Let $\mathbf{1}_d$ denote the $d$-dimensional vector of ones.
If $\boldsymbol{\lambda} = \lambda\,\boldsymbol{1}_d$, $\boldsymbol{\alpha} = \alpha\,\boldsymbol{1}_{|\mathcal{E}|}$ with $\lambda > 0$, and $\alpha \in [0,1]$, and all $B_v$ are identically distributed, the TVaR contributions in Corollary~\ref{cor:sizebiasedcontribution} follow the same ordering as in Proposition~1 of \cite{cote2024centrality}, bearing connection to the theory of network centrality. See Algorithm~3 in Supplement~\ref{sup:algo-alloc} for a procedure on computing expected allocations. It relies on the efficiency of the FFT algorithm and scales well to high-dimensional computations.

\subsection{Asymptotic results on linear risk sharing}
\label{sect:asymptotic}
In this section, we investigate the asymptotic behavior of linear risk-sharing under the MPMRF risk model. To this aim, we first formalize, in the following proposition, how the covariance between vertices varies as a function of their distance in the graph. 
\begin{proposition}\label{prop:cor}
Let $\boldsymbol{X}$ follow the risk model in \eqref{eq:model}, with $\boldsymbol{N} \sim \text{MPMRF}(\boldsymbol{\lambda}, \boldsymbol{\alpha}, \mathcal{T})$ as in Theorem~\ref{th:flexible}. For all distinct $v, w \in \mathcal{V}$ and letting $\prod_{e\in \emptyset} \alpha_e =1$ the covariance between $N_v$ and $N_w$ is given by
\begin{equation*}
\mathrm{Cov}(N_v, N_w) = \sqrt{\lambda_v\lambda_w} \!\! \prod_{e\in\mathrm{path}(v,w)} \alpha_e; ~~ \text{and thus,}~~ \mathrm{Cov}(X_v, X_w) = \mathbb{E}[B_v]\mathbb{E}[B_w] \sqrt{\lambda_v\lambda_w} \!\!\prod_{e\in\mathrm{path}(v,w)} \alpha_e.
\end{equation*}
\end{proposition}

The dependence between risks $X_v$ and $X_w$ decays exponentially with the length of the path between $v$ and $w$. Consequently, as trees grow in radius, correlations between distant vertices are naturally attenuated. We will show that this leads to the law of large numbers being applicable to the average risk $S/d$.

\begin{figure}[t]
\centering
\begin{subfigure}[b]{0.3\textwidth}
\centering
\includegraphics[width=0.85\textwidth]{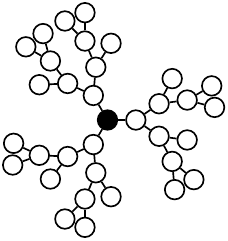}
\caption{Cayley tree $\mathcal{C}^{(3,3)}$}
\label{fig:cayley-3-4}
\end{subfigure}
\begin{subfigure}[b]{0.3\textwidth}
\centering
\includegraphics[width=\textwidth]{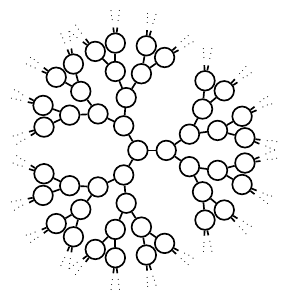}
\caption{Bethe lattice $\mathcal{B}^{(3)}$}
\label{fig:bethe-3}
\end{subfigure}
\caption{Illustration of a Cayley tree and a Bethe lattice, both of degree 3.}
\label{fig:cayley-4-5}
\end{figure}

To have regularly growing trees of infinite vertices, we use Bethe lattices, the infinite analog of the Cayley tree.
A Cayley tree $\mathcal{C}^{(\chi, \xi)}$ is a tree such that, with respect to some root $r\in\mathcal{V}$, each non-leaf vertex is connected to $\chi$ neighbors, $\chi \geq 2$, also referred to as such a tree's degree, and $\xi$ denotes the length of the shortest path between a root and a leaf. Figure~\ref{fig:cayley-4-5} provides an illustration of a Cayley tree $\mathcal{C}^{(3,3)}$.  More precisely, a Bethe lattice $\mathcal{B}^{(\chi)}$ with degree $\chi$ is obtained by letting the length of the shortest path between a root and a leaf $\xi$ tend to $\infty$.  These trees offer a natural framework for studying asymptotic regimes on infinite and expanding tree structures \citep{ostilli2012cayley} and allow to derive tractable results. These structures are of particular interest in computational statistics; see \cite{baxter2016exactly}.
Figure~\ref{fig:bethe-3} shows a portion of a Bethe lattice with degree $\chi = 3$.

 The convergence of $S/d$ defined on a Bethe lattice is explicited in the following proposition. 
For all $v \in \mathcal{V}$, let $\xi_v = |{\mathrm{path}(r,v)}|$ denote the distance between the root $r$ and vertex $v$. 
For any $\xi\in\mathbb{N}$, we write $\mathcal{T}^{[\xi]} = (\mathcal{V}^{[\xi]}, \mathcal{E}^{[\xi]})$ for a subtree of $\mathcal{T}$ made of all $v\in\mathcal{V}$ such that $\xi_v \leq \xi$, for some root $r$. Let $d^{[\xi]} = |{\mathcal{V}^{[\xi]}}|$.  Note that $\mathcal{V}^{[0]} = \{r\}$.

\begin{proposition}[Weak law of large numbers]
\label{th:LawLargeNumbers}
    Let $\{\mathcal{B}^{(\chi)[\xi]},\,\xi\in\mathbb{N}\}$ be a sequence of truncated Bethe lattices of degree $\chi>0$
    and \mbox{$\{\boldsymbol{X}^{[\xi]},\, \xi \in \mathbb{N}\}$} be a sequence of portfolios, each defined as in \eqref{eq:model}, where  $\boldsymbol{N}^{[\xi]} \sim \text{MPMRF}(\boldsymbol{\lambda},\boldsymbol{\alpha}, \mathcal{B}^{(\chi)[\xi]})$ and $\sup_{v\in\mathcal{V}}\mathbb{E}[B_v^2]<\infty$. Also assume $\boldsymbol{\alpha}$ and $\boldsymbol{\lambda}$ are uniformly upper bounded, meaning there exists $\lambda_{\mathrm{sup}} := \mathrm{sup}_{v \in \mathcal{V}} \lambda_v <  \infty$ and  $\alpha_{\mathrm{sup}} := \mathrm{sup}_{e \in \mathcal{E}}\alpha_e \in[0,1)$. Let $S^{[\xi]} = \sum_{v\in\mathcal{V}^{[\xi]}}X_{v}^{[\xi]}$, for every $\xi\in\mathbb{N}$. The sequence of random variables $\{W^{[\xi]}, \, \xi \in \mathbb{N}\} = \{\frac{1}{d^{[\xi]}}S^{[\xi]},\,\xi\in\mathbb{N}\}$ converges in probability to $\mathbb{E}[W^{[\xi]}] < \infty$
    as $\xi \to \infty$.
\end{proposition}

Proposition~\ref{th:LawLargeNumbers} implies that, regardless of the local dependence strength, the MPMRF compound distribution will seemingly lead, at the macroscopic level, to an average claim amount with the same behavior as in the independence case.  The following theorem shows that, as the number of participants in a portfolio grows, linear risk sharing rules converge in probability to the pure premium.
\begin{theorem}  \label{th:convlinRS}
Consider the setting of Proposition~\ref{th:LawLargeNumbers}.
For every $v \in \mathcal{V}$, if {$a_{v,d^{[\xi]}} =\mathcal{O}({1}/{d^{[\xi]}})$},
then $\lim_{\xi \rightarrow \infty} h_{v,d^{[\xi]}}^{\text{lin}}(S^{[\xi]})  =  {\mathbb{E}[{X_{v}^{[\xi]}}]}$ in probability.  
\end{theorem}

Encrypting the dependence structure on a Bethe lattice serves as a reference basis for the convergence results. The following corollary extends the results to general growing tree structures. Let $\mathrm{deg}(v)$ denote the degree of vertex $v$, that is the number of neighbors of $v$ in the graph.

\begin{corollary}
\label{cor:domBethe}
Let \mbox{$\{\boldsymbol{X}^{[\xi]},\, \xi \in \mathbb{N}\}$} be a sequence of portfolios, each defined as in \eqref{eq:model}, with $\boldsymbol{N}^{[\xi]}  \sim \text{MPMRF}(\boldsymbol{\lambda}, \boldsymbol{\alpha}, \mathcal{T}^{[\xi]})$, with $\sup_{v \in \mathcal{V}^{[\xi]}} \mathrm{deg}(v) = m < \infty$ for all $\xi\in\mathbb{N}$. Also assume $\boldsymbol{\alpha}$ and $\boldsymbol{\lambda}$ are uniformly upper bounded, meaning there exists $\lambda_{\mathrm{sup}} := \mathrm{sup}_{v \in \mathcal{V}} \lambda_v <  \infty$ and  $\alpha_{\mathrm{sup}} := \mathrm{sup}_{e \in \mathcal{E}}\alpha_e \in[0,1)$. As $\xi \to \infty$, the variance of the average claim amount $S^{[\xi]}/d^{(\xi)}$ is asymptotically upper bounded by that of $S^{*[\xi]}/d^{(\xi)}$, which is defined on $\mathcal{B}^{(m)}$, with parameters $\lambda_{\mathrm{sup}}$ for all vertices and $\alpha_{\mathrm{sup}}$ for all edges. 
\end{corollary}
This result shows that, in large and regularly growing trees with bounded degree, local dependence does not generate clustering strong enough to alter the aggregate behavior of risks, and thus, the latter is akin to that of an independent system for risk-sharing purposes. We illustrate an application of Corollary~\ref{cor:domBethe} in the following example. 
\begin{figure}[t]
        \begin{subfigure}{0.52\textwidth}
            \centering
            \resizebox{0.6\textwidth}{!}{    
        
    \begin{tikzpicture}[every node/.style={circle, draw, inner sep=2pt, minimum size=6mm}, scale = 0.6, transform shape]
    \tikzstyle{vertex}=[circle, draw=black, fill=white, minimum size=10pt, inner sep=0pt]

    \node[vertex, fill=black] (G-1) at (3,3.5) {};

    \node[vertex] (G-2) at (3,4.5) {};
    \node[vertex] (G-3) at (3,2.5) {};

    \node[vertex] (G-4) at (2,4.5) {};
    \node[vertex] (G-5) at (4,4.5) {};
    \node[vertex] (G-6) at (2,2.5) {};
    \node[vertex] (G-7) at (4,2.5) {};
    \foreach \label/\x/\y in {
        8/1/5, 9/1/4, 10/5/5, 11/5/4, 12/1/3, 13/1/2, 14/5/3, 15/5/2,
        16/0/5.3, 17/0/4.7, 18/0/4.3, 19/0/3.7, 20/6/5.3, 21/6/4.7,
        22/6/4.3, 23/6/3.7, 24/0/3.3, 25/0/2.7, 26/0/2.3, 27/0/1.7,
        28/6/3.3, 29/6/2.7, 30/6/2.3, 31/6/1.7}
        \node[vertex] (G-\label) at (\x,\y) {\makebox[0.35cm][c]{}};

    \foreach \from/\to in {
        1/2, 1/3, 2/4, 2/5, 3/6, 3/7, 4/8, 4/9, 5/10, 5/11, 6/12, 6/13,
        7/14, 7/15, 8/16, 8/17, 9/18, 9/19, 10/20, 10/21, 11/22, 11/23,
        12/24, 12/25, 13/26, 13/27, 14/28, 14/29, 15/30, 15/31}
        \draw (G-\from) -- (G-\to);
\end{tikzpicture}}
            \caption{Binary tree $\mathcal{T}_{\!\!\bullet}^{\text{binary}}$}
            \label{subfig:tree_binary}
        \end{subfigure}
        \hfill
                \begin{subfigure}{0.45\textwidth}
            \centering
            \resizebox{0.6\textwidth}{!}{\begin{tikzpicture}[every node/.style={circle, draw, inner sep=2pt, minimum size=6mm}, scale = 0.6, transform shape]
        \tikzstyle{vertex}=[circle, draw = black,fill=white,minimum size=10pt,inner sep=0pt]
        
\foreach \label in {2,3,4,5,6,7,8,9}
{
\draw  (0,0) -- ({(\label-2)*12}:2cm);
\node[vertex] (\label) at ({(\label-2)*12}:2cm) {\makebox[0.35cm][c]{\makebox[0.35cm][c]{}}};
        }
\foreach \label in {10,11,12,13,14,15,16,17,18,19,20,21,22,23,24,25,26,27,28,29,30,31}
{\draw  (0,0) -- ({(\label-2)*12}:2cm);
\node[vertex] (\label) at ({(\label-2)*12}:2cm) {\makebox[0.35cm][c]{}};
        }
\node[vertex, fill = black] (base) at (0,0) {};
    \end{tikzpicture}}
            \caption{Star tree $\mathcal{T}_{\!\!\bullet}^{\text{star}}$}
            \label{subfig:tree_star}
        \end{subfigure}
        \caption{Tree structures from Example~\ref{ex:convergence-fmp-sd}.}
        \label{fig:31tree}
\end{figure}
\begin{figure}[ht]
\centering
\begin{subfigure}[b]{0.45\textwidth}
\hspace{0.6cm}
\includegraphics[width=0.7\textwidth]{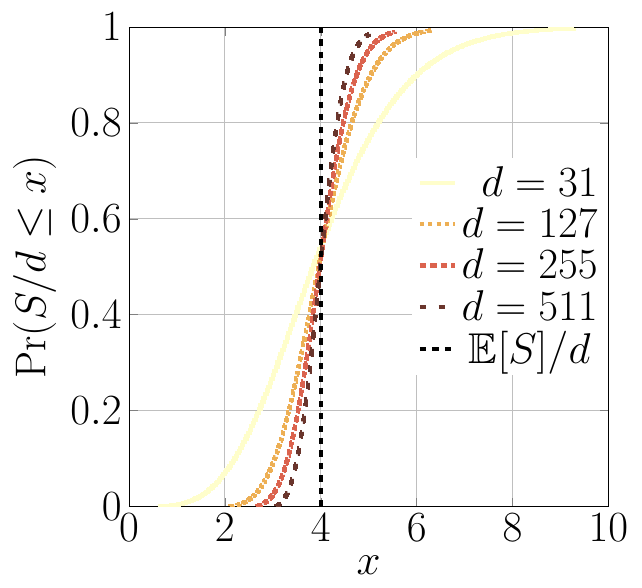}
\caption{Cdf of $S^{\text{binary}}/d$}
\label{subfig:cdf_binary}
\end{subfigure}
\begin{subfigure}[b]{0.45\textwidth}
\hspace{0.6cm}
\includegraphics[width=0.7\textwidth]{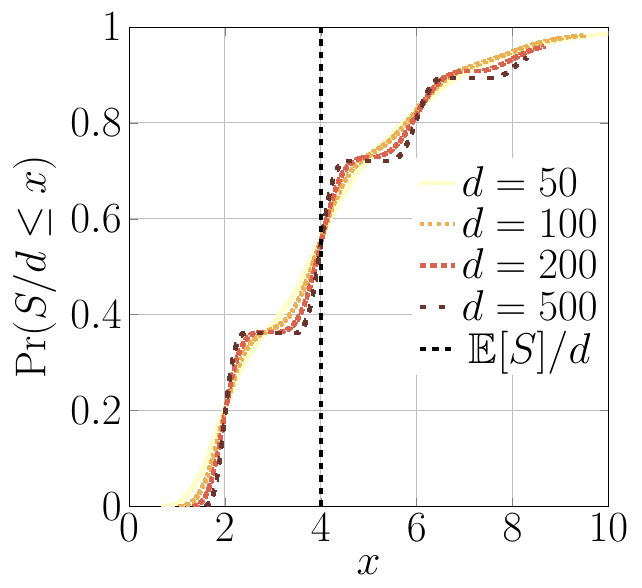}
\caption{Cdf of $S^{\text{star}}/d$}
\label{subfig:cdf_star}
\end{subfigure}
\caption{Cdfs of $S/d$ for Example~\ref{ex:convergence-fmp-sd}.}
\label{fig:asymptotic-s}
\end{figure}

\begin{example}[Asymptotic behavior of $S/d$]
\label{ex:convergence-fmp-sd}
Consider $\mathcal{T}^{\mathrm{binary}}$, a binary tree (as in Figure~\ref{subfig:tree_binary}), and a portfolio of risks $\boldsymbol{X}$ where $\boldsymbol{N}\sim \text{MPMRF}(\boldsymbol{\lambda}, \boldsymbol{\alpha}, \mathcal{T}^{\mathrm{binary}})$, with $\boldsymbol{\lambda} = \mathbf{1}_d$ and $\boldsymbol{\alpha} = 0.5 \mathbf{1}_{d-1}$, and $B_v \sim \text{NBinom}(2,1/3)$ such that $\mathbb{E}[B_v]=4$,
 We let $\mathcal{T}^{\mathrm{binary}}$ grow regularly by adding levels to the binary tree structure.
By Corollary~\ref{cor:domBethe}, the variance of $S/d$ is asymptotically dominated by that of $S^*/d$, defined on $\mathcal{B}^{(3)}$. Thus, by Proposition~\ref{th:LawLargeNumbers}, the law of large numbers applies. Figure~\ref{subfig:cdf_binary} shows the cdfs of $S/d$ for progressively larger versions of the trees: we see that the distribution of $S/d$ becomes increasingly concentrated around its theoretical mean $\mathbb{E}[S]/d = 4$.
Next, consider $\mathcal{T}^{\mathrm{star}}$, a star tree (as in Figure~\ref{subfig:tree_star}), and let it grow by adding vertices directly to the central node (the dark vertex in Figure~\ref{subfig:tree_star}). Since the resulting sequence of trees does not have uniformly bounded degree, we cannot apply Corollary~\ref{cor:domBethe}. Incidentally, the cdf of $S/d$, in Figure~\ref{subfig:cdf_star}, with $\boldsymbol{X}$ defined using the same parameters as before, suggests that the law of large numbers is not applicable in this case. 
\end{example}

Example~\ref{ex:convergence-fmp-sd} illustrates that $S/d$ may converge to a non-degenerate distribution as $d \to \infty$, which has nontrivial implications for risk pooling and diversification (see, e.g., \cite{denuit2021risk}).

\section{Estimation procedure}
\label{sec:estimation}
We describe the estimation procedure for the MPMRF frequency model, which involves constructing a correlation-based maximum spanning tree (MST) to estimate the tree structure and obtaining the maximum likelihood (ML) estimates for $\boldsymbol{\lambda}$ and $\boldsymbol{\alpha}$ given the resulting tree.

\subsection{Tree structure estimation}
To estimate the tree structure, we construct a MST from the matrix of empirical correlations; 
one may use Kruskal's algorithm for this construction \citep{kruskal1956shortest}. 

For Ising models with symmetric marginal distributions (i.e., with no external fields), \cite{bresler2020learning} established that the Chow-Liu dependence tree \citep{chow1968approximating}, which maximizes likelihood, coincides with the tree maximizing empirical correlations. This equivalence reflects the fact that, in that setting, pairwise correlation is a strictly monotone function of the Ising's underlying interaction parameter. Consequently, ordering edges by mutual information or by correlation yields the same dependence tree, and the corresponding approximation minimizes the Kullback-Leibler divergence to the true distribution.

An analogous simplification arises in our context. Although the mutual information of the bivariate Poisson distribution does not admit a closed-form expression, the approximation 
$I(X_1, X_2) \approx -\ln(1 - \rho_P(X_1,X_2)),$
is strictly increasing in $\rho_{P}(X_1,X_2)$ and very accurate for $\rho<0.7$ \citep{guerrero1995entropy}. Ranking edges by empirical mutual information is thus approximately equivalent to ranking them by empirical correlations. This monotonicity allows us to recover the Chow-Liu tree using correlations.

A further justification arises from the small sample sizes in our numerical illustration~: a sensitivity analysis (Supplement~\ref{sup:tree-sensitivity}) shows that the true tree structure is recovered more reliably if empirical correlations are used as edge weights. This behaviour is consistent with the design of the MPMRF model: edge-wise dependence is parameterized through correlations. We theoretically investigate the sample size required to recover the Chow-Liu MST in a separate working paper.

\subsection{MPMRF maximum likelihood estimators}
We estimate the frequency model parameters $(\boldsymbol{\lambda}, \boldsymbol{\alpha})$ by maximizing the log-likelihood derived from Proposition~\ref{th:propN}\ref{item:pmfN}, under the constraints of Theorem~\ref{th:flexible}, using numerical optimization. The maximization may be carried out using base R \texttt{optim} quasi-Newton BFGS algorithm, with parameter bounds enforced through a scaled logit reparametrization and a logarithmic reparametrization for the dependence parameters and marginal mean parameters, respectively. 
The structure of $\boldsymbol{N}$ facilitates the ML estimation task: its stochastic construction separates marginal and dependence parameters, and the joint pmf \eqref{eq:jointpmf-vecteurN} factorizes over cliques of the tree (pairs of connected vertices). As a result, ML estimation naturally aligns with the sequential procedure frequently applied in copula modelling, itself a specific instance of composite likelihood \citep{varin2011overview}. Initialization exploits the separation between frequency parameters and dependence parameters in the likelihood. Since the marginal intensities can be estimated independently of the dependence structure, the frequency parameters $\boldsymbol{\lambda}$ are first obtained by maximizing the marginal likelihood. Conditional on these estimates, the dependence parameters $\boldsymbol{\alpha}$ are initialized through sequential bivariate likelihood maximization. These initial values can serve as starting points for the joint maximum-likelihood estimation of the full model. Since the likelihood function is invariant to the choice of root, the optimization results do not depend on a selected root vertex.

\section{Data illustration}
\label{sect:data}
We examine the applicability and performance of the MPMRF model by analyzing yearly rainfall measurements (in millimeters) from extreme events collected at weather stations in Nova Scotia. The data was sourced from the archives of Environment and Climate Change Canada (ECCC) using the \texttt{weathercan} package \citep{lazerte2018weathercan}. Our analysis is inspired by the work of \cite{murphy2025multivariate}, who proposed a multivariate Poisson distribution based on a triangular comonotonic shock construction (MPTCS) and applied it to annual extreme event count data from three weather stations. We compare the performance of the MPMRF frequency model, as proposed in Theorem~\ref{th:flexible}, with the MPTCS model. Both models feature marginal Poisson distributions and account for positive dependence. Additionally, we extend our analysis to jointly model the frequency and severity of rainfall events across a portfolio of ten weather stations using the MPMRF risk model.

To conduct our analysis, we work with three datasets.
The first dataset, sourced from \cite{murphy2025multivariate}, consists of 71 yearly trivariate observations of extreme events recorded at three stations covering the period from 1919 to 2000.  We create the second and third datasets based on three other stations and ten stations, respectively, with data spanning from 1949 to 2000. We provide in Table~\ref{tab:weatherstations} the complete list of weather stations included in this study. Note that some stations were moved during these periods, explaining their two climate identification numbers. The third dataset also records the rainfall amounts associated with each extreme event. 

\subsection{Datasets 2 and 3 construction}
We use the \textit{peak-over-threshold} (POT) approach to model extreme events; see \cite{embrechts1997modelling}, Chapter~6. By Theorem~7.1 of \cite{coles2001classical}, counts of extreme events, over a sufficiently high threshold $u$, are Poisson distributed.  The excesses $Y = X-u | X > u$ follow a Generalized Pareto Distribution (GPD) \citep{chavez2005generalized}, whose cdf is given by
\begin{equation*}
F_Y(y)=
\begin{cases}
1 - \left(1+{\xi y}/{\sigma}\right)^{-1/\xi}, & \xi \neq 0,\\
1 - \mathrm{e}^{-{y}/{\sigma}}, & \xi=0,
\end{cases}
\end{equation*}
where $\sigma >0$, and $\xi \in \mathbb{R}$ denote the scale and shape parameters of the GPD.

The construction of Datasets~2 and~3 requires selecting a threshold such that exceedances' frequency and their corresponding excesses follow a Poisson and a GPD distribution, respectively, for each station. Following \cite{thiombiano2017nonstationary}, we relied on the mean-excess plot and the stability of the estimated GPD shape parameter to guide threshold selection. A chi-squared goodness-of-fit test was then applied to the excesses above the chosen threshold. We grouped exceedance-count classes so that all expected frequencies were at least 5.  For all stations, the null hypothesis of a GPD distribution could not be rejected at the 20\% level. We also performed Anderson-Darling tests, which confirm these conclusions. Because the threshold determines both the number of POT events and the validity of modelling assumptions, we also verified the independence of exceedances. A one-day declustering rule was applied to daily precipitation data, after which the autocorrelation function and Ljung-Box test ($p$-value of 5\%) confirmed negligible residual dependence. For the frequency component, independence of annual exceedance counts was examined using the same diagnostics, and a chi-squared goodness-of-fit test showed that the null hypothesis of a Poisson distribution could not be rejected at the 30\% level. The selected thresholds, together with the distribution of cluster sizes for each station used in Dataset~2 and~3, are reported in Table~\ref{tab:cluster_sizes}. The complete dataset construction procedure, along with diagnostic plots and statistical tests are provided in Supplement~\ref{sup:datasets}.

\begin{figure}[t]
\begin{minipage}[t]{0.4\textwidth}
  \centering
  \small
  \captionof{table}
  {
  Vertex numbers, meteorological stations, and climate ID suffixes, and the datasets in which they are used.
  }
  \label{tab:weatherstations}
  \begin{threeparttable}
    \begin{tabular}{clrr}
      \toprule
      ID & Station            & ID suffix\tnote{a}   & Set\\
      \midrule
      1 & Baddeck             & 0300, 0301 &3           \\
      2  & Upper Stewiacke     & 6200 &3                 \\
      3  & Mount Uniacke       & 3600 &2,3                 \\
      4  & Salmon Hole         & 5000 &2,3                 \\
      5  & St Margaret's Bay   & 4800 &3                 \\
      6  & Greenwood A   & 2000 &3                 \\
      7 & Shearwater A       & 5090 &3                 \\
      8  & Springfield         & 5200 &1,3                 \\
      9 & LiverPool Big Falls & 3001, 3100 &2,3           \\
      10 & Yarmouth            & 6490, 6500 &3           \\
      11 & Annapolis Royal  &  0100 &1 \\
      12 & Kentville CDA    & 2800, 2810 &1 \\
      \bottomrule
    \end{tabular}
    \begin{tablenotes}
      \footnotesize
      \item[a] All full station IDs are of the form \texttt{820xxxx}.
    \end{tablenotes}
  \end{threeparttable}
\end{minipage}
\hfill
\begin{minipage}[t]{0.52\textwidth}
\small
\captionof{table}{Extreme precipitation events threshold $u$ and distribution of cluster sizes (in days) after declustering.\\
}
\label{tab:cluster_sizes}
\centering
\begin{threeparttable}
\begin{tabular}{lrrrr|r}
\toprule
ID & $u$ & 1 day & 2 days & 3 days & Total \\
\midrule
1 & 37.6 & 143 &   6 & 0 & 149 \\ 
  2 & 23.9 & 389 &  18 &  2 & 409 \\ 
  3 & 34 & 289 &  12 &  0 & 301 \\ 
  4 & 33 & 236 &  11 &   1 & 248 \\ 
  5 & 31 & 282 &  12 &  0 & 294 \\ 
  6 & 22.8   & 284 &  13 &  1 & 298 \\
  7 & 30.7 & 322 &  10 &   1 & 333 \\ 
  8 & 27.9 & 345 &  20 &  0 & 365 \\ 
  9 & 31.2 & 360 &  14 &   1 & 375 \\ 
  10 & 24.6 & 428 &  15 &   2 & 445 \\ 
\bottomrule
\end{tabular}
\end{threeparttable}
\end{minipage}
\end{figure}

\subsection{Frequency models comparison}\label{subsec:frequency}
We compare the performance of the MPMRF and MPTCS frequency models for the three-station datasets.  The ML estimates of the frequency means and the estimated Pearson correlations for both datasets and models are reported in Tables~\ref{tab:lambda_compare_all} and \ref{tab:rho_compare_all}.  For both datasets, the means of the MPMRF marginal distributions are exactly the empirical means, but not for MPTCS; this is because MPMRF is able to dissociate marginal and dependence modelling.  In terms of dependence, for Dataset~1, the MPMRF model yields correlations that are closer to the empirical values for the edges of its tree,  given by $(2)$--$(1)$--$(3)$. In contrast, the MPTCS model performs better for the pair $(2,3)$. For Dataset~2, the inferred dependence structure forms the tree $(1)$--$(2)$--$(3)$, with MPMRF outperforming MPTCS on the edge $(1,2)$.  MPTCS is slightly better on edge $(2,3)$. However, MPTCS is better at capturing the correlation within $(1,3)$. This highlights a structural limitation of the MPMRF model: correlations between non-adjacent vertices necessarily factorize along the unique path of the tree, which restricts the maximal dependence the model can reproduce. The tree structure identifies the most influential relationships within the complete network.

\begin{table}[t]
\centering
\begin{minipage}[t]{0.48\textwidth}
\centering
\small
\begin{threeparttable}
\caption{Event counts estimates for Datasets 1 and 2.\\ \phantom{a}}
\label{tab:lambda_compare_all}
\begin{tabular}{llrrrr}
\toprule
Dataset & ID & Station & $\hat{\lambda}^{E}$ & $\hat{\lambda}^{A}$ & $\hat{\lambda}^{B}$ \\
\midrule
1 & 1 & Annapolis Royal     & 7.37 & 7.37 & 7.36 \\
  & 2 & Springfield         & 13.41 & 13.41 & 13.37 \\
  & 3 & Kentville CDA       & 10.66 & 10.66 & 10.61 \\
\midrule
2 & 1 & Liverpool Big Falls &  8.72 & 8.72 & 8.72 \\ 
  & 2 & Mount Uniacke       & 7.00 & 7.00 & 7.01\\ 
  & 3 & Salmon Hole         & 5.77 & 5.77 & 5.73 \\
\bottomrule
\end{tabular}
\begin{tablenotes}
\footnotesize
\item Empirical (E), MPMRF (A), \cite{murphy2025multivariate} (B).
\end{tablenotes}
\end{threeparttable}
\end{minipage}
\hfill
\begin{minipage}[t]{0.35\textwidth}
\centering
\small
\caption{Pairwise Pearson correlations for Datasets 1 and 2.}
\label{tab:rho_compare_all}
\begin{threeparttable}
\begin{tabular}{lrrr}
\toprule
Pair & $\hat{\rho}_P^E$ & $\hat{\rho}_P^A$ & $\hat{\rho}_P^B$ \\
\midrule
(1,2) & 0.625 & 0.585 & 0.533 \\
  (1,3) & 0.570 & 0.569 & 0.548 \\
  (2,3) & 0.494 & 0.333 & 0.517 \\
\midrule
(1,2) & 0.604 & 0.541 & 0.557 \\
  (1,3) & 0.460 & 0.305 &  0.368 \\
  (2,3) & 0.627 & 0.564 & 0.406 \\
\bottomrule
\end{tabular}
\begin{tablenotes}
\footnotesize
\item Empirical (E), MPMRF (A), \cite{murphy2025multivariate} (B).
\end{tablenotes}
\end{threeparttable}
\end{minipage}
\end{table}

To compare the models formally, we use the Bayesian Information Criterion (BIC) and the corrected Akaike Information Criterion (AICc). The AICc adjusts the traditional AIC for small-sample bias, which is particularly relevant given our relatively small sample sizes \citep{hurvich1989regression}. It is defined as $\mathrm{AICc} = \mathrm{AIC} + {2k(k+1)}/{(n - k - 1)}$. Table~\ref{tab:aicbic_all} presents the BIC and AICc values for each model across both datasets. For Dataset 1, the MPTCS model shows lower values for both criteria, indicating a better statistical fit. In contrast, for Dataset 2, the MPMRF model achieves lower values. These findings suggest that neither model consistently outperforms the other across different trivariate datasets.

\begin{table}[t]
\centering
\caption{Model comparison using BIC and AICc criteria for Datasets 1 and 2.}
\begin{tabular}{llrr}
\toprule
Dataset & Criterion & MPMRF model & \cite{murphy2025multivariate} \\
\midrule
1 & BIC  & 1063.07 & \textbf{1058.08} \\
 & AICc & 1052.68 & \textbf{1045.82} \\
\midrule
2 & BIC  &  \textbf{622.51} &  626.12 \\
 & AICc &  \textbf{615.33} &  617.89 \\
\bottomrule
\end{tabular}
\label{tab:aicbic_all}
\end{table}

In terms of computational performance, the MPTCS model encounters challenges in parameter estimation as the dimension increases. As discussed in Section~4 of \cite{murphy2025multivariate}, non-convergence may occur, particularly in scenarios with smaller sample sizes. To improve stability, the authors use a parameter grid search to initialize the sequential likelihood estimation and recommend employing multiple parameter initializations. For the three-dimensional MPTCS estimation in Dataset 2, we followed this strategy by generating 1000 random initializations of the free parameters and selecting the configuration that attained the highest sequential likelihood value. The runtime was approximately 100 seconds. Starting from the selected initialization, the two-step ML procedure converges in roughly 18 seconds, whereas the full ML estimation requires about 32 seconds. While this grid-search approach is feasible in lower dimensions, it becomes increasingly complex in higher dimensions. For example, achieving estimation at a dimension of $d = 10$ is unfeasible on a personal computer.

In contrast, both the sequential likelihood estimation and the ML estimation of the MPMRF model on Dataset 2 require less than one second. The MPMRF frequency model scales well with dimension and retains interpretability in its dependence structure, as demonstrated in the subsequent risk model analysis.

\subsection{MPMRF risk model on the 10-station dataset}
\label{subsec:analysis}

To showcase the applicability of the MPMRF model for risk modelling in greater dimensions, we now turn to Dataset~3. We fit a risk model as in \eqref{eq:model} for the rain excesses (not just the frequency of exceedances). For the frequency model, we compute the parameters' estimates using the  procedure outlined in Section~\ref{sec:estimation}; severities are modeled separately, given their independence with the frequency. The total runtime is less than two seconds. 

For the marginal distributions of $\boldsymbol{X}$, the values of the ML estimates are reported in Table~\ref{tab:station_parameters}.  For the severity, ML estimation were produced using the \texttt{QRM} package of  \cite{mcneil2015quantitative}. For stations where the profile likelihood 95\% confidence interval for the shape parameter $\xi$ included zero, we set $\hat{\xi}=0$ and fitted an exponential distribution, following the principle of parsimony. Salmon Hole (ID~4) and Liverpool Big Falls (ID~9) exhibit high severity means and variances, making them significant from a risk modelling perspective. For diagnostics on the marginal fits, see Supplement~\ref{sup:datasets}.

\begin{figure}[t]
\centering
\begin{minipage}{0.5\textwidth}
    \centering
        \includegraphics[width=\linewidth]{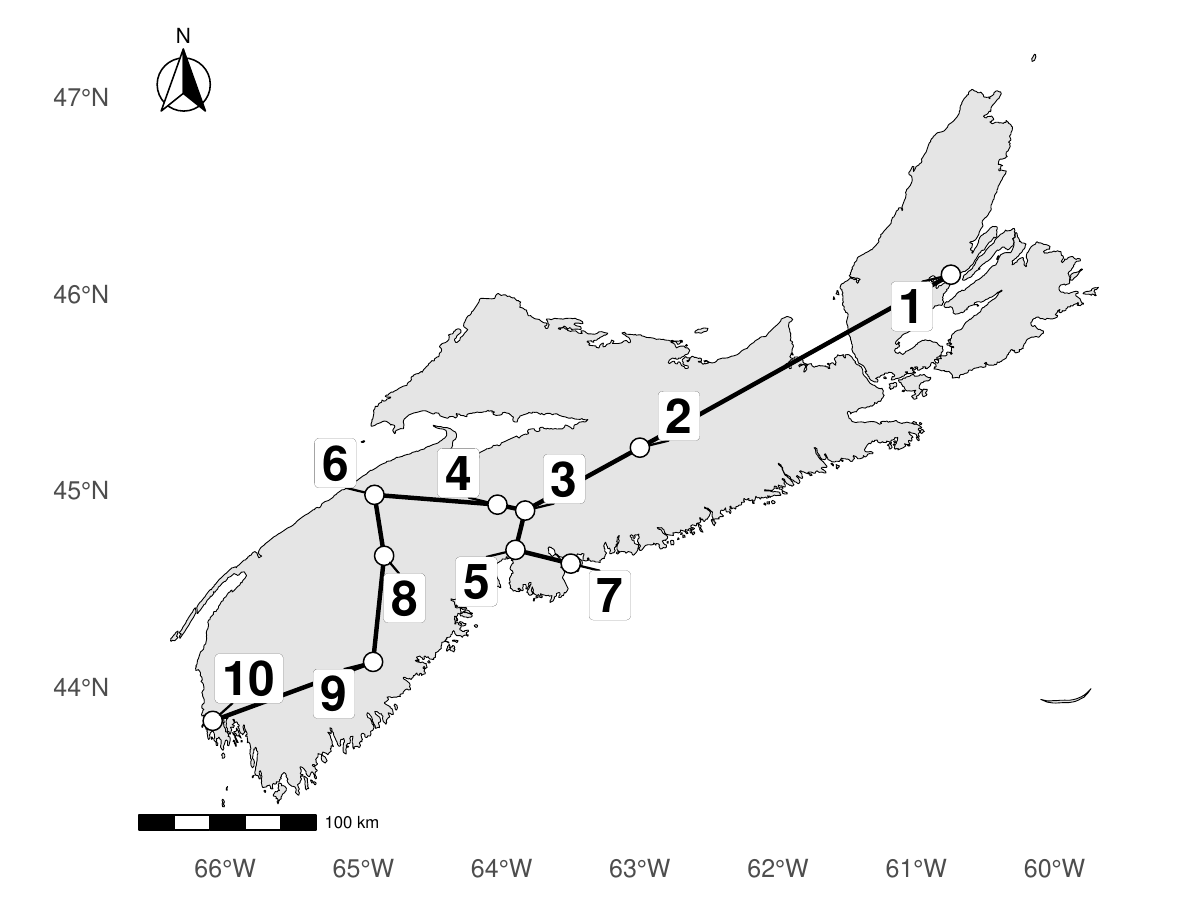} 
\end{minipage}
\begin{minipage}{0.4\textwidth}
    \centering
    \begin{tabular}{cl}
        \toprule
        Vertex & Station Name \\
        \midrule
        1 & Baddeck \\
        2 & Upper Stewiacke \\
        3 & Mount Uniacke \\
        4 & Salmon Hole \\
        5 & St Margaret's Bay \\
        6 & Greenwood A \\
        7 & Shearwater A \\
        8 & Springfield \\
        9 & Liverpool Big Falls \\
        10 &  Yarmouth \\
        \bottomrule
    \end{tabular}
\end{minipage}
\captionof{figure}{Correlation-based maximum spanning tree of 10 meteorological stations in Nova Scotia}
    \label{fig:mst}
\end{figure}

\begin{table}[t]
\centering
\caption{Parameter estimates for the frequency, severity and dependence structure. Bootstrap standard deviations on 1000 samples are provided for $\boldsymbol{\lambda}$ and $\boldsymbol{\alpha}$.}
\label{tab:all_parameters}
\begin{subtable}[t]{0.59\textwidth}
\small
\centering
\subcaption{Estimated frequency and severity parameters with characteristics.}
\begin{tabular}{crrrrrr}
  \toprule
    & \multicolumn{1}{c}{Frequency} & \multicolumn{3}{c}{Severity} \\
  \cmidrule(lr){2-2} \cmidrule(lr){3-6}
  ID & $\hat{\lambda}$ (sd) &  $\hat{\sigma}$ & $\hat{\xi}$ 
  & $\widehat{E}[B_v]$ & $\widehat{\mathrm{Var}}(B_v)$ \\ 
  \midrule
1 & 3.47 (0.31) & 12.85 & 0 & \textbf{50.45} & 165.12 \\ 
  2 & 9.51 (0.48) & 11.16 & 0 & 35.06 & 124.55 \\ 
  3 & 7.00 (0.46) & 13.51 & 0 & 47.51 & 182.52 \\ 
  4 & 5.77 (0.41) & 11.79 & 0.19 & \textbf{47.56} & \textbf{341.72} \\ 
  5 & 6.84 (0.44) & 13.18 & 0 & 44.18 & 173.71 \\ 
  6 & 6.93 (0.43) & 10.44 & 0.12 & 37.26 & 185.19 \\
  7 & 7.67 (0.48) & 13.94 & -0.08 & 43.60 & 143.62 \\ 
  8 & 8.49 (0.50) & 13.68 & 0 & 41.58 & 187.14 \\ 
  9 & 8.72 (0.44) & 15.05 & 0 & \textbf{46.25} & \textbf{226.50} \\ 
  10 & \textbf{10.35} (0.46) & 10.84 & 0.18 & 37.82 & \textbf{273.06} \\ 
\bottomrule
\end{tabular}
\label{tab:station_parameters}
\end{subtable}
\hfill
\begin{subtable}[t]{0.38\textwidth}
\centering
\subcaption{Estimated dependence parameters.}
\small
\begin{tabular}{crrr}
\toprule
$(u, v)$ & $\hat{\alpha}_{(u,v)}$ (sd) & ${\alpha^{\max}_{(u,v)}}$ & $\frac{\hat{\alpha}_{(u,v)}}{\alpha^{\max}_{(u,v)}}$\\
\midrule
(8,9)  &  0.625 (0.08)  &  0.987  &  0.633  \\
(2,3)  &  0.622 (0.08)  &  0.858  &  0.725  \\
(4,6)  &  0.579 (0.09)  &  0.912  &  0.635  \\
(3,5)  &  0.554 (0.08)  &  0.988  &  0.561  \\
(3,4)  &  0.564 (0.08)  &  0.908  &  0.621  \\
(6,8)  &  0.586 (0.10)  &  0.904  &  0.649  \\
(5,7)  &  0.488 (0.08)  &  0.944  &  0.517  \\
(9,10)  &  0.549 (0.08)  &  0.918  &  0.598  \\
(1,2)  &  0.512 (0.08)  &  0.604  &  0.849  \\
\bottomrule
\end{tabular}
\label{tab:dependencemles}
\end{subtable}
\end{table}

\begin{figure}[t]
\centering
\begin{subfigure}{0.45\textwidth}
    \centering
    \includegraphics[width=\textwidth]{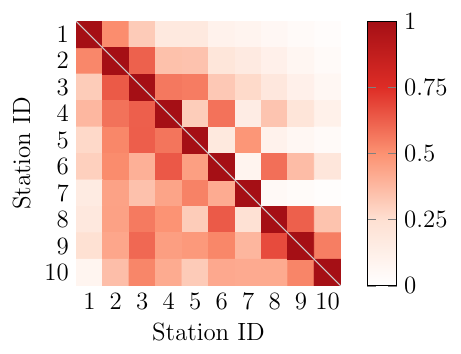}
    \subcaption{Extreme-event-count correlations}
    \label{fig:cor_count}
\end{subfigure}
\begin{subfigure}{0.45\textwidth}
    \centering
    \includegraphics[width=\textwidth]{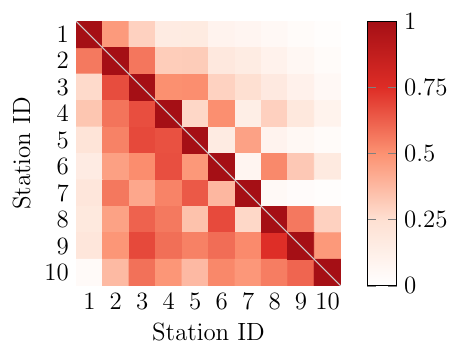}
    \subcaption{Extreme-event random variable correlations}
    \label{fig:cor_x}
\end{subfigure}
\hfill
\caption{Correlation heatmaps of yearly extreme precipitation events counts (left) and amounts (right) comparing empirical (lower triangle) and theoretical (upper triangle).}
\label{fig:correlation_heatmaps}
\end{figure}

Regarding dependence modelling, the MST constructed from the empirical correlations is diplayed in Figure~\ref{fig:mst}. Dependence parameters are reported in Table~\ref{tab:dependencemles} following the ordering induced by the MST.  Column~4 of Table~\ref{tab:dependencemles} reports, for each edge, the ratio of the estimated dependence parameter to its corresponding maximal admissible value, from the constraint $\alphavj{u}{v} \in (0, \alphavj{u}{v}^{\max}]$, 
with $\alphavj{u}{v}^{\max}:=\min(\sqrt{\lambda_u/\lambda_v}, \sqrt{\lambda_v/\lambda_u})$, $(u,v)\in\mathcal{E}$. We note that, for most edge combinations, the maximal admissible value largely exceeds the ML estimate. In Figure~\ref{fig:cor_count}, we present the correlation for the frequency, with empirical correlations (lower triangle) and  model-induced correlations (upper triangle). Figure~\ref{fig:cor_x} shows the analogous comparison for annual excesses $\boldsymbol{X}$. Under the independence assumption for claim severities in the MPMRF risk model, the correlations ratio between $\boldsymbol{X}$ and $\boldsymbol{N}$ is $\rho_{\boldsymbol{X}}(X_u,X_v)/\rho_{\boldsymbol{N}}(N_u,N_v) = \mathbb{E}[B_u]\mathbb{E}[B_v]/\sqrt{\mathbb{E}[B_u^2]\mathbb{E}[B_v^2]}$ for vertices $u,v \in \mathcal{V}$, which ranges from 0.8541 to 0.9344. The comparison of theoretical (lower triangle) and empirical (upper triangle) pairwise correlations in Figure~\ref{fig:cor_count} illustrates the effect of path-based correlations on a tree structure. For both heatmaps, we observe clearly how the tree-structure shows through the dependence captured by the model: close to the diagonal, where most edges lie, the model's correlations are very similar to the empirical ones. However, as paths between components lengthens, the model cannot capture as much dependence due to the correlation decay feature of tree-structured MRFs: correlations between non-adjacent vertices are forced to factorize multiplicatively along the unique path of the tree.  

Despite the cost of reduced flexibility in capturing long-range or higher-order interactions, the tree structure confers substantial analytical tractability, enabling closed-form likelihood evaluation, scalable inference, and efficient computation of portfolio-level risk measures. It is also interpretable: Figure~\ref{fig:mst} confirms the nontrivial spatial-dependence channels through which yearly extreme events co-vary. 

We next proceed to perform the two risk management tasks discussed in Sections~\ref{sect:aggregate} and~\ref{sect:allocation} for the obtained MPMRF model. The first risk management task is to assess the distribution of $S$ and measure the risk. We proceed using discretization methods.Let $\widetilde{B}$ be the discretized version of a continuous severity random variable $B$. If $B \sim \mathrm{GPD}(\xi, \sigma; u)$, then $\widetilde{B}$ follows a discretized generalized Pareto distribution (DGPD) with parameters $(\xi, \sigma; u)$, with 
$
p_{\widetilde{B}}(x) = \overline{F}_{B}(x h) - \overline{F}_{B}((x + 1)h)$,
for every
$x \in \{u + h: h\in\mathbb{N}\}
$
where $\overline{F}$ denotes the survival function  and $h$ is the discretization step. For the use of the DGPD in a practical context, see \cite{prieto2014modelling}. We use this specification for the severity component in our portfolio analysis with $h = 0.1$, as the dataset's total rainfall is reported on a decimal scale. Using the FFT algorithm, we obtain the distribution of the discretized aggregate loss, denoted $\widetilde{S}$.  Table~\ref{tab:tvar_mean_var} summarizes key portfolio-level statistics and compares them with those obtained using the assumption of independent compound Poisson distributions. 
{The marked differences illustrate the critical role of dependence in the frequency component for an accurate assessment of portfolios' risks.}

\begin{table}[t]
\centering
\caption{Characteristics and risk measures for $\widetilde{S}$, with comparison to the independent case, $\widetilde{S}^{\perp\!\!\!\perp}$. Values are rounded to the nearest whole number.}
\label{tab:tvar_mean_var}
\small
\begin{tabular}{lcccccccc}
\toprule
& \multicolumn{3}{c}{{Characteristics}} & & \multicolumn{4}{c}{{TVaR$_\kappa(Z)$}} \\
\cmidrule{2-4} \cmidrule{6-9}
& $\mathbb{E}[Z]$ & $\text{Var}(Z)$ & CV & & $\kappa=0.80$ & $\kappa=0.90$ & $\kappa=0.95$ & $\kappa=0. 99$ \\
\midrule
$Z = \widetilde{S}^{\perp\!\!\!\perp}$ & 3155 & 149\;798 & 0.12 & & 3707 & 3854 & 3984 & 4243 \\
$Z = \widetilde{S}$ & 3155 & 442\;542 & 0.21 & & 
4124 &  4396 & 4639 & 5133 \\
\bottomrule
\end{tabular}
\end{table}

\begin{figure}[t]
    \centering
    \begin{subfigure}[t]{0.48\textwidth}
        \centering
        \includegraphics[width=0.75\linewidth]{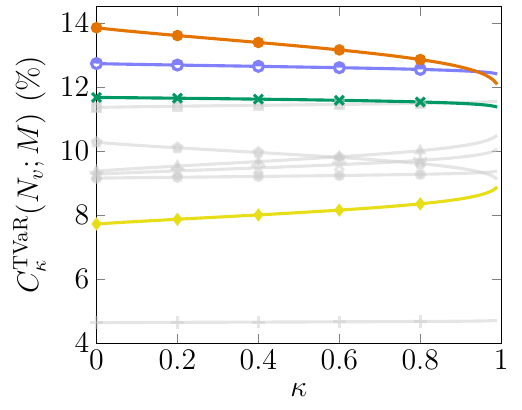}
        \caption{Relative contributions under $\boldsymbol{N}$.}
        \label{fig:rel_n}
    \end{subfigure}
    \hfill
        \begin{subfigure}[t]{0.48\textwidth}
        \centering
        \includegraphics[width=0.86\linewidth]{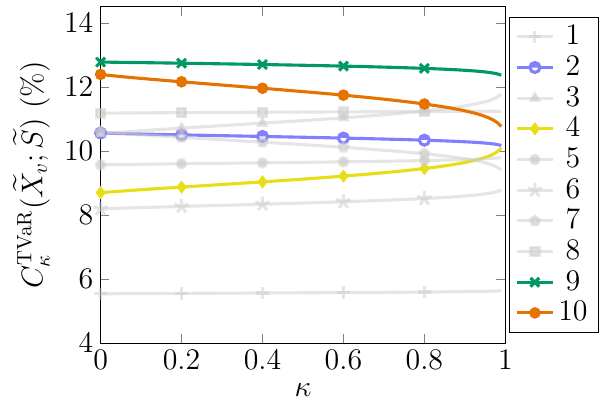}
        \caption{Relative contributions under $\boldsymbol{X}$.}
        \label{fig:rel_x}
    \end{subfigure}
    \caption{Relative contributions to $\text{TVaR}_{\kappa}(Z)$, using  the frequency vector $\boldsymbol{N} \; (Z = M)$ and the risk model vector $\boldsymbol{X} \; (Z = S)$.}
    \label{fig:contribution_summary}
\end{figure}

The second risk management task is to analyze how each station individually contributes to the total risk. We use the TVaR allocation principle in \eqref{eq:contribTVaR}. 
We compute the TVaR contributions for each vertex in the frequency and compound model using the method based on OGFEAs from Section~\ref{sect:allocation}. These contributions are denoted as $ C^{\mathrm{TVaR}}_{\kappa}(N_v;M) $ and $ C^{\mathrm{TVaR}}_{\kappa}(\widetilde{X}_v;\widetilde{S}) $, respectively. Figure~\ref{fig:contribution_summary} illustrates the relative contributions to $ \mathrm{TVaR}_{\kappa}(M) $ and $ \mathrm{TVaR}_{\kappa}(\widetilde{S}) $ from each station.

By examining Figure~\ref{fig:rel_n}, we observe that stations Upper Stewiacke (ID~2) and Yarmouth (ID~10)  have a more significant impact on the overall frequency distribution, which is consistent with their high frequency  mean parameters.  For Figure~\ref{fig:rel_x}, which incorporates both the frequency and severity components, Liverpool Big Falls (ID~9) emerges as the dominant contributor across all values of~$\kappa$. This behaviour is driven by the combination of its relatively high mean frequency parameter and its large severity mean and variance (compared to other stations), as reported in Table~\ref{tab:station_parameters}. Indeed, a comparison between Figures~\ref{fig:rel_n} and~\ref{fig:rel_x} reveals that, while the shape of the relative contribution curves remains consistent, the curves are shifted vertically depending on the combined frequence-severity distributions.

The relative contribution to aggregate risk of certain random variables decreases as the $ \kappa $ value increases. Locations connected to the tree through weaker (and fewer) edge correlations exhibit declining relative importance as $\kappa$ increases, reflecting their limited  dependence with other vertices.
Conversely, locations connected by strong edges and occupying central positions in the MST show increasing contributions at higher $\kappa$ values.  This observation offers critical insights for modelling. By examining the correlation edges of each vertex within the dependence structure, risk modelers can predict which components of the portfolio will see increased relative contributions as they move further into the tail of the aggregate risk distribution.  Most notably, Salmon Hole (ID~4) demonstrates the steepest upward trend across  $\kappa$ values.
As a consequence of both high severity and its globally influential correlations within the tree (see Table~\ref{tab:all_parameters}), Salmon Hole's ranking in relative TVaR$_{0.99}$ contribution rises from 9th (frequency alone) to 6th (frequency-severity).

\begin{figure}[t]
    \centering
    \begin{subfigure}[b]{0.3\textwidth}
        \centering
        \includegraphics[width = 0.8\textwidth]{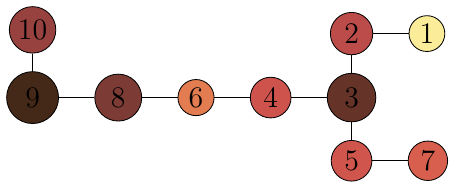}
        \caption{$C^{\mathrm{Cov}}_{0.99}(X_v, S)$ scaling,  $v \in \mathcal{V}$.}
        \label{fig:mlesdep2}
    \end{subfigure}
    \hfill
    \begin{subfigure}[b]{0.33\textwidth}
        \centering
        \includegraphics[width = 0.72\textwidth]{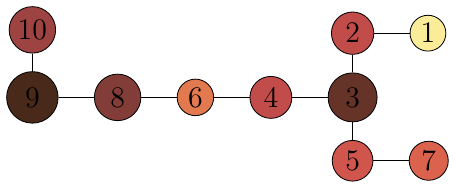}
        \caption{$C^{\mathrm{TVaR}}_{0.99}(\widetilde{X}_v, \widetilde{S})$ scaling,  $v \in \mathcal{V}$.}
        \label{fig:mlesdep3}
    \end{subfigure}
    \hfill
    \begin{subfigure}[b]{0.3\textwidth}
        \centering
        \includegraphics[width = 0.8\textwidth]{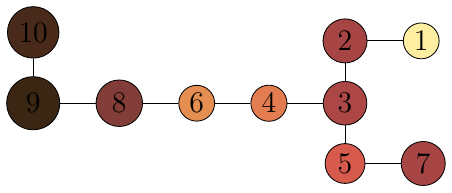}
        \caption{$\mathbb{E}[X_v]$ scaling,  $v \in \mathcal{V}$.}
        \label{fig:mlesdep1}
    \end{subfigure}
    \resizebox{0.35\textwidth}{!}{
    \begin{tikzpicture}
    \begin{axis}[
        hide axis,
        scale only axis,
        height=0.2cm,
        width=0.7\textwidth,
        colormap name=batlow,
        colorbar,
        colorbar horizontal,
        point meta min=5,
        point meta max=15,
        colorbar style={
            title={Contribution \% scaling},
            title style={at={(0.5,-1.8)}, anchor=north, font=\huge},
            tick label style={font=\Large},
            xtick={4,6,8,10,12,14}
        }
    ]
    \addplot [draw=none] coordinates {(8,0) (16,0)};
    \end{axis}
    \end{tikzpicture}
    }
    \caption{$\mathcal{T}$ with vertex sizes scaled based on (a) $C^{\mathrm{Cov}}_{0.99}(X_v, S)$, (b) $C^{\mathrm{TVaR}}_{0.99}(\widetilde{X}_v, \widetilde{S})$, and (c) $\mathbb{E}[X_v]$.}
    \label{fig:scaling_comparisontext}
\end{figure}

To compare allocations to vertices using different principles, we evaluate the contribution of each risk under both covariance-based and TVaR-based allocation rules. This comparison is illustrated in Figure~\ref{fig:scaling_comparisontext}, which shows the contributions of each station by scaling vertex sizes based on: (a) the covariance-based allocation $ C^{\mathrm{Cov}}_\kappa({X_v}, {S}) $ and (b) the TVaR-based allocation $ C^{\mathrm{TVaR}}_\kappa(\widetilde{X}_v, \widetilde{S}) $, for each $ v \in \mathcal{V} $ and $ \kappa = 0.99 $. In both cases, scaling is performed relative to the portfolio's TVaR. As a benchmark, panel (c) displays the relative contribution of each station's marginal mean to the portfolio expectation $ \mathbb{E}[{S}] $. Note that we compute $ C^{\mathrm{Cov}}_\kappa(X_v, S) $ exactly using Proposition~\ref{prop:cor}.  Supplement~\ref{sup:comp} complements 
Figure~\ref{fig:scaling_comparisontext}
by reporting the contributions and their differences, confirming that $C^{\mathrm{TVaR}}_\kappa(X_v, S)$ and $C^{\mathrm{Cov}}_\kappa(X_v, S)$ are close.

\section{Conclusion}
We have examined the risk model in \eqref{eq:model} wherein we introduced dependence between the claim counts using distributions from the family $\mathbb{MPMRF}$.  We established that $\mathbb{MPMRF}$ is a subset of $\mathbb{MPCS}$, but its specific parameterization enables more tractable analysis of high-dimensional models with Poisson marginals. 
We discussed two risk management tasks. First, we provided tools for analyzing the aggregate claim amount $S$ of a portfolio. Second, we investigated risk allocation, providing analytical expression of expected allocations and methods for exact computations of risk allocations. We discussed linear risk sharing for infinite-size portfolios and showed convergence in probability of the average risk amount if the tree's radius becomes large. We described estimation procedures for the model's parameter. Finally, we illustrated our findings through a real data analysis. 

Further research can be undertaken on this family of risk models. 
Tighter and more general asymptotic results for risk-sharing rules, in the spirit of \cite{denuit2022risk}, could be established. Incorporating dependence among the severity random variables could further enhance the risk model's use. Focusing on the frequency component, the separation between marginal distributions and the dependence structure naturally lends itself to extensions within the framework of generalized linear models (e.g., Poisson regression with log-linear link). The framework could also be broadened to other distributional families.

A potential limitation of the proposed risk model is the underestimation of correlations between non-adjacent vertices in real-data applications. Analysis of the statistical frequency model indicated that, while the model performs well on edges, it tends to underestimate correlations for non-adjacent pairs. Non-adjacent correlations may be better captured by modelling residual dependence through an auxiliary tree. 

\paragraph{Acknowledgments} We would like to thank three anonymous referees and the handling editor for precious comments that helped improve the quality of the paper. 
This work was partially supported by the Natural Sciences and Engineering Research Council of Canada (Cossette: RGPIN-2025-04077; Marceau: RGPIN-2020-05605; Côté: 581589859, Dubeau: BESC-M), and by the Chaire en actuariat de l'Université Laval (Marceau).

\bibliographystyle{apalike}
\bibliography{bibliography}

\appendix

\section{Proofs}
\label{sect:proofs}

\subsection{Proof of Theorem~\ref{th:flexible}}
First, we argue that $\boldsymbol{N}$ is a MRF. The construction given in \eqref{eq:StoConstruct} is akin to the one presented in Theorem~1 of \cite{cote2025tree} about the stochastic dynamics at play. The arguments provided for the proof of that theorem remain relevant: the maximum information about a random variable $N_v$, $v \in \mathcal{V}$, is obtained by knowing the value of its neighbors. Thus, it satisfies the local Markov property, meaning $\boldsymbol{N}$ is a MRF. 

We next prove by induction that $N_v\sim\text{Poisson}(\lambda_v)$ for all $v \in \mathcal{V}$, with the root $r$ as the starting point; evidently, \mbox{$N_r\sim \text{Poisson}(\lambda_r)$}. We next suppose the statement holds true for $N_{\mathrm{pa}(w)}$, $w \in \mathcal{V} \backslash \{r\}$, and prove $N_w\sim\text{Poisson}(\lambda_w)$. Following the construction in \eqref{eq:StoConstruct}, $L_w$ is independent of all $L_v,\,v \in \mathcal{V}\backslash\{w\}$ and of $N_{\pa{w}}$, since $w \not\in \mathrm{path}(\mathrm{pa}(w), r)$. It follows that the pgf of $N_w$ is given by
\begin{equation*}
\mathcal{P}_{N_w}(t) = \mathcal{P}_{\left(\alphav{w}\sqrt{{\lambda_w}/{\lambda_{\mathrm{pa}(w)}}} \right)\circ N_{\mathrm{pa}(w)}}(t)  \mathcal{P}_{L_w}(t), \quad t \in [-1, 1].
\end{equation*}
From the properties of the binomial thinning operator (one may refer to Theorem~11(d) of \cite{cote2025tree}), we have 
\begin{align}
\notag
\mathcal{P}_{N_w}(t) &= \mathcal{P}_{N_{\mathrm{pa}(w)}}\Big(1 - \alphav{w}\sqrt{{\lambda_w}/{\lambda_{\mathrm{pa}(w)}}} + \alphav{w}\sqrt{{\lambda_w}/{\lambda_{\mathrm{pa}(w)}}}t\Big)  \mathcal{P}_{L_w}(t)\\
&= \mathrm{e}^{\lambda_{\mathrm{pa}(w)}\left(1 + \alphav{w}\sqrt{{\lambda_w}/{\lambda_{\mathrm{pa}(w)}}}(t - 1) - 1\right)}  \mathrm{e}^{\left(\lambda_w - \alphav{w}\sqrt{\lambda_{\mathrm{pa}(w)}\lambda_w}\right)(t - 1)}, \quad t \in [-1, 1], \label{eq:pgf-nw}
\end{align}
from the respective pgfs of $L_w$ and $N_{\mathrm{pa}(w)}$ given the induction hypothesis. Simplifying \eqref{eq:pgf-nw} provides $\mathcal{P}_{N_w}(t) = \mathrm{e}^{\lambda_w(t - 1)}$, $t \in [-1, 1]$; thus, $N_w \sim \mathrm{Poisson}(\lambda_w)$. The assertion is validated for both the case of the root and the parent-child inductive case; we conclude $N_v\sim\text{Poisson}(\lambda_v)$ for every $v\in\mathcal{V}$. 

Conditioning on $N_{\mathrm{pa}(v)}$ and using \eqref{eq:StoConstruct}, the joint pgf of two neighbors $(N_{\pa{v}}, N_v)$ is given by
\begin{equation*}
\mathcal{P}_{N_{\pa{v}}, N_v}(t_{\pa{v}}, t_v) = \mathrm{e}^{\lambda_{\pa{v}}(t_{\pa{v}} - 1) + \lambda_v(t_v - 1) + \alphav{v}\sqrt{\lambda_{\pa{v}}\lambda_v}(t_v - 1)(t_{\pa{v}} - 1)}, \quad  t_{\pa{v}}, t_v \in [-1,1]. 
\end{equation*}

Since $\mathcal{P}_{N_{\pa{v}}, N_v}(t_{\pa{v}}, t_v)$ is symmetric regarding $N_{\pa{v}}$ and $N_v$, the stochastic dynamics on an edge are reversible. Given the local Markov property,  this reversibility extends to the stochastic dynamics on a path. Moreover, note that choosing a root $r'\neq r$ only affects the parent-child relationships of the vertices on $\mathrm{path}(r, r')$. Other vertices remain children to their parent, with unchanged stochastic dynamics.

\subsection{Proof of Proposition~\ref{th:propN}}
The proof resembles those of Theorems~3 and~4 of \cite{cote2025tree} and is thus omitted.

 \subsection{Proof of Proposition~\ref{th:StoRepCommonShocks}}
The choice of the root has no incidence on the distribution of $\boldsymbol{N}$ (Theorem~\ref{th:flexible}). For all $v\in\mathcal{V}$, we have 
\begin{equation}
    \eta_v^{\mathcal{T}_{\!\!r}}(\boldsymbol{t}_{v\mathrm{dsc}(v)};\boldsymbol{\theta}_{\mathrm{dsc}(v)}) = t_v \sum_{\mathcal{C} \in \mathscr{P}(\mathrm{ch}(v))} \left(\prod_{j\in \mathrm{ch}(v)\backslash \mathcal{W}} (1-\theta_j) \right) \left(\prod_{i \in \mathcal{W}} \theta_i \eta_i^{\mathcal{T}_{\!\!r}}(\boldsymbol{t}_{i\mathrm{dsc}(i)};\boldsymbol{\theta}_{\mathrm{dsc}(i)}) \right).  
    \label{eq:proof-common-shock-1}
\end{equation}

Let $\Xi_v = \{\mathcal{W}\cap (\{v\}\cup \mathrm{dsc}(v)) : \mathcal{W}\in\Theta_v\}$. Write $\mathrm{ch}^{i}(v)$ to refer to the $i$th generation of the descendants of $v$, $i\in\mathbb{N}$; for instance, $\mathrm{ch}^2(v)$ are the children of the children of $v$. Let $\Xi_v^{[i]} = \{\mathcal{W}\cap \mathrm{ch}^{i}(v) : \mathcal{W}\in\Xi_v\}$. From \eqref{eq:proof-common-shock-1}, we use the result recursively to further develop $\eta_{v}^{\mathcal{T}_{\!\!v}}$, yielding after one iteration
\begin{align}
        &\eta_v^{\mathcal{T}_{\!\!r}}(\boldsymbol{t}_{v\mathrm{dsc}(v)};\boldsymbol{\theta}_{\mathrm{dsc}(v)})\notag \\
        &= t_v \sum_{\mathcal{W} \in \Xi_v^{[2]}} \left(\prod_{j\in \mathrm{ch}(v)\backslash \mathcal{C}} (1-\theta_j) \right) \!\! \left(\prod_{i \in \mathcal{C}} \theta_i t_i  \left(\prod_{k\in \mathrm{ch}(i)\backslash \mathcal{W}} (1-\theta_k) \right) \!\!\left(\prod_{\ell \in  \mathrm{ch}(i)\cup \mathcal{W}} \theta_\ell \eta_i^{\mathcal{T}_{\!\!r}}(\boldsymbol{t}_{i\mathrm{dsc}(i)};\boldsymbol{\theta}_{\mathrm{dsc}(i)}) \right) \!\! \right)\notag\\
        &= t_v \sum_{\mathcal{W}\in \Xi^{[2]}_v}  \left(\prod_{(\mathrm{pa}(j),j)\in\mathcal{E}^{\dagger}_{\mathcal{W}}} (1-\theta_j) \right) \left(\prod_{(\mathrm{pa}(i),i)\in\mathcal{E}_{\mathcal{W}}} \theta_i t_i\right) \left(\prod_{\ell\in \mathrm{ch}^2(i)} \frac{1}{t_\ell}  \eta_{\ell}^{\mathcal{T}_{\!\!r}}(\boldsymbol{t}_{\ell\mathrm{dsc}(\ell)};\boldsymbol{\theta}_{\mathrm{dsc}(\ell)})\right) \label{eq:proof-common-shock-3}
\end{align}

Continuing recursively in this fashion down to the bottom of the tree (say it takes $m\in\mathbb{N}$ iterations), we have $\Xi^{[m]}_v = \Xi_v$ , and \eqref{eq:proof-common-shock-3} becomes
\begin{equation}
     \eta_v^{\mathcal{T}_{\!\!r}}(\boldsymbol{t}_{v\mathrm{dsc}(v)};\boldsymbol{\theta}_{\mathrm{dsc}(v)}) = t_v \sum_{\mathcal{W}\in\Xi_v} \left(\prod_{(\mathrm{pa}(j),j)\in\mathcal{E}^{\dagger}_{\mathcal{W}}} (1-\theta_j) \right) \left(\prod_{(\mathrm{pa}(i),i)\in\mathcal{E}_{\mathcal{W}}} \theta_i t_i\right),
     \label{eq:proof-common-shock-4}
\end{equation}
since $\eta_{\ell}^{\mathcal{T}_{\!\!r}}(\boldsymbol{t}_{\ell\mathrm{dsc}(\ell)};\boldsymbol{\theta}_{\mathrm{dsc}(\ell)}) = t_{\ell}$ for every $\ell\in \xi_v^{[m]}$. Inserting \eqref{eq:proof-common-shock-4} into \eqref{eq:jointpgf}, 
$\mathcal{P}_{\boldsymbol{N}}$ is thus given by
\begin{equation}
    \mathcal{P}_{\boldsymbol{N}}(\boldsymbol{t}) = 
        \prod_{v\in\mathcal{V}}
    \prod_{\mathcal{W}\in \Xi_v}
    \mathrm{e}^{
    \zeta_{L_v}
    \left(\prod_{(\mathrm{pa}(j),j)\in\mathcal{E}^{\dagger}_{\mathcal{W}}} (1-\theta_j) \right)
    \left(\prod_{(\mathrm{pa}(i),i)\in\mathcal{E}_{\mathcal{W}}} \theta_i \right)
    \left(
    \prod_{i\in\mathcal{W}} t_i - 1
    \right)},
    \quad \boldsymbol{t}\in[-1,1]^{d},
      \label{eq:proof-common-shock-5}
\end{equation}
where $\{\zeta_{L_v}, \, v\in\mathcal{V}\}$ and $\{\theta_v,\, v\in\mathcal{V}\}$ are defined as in Proposition~\ref{th:propN}. Note that from the construction of $\Xi_v$, no combination of $t_v$'s is repeated across the product operations. One recognizes in \eqref{eq:proof-common-shock-5} the joint pgf of the sum of independent Poisson random vectors of the form $\boldsymbol{Y}_{\mathcal{W}} = (Y_{\mathcal{W}}\mathbbm{1}_{v\in\mathcal{W}} ,\, v\in\mathcal{V})$, one for each element of $\bigcup_{v\in\mathcal{V}} \Xi_v = \bigcup_{v\in\mathcal{V}} \Theta_v$,  where the frequency parameter associated to $Y_{\mathcal{W}}$ is 
\begin{equation}
    \gamma_{\mathcal{W}} = \zeta_{L_v} \left(\prod_{(\mathrm{pa}(j),j)\in\mathcal{E}^{\dagger}_{\mathcal{W}}} (1-\theta_j) \right) \left(\prod_{(\mathrm{pa}(i),i)\in\mathcal{E}_{\mathcal{W}}} \theta_i\right), \quad \mathcal{W} \in \bigcup_{v\in\mathcal{V}} \Theta_v,  \label{eq:proof-common-shock-6}
\end{equation}
where $v$ is the utmost vertex in $\mathcal{W}$ according to the rooting in $r$. Equation \eqref{eq:proof-common-shock-6} is equivalent to \eqref{eq:gammaparamMPMRF} after substituting back to the orginal parameters.

\subsection{Proof of Remark \ref{rem:tls-mixederlang}}\label{app:mixederlang}
The techniques for that matter are inspired from the work in \cite{willmot2011risk}. The cdf and LST of $B_v$, $v \in \mathcal{V}$, are given respectively by
\begin{equation*}\label{eq:cdf-lst-B}
F_{B_v}(x) = \sum_{k=1}^{\infty} \pi_{v,k} H(x; k, \beta_v) \quad  x \geq 0,
\quad \quad
\mathcal{L}_{B_v}(t) = \sum_{k = 1}^{\infty} \pi_{v,k} \left(\frac{\beta_v}{\beta_v + t}\right)^k, \quad  t > 0, 
\end{equation*}
where $H(x; k, \beta_v) = 1 - \mathrm{e}^{-\beta_v x} \sum_{l=0}^{k-1} \frac{(\beta_v x)^l}{l!}$, $x \geq 0$, is the cdf of the $k$th Erlang distribution with rate $\beta_v$.

To reformulate $\mathcal{L}_S$ in \eqref{eq:lst-S-Poisson}, we first express every 
$\mathcal{L}_{B_v}$ according to the maximum rate parameter \mbox{$\beta_{\max} :=\max\{\beta_v : v\in \mathcal{V}\}$}.
Let \mbox{$v_{\max} := \mathrm{argmax}\{\beta_v : v\in \mathcal{V}\}$}. For $v \in \mathcal{V} \backslash \{v_{\max}\}$, 
 we can express $\mathcal{L}_{B_v}$ as
\begin{equation}\label{eq:mixed_modif}
    \mathcal{L}_{B_v}(t) = \sum_{k = 1}^{\infty} \pi_{v, k} \left[ \frac{q_v}{1 - (1 - q_v) \left(\frac{\beta_{\max}}{\beta_{\max} + t}\right)}\left(\frac{\beta_{\max}}{\beta_{\max}+ t}\right) \right]^k = \sum_{k = 1}^{\infty} \pi_{v, k} \mathcal{P}_{K_{v,k}}\left(\frac{\beta_{\max}}{\beta_{\max}+ t}\right), \quad  t \geq 0,
\end{equation}
where $K_{v, k}$ follows a negative binomial distribution with number of successful trials $k$ and success probability {$q_v ={\beta_v}/{\beta_{\max}}$}. For all $v \in \mathcal{V}$, we may write $\mathcal{L}_{B_v}(t) = \mathcal{P}_{\widetilde{K}_v}(\mathcal{L}_{B_{\max}}(t))$, corresponding to a compound distribution with primary distribution given by the random variable $\widetilde{K}_v$ with pmf  $p_{\widetilde{K}_v}(x) = \sum_{k=1}^\infty \pi_{v,k} \, p_{K_{v,k}}(x)$, $x \in \mathbb{N}_1$, and secondary distribution $B_{\max} \sim \mathrm{Exp}(\beta_{\max})$. The result follows inserting \eqref{eq:mixed_modif} in \eqref{eq:lst-S-Poisson}.

\subsection{Proof of Theorem~\ref{th:OGFEAX}}
From Theorem~3.4 of \cite{blier2025efficient} and \eqref{eq:jointtls-vecteurX}, taking $v$ as the root for the joint pgf of $\boldsymbol{X}$,
\begin{align}\notag
    \mathcal{P}_S^{[v]}(t) &= \left.\left[t_v \frac{\partial}{\partial t_v}\mathcal{P}_{\boldsymbol{X}}(\boldsymbol{t}) \right]\right|_{\boldsymbol{t}=t\boldsymbol{1}_d} \\
    &= t  \left.\left[ \frac{\partial}{\partial t_v}\mathrm{e}^{\lambda_v (\eta_v^{\mathcal{T}_{\!\!v}}(\boldsymbol{\mathcal{P}}_{B_v}(\boldsymbol{t}_{v\mathrm{dsc}(v)}); \boldsymbol{\theta}_{\mathrm{dsc}(v)})-1)} \prod_{j\in\mathcal{V}\backslash\{v\}} \mathrm{e}^{\zeta_{L_j}
 (\eta_j^{\mathcal{T}_{\!\!v}}(\boldsymbol{\mathcal{P}}_{B_j}(\boldsymbol{t}_{j\mathrm{dsc}(j)}); \boldsymbol{\theta}_{\mathrm{dsc}(j)}) -1)}\right]\right|_{\boldsymbol{t}=t\boldsymbol{1}_d},
 \label{eq:proofOGFEA-1}
\end{align}
for $t\in[-1,1]$, where $\zeta_{L_j} = \lambda_j (1-\alpha_{(\mathrm{pa}(j),j)}\sqrt{\lambda_{\mathrm{pa}(j)}/ \lambda_j} )$. We choose $v$ as the root for  $\mathcal{P}_{\boldsymbol{X}}$ as it simplifies the differentiation and has no incidence on the result. All the multiplicands  in \eqref{eq:proofOGFEA-1} are free of $t_v$ since, if $v$ is the root, $v\not\in j\mathrm{dsc}(j)$ for every other $j\in\mathcal{V}\backslash\{v\}$. Performing the differentiation in \eqref{eq:proofOGFEA-1} yields
\begin{align*}
\mathcal{P}_{S}^{[v]}(t) &=   \lambda_v t  \left.\left[ \frac{\partial}{\partial t_v}\eta_v^{\mathcal{T}_{\!\!v}}(\pmb{\mathcal{P}}_{B_v}(\boldsymbol{t}_{v\mathrm{dsc}(v)}); \boldsymbol{\theta}_{\mathrm{dsc}(v)})\right]\right|_{\boldsymbol{t}=t\boldsymbol{1}_d} \prod_{j\in\mathcal{V}} \mathrm{e}^{
\zeta_{L_j}
 (\eta_j^{\mathcal{T}_{\!\!v}}(\pmb{\mathcal{P}}_{B_j}(t \, \boldsymbol{1}_{j\mathrm{dsc}(j)}); \boldsymbol{\theta}_{\mathrm{dsc}(j)})-1)}\\
 &= \lambda_v t  \left[\frac{\mathrm{d}}{\mathrm{d} t} \mathcal{P}_{B_v}(t)\right]  \frac{1}{\mathcal{P}_{B_v}(t)} \eta_v^{\mathcal{T}_{\!\!v}}(\pmb{\mathcal{P}}_{B_v}(t \, \mathbf{1}_{|\{v\} \cup \mathrm{dsc}(v)|}); \boldsymbol{\theta}_{\mathrm{dsc}(v)}) \mathcal{P}_S(t), \quad t\in[-1,1],
\end{align*}
from $\eta_v^{\mathcal{T}_{\!\!v}}$ given in \eqref{eq:eta}, with the vector \mbox{$\pmb{\mathcal{P}}_{B_v}(\boldsymbol{t}_{v\mathrm{dsc}(v)}) = (\mathcal{P}_{B_{v,j}}(t_j), j \in \{v\} \cup \mathrm{dsc}(v))$}. 

\subsection{Proof of Corollary~\ref{cor:OGFEAstorep}}
The pgf of ${B^*_v}$ is given by $\mathcal{P}_{B^*_v}(t) = \tfrac{t}{\mathbb{E}[B_v]}\tfrac{\mathrm{d}}{\mathrm{d}t}\mathcal{P}_{B_v}(t)$. Hence, the OGFEA in (\ref{eq:OGFEA}) is rewritten as
\begin{equation*}
        \mathcal{P}_S^{[v]}(t) = \lambda_v  \mathbb{E}[B_v]   \mathcal{P}_{K^{(v)}}(t)  \mathcal{P}_S(t) =  \lambda_v  \mathbb{E}[B_v]  \mathcal{P}_{K^{(v)}+S}(t) = \sum_{k=0}^{\infty} \left(\lambda_v \mathbb{E}[B_v] p_{K^{(v)}+S}(k) \right)\, t^{k},   
    \end{equation*}
$t\in[-1,1]$, with the second equality following from the independence of $K^{(v)}$ and $S$. From the OGFEA definition in (\ref{eq:OGFEA}),
the expected allocations for $k\in\mathbb{N}$ are given by the polynomial's coefficients.

\subsection{Proof of Corollary~\ref{cor:sizebiasedcontribution}}
The result follows directly by inserting \eqref{eq:expealloc} into \eqref{eq:contribTVaR}. 

\subsection{Proof of Proposition~\ref{prop:cor}}
 The proof is straightforward for $\mathrm{Cov}(N_v, N_w)$ if $v=w$. We now suppose $v=\mathrm{pa}(w)$; then, given  \eqref{eq:StoConstruct}, 
\begin{align}
\mathrm{Cov}(N_{v}, N_{w}) &= \mathrm{Cov}\left(N_{v}, \left(\alphavj{v}{w}\sqrt{{\lambda_{w}}/{\lambda_{v}}}\right)\circ N_{v} + L_{w}\right)\overset{\perp\!\!\!\perp}{=} \mathrm{Cov}\left(N_{v}, \left(\alphavj{v}{w}\sqrt{{\lambda_{w}}/{\lambda_{v}}}\right)\circ N_{v}\right).
\label{eq:CovNvNw-1}
\end{align}
From the properties of the binomial thinning operator, (\ref{eq:CovNvNw-1}) becomes
\begin{equation}
\mathrm{Cov}(N_v,N_w) = \alphavj{v}{w}\sqrt{{\lambda_{w}}/{\lambda_{v}}}\mathrm{Var}(N_v) = \sqrt{\lambda_v\lambda_w}\alphavj{v}{w},
\label{eq:CovNvNw-2}
\end{equation}
which corresponds to our result for the case $v=\mathrm{pa}(w)$. The general result for every $v,w\in\mathcal{V}$ is then obtained by using \eqref{eq:CovNvNw-2} and the same \emph{modus operandi} as in the proof of Theorem~5 of \cite{cote2025tree} -- that is, by iterative conditioning on every successive vertex on the path from $v$ to $w$.  Conditioning on both claim count random variables, 
given that $\{B_{v,j},\, j\in\mathbb{N}_{1}\}$ and $\{B_{w,j},\, j\in\mathbb{N}_1\}$ are independent sequences of independent and identically distributed random variables, yields the desired result. 

\subsection{Proof of Proposition~\ref{th:LawLargeNumbers}}
The number of vertices in $\mathcal{B}^{(\chi)[\xi]}$ is given by
\begin{equation}
d^{[\xi]} = |\mathcal{V}^{[\xi]}| = 1 + \sum_{i=0}^{\xi - 1} \chi(\chi-1)^i = 1 + \chi\frac{(\chi-1)^{\xi}-1}{\chi-2}, \quad  \xi\in\mathbb{N}.
\label{eq:proofWLLN-d}
\end{equation}
Let $\mathcal{V}_i$ be the set of vertices in the $i$th level of $\mathcal{B}^{(\chi)[\xi]}$, $i\in\{0,1,\ldots,\xi\}$. For $u\in\mathcal{V}_{0}$, the vertex at the center of the Cayley tree, we obtain, from Proposition~\ref{prop:cor},
\begin{equation}\label{eq:covasymp1}
    \mathrm{Cov}(X_u^{[\xi]}, S^{[\xi]}) = 
    \sum_{i=0}^{\xi}\sum_{v\in\mathcal{V}_i}\mathrm{Cov}(X_v^{[i]},X_u^{[i]})  = \lambda_u \mathbb{E}[B_u^2] + \sum_{i=1}^{\xi} \mathbb{E}[B_u] \mathbb{E}[B_v]
    \sqrt{\lambda_u \lambda_v} \prod_{e \in \mathrm{path}(u,v)} \alpha_e.
\end{equation}
Bounding \eqref{eq:covasymp1} using $\lambda_{\mathrm{sup}}$, $\alpha_{\mathrm{sup}}$ and a constant $C = \max_{u, v \in \mathcal{V}}(\mathbb{E}[B_u^2],\mathbb{E}[B_u]\mathbb{E}[B_v])$, we obtain
\begin{align}\notag
    \mathrm{Cov}(X_u^{[\xi]}, S^{[\xi]}) &\leq C \left( \lambda_{\mathrm{sup}} + \sum_{i=1}^{\xi}\sum_{v\in\mathcal{V}_i} \sqrt{\lambda_{\mathrm{sup}} \lambda_{\mathrm{sup}}} \alpha_{\mathrm{sup}}^{|\mathrm{path}(u,v)|} \right) 
    \notag
    \\
     &= C \left(\lambda_{\mathrm{sup}} + \lambda_{\mathrm{sup}} \sum_{i=1}^{\xi}\chi \alpha_{\mathrm{sup}} ((\chi-1) \alpha_{\mathrm{sup}})^{i-1} \right) \notag
    \\
    &= C \lambda_{\mathrm{sup}} \left(1+\chi\alpha_{\mathrm{sup}} \frac{((\chi-1)\alpha_{\mathrm{sup}})^{\xi-1}-1}{(\chi-1)\alpha_{\mathrm{sup}} - 1} \right), 
    \label{eq:proofWLLN-1}
\end{align}
for $\xi\in\mathbb{N}$. The topology in a Bethe lattice is such that the structure around every vertex remains identical, regardless of the chosen vertex. Hence, for every $v\in\mathcal{V}$, 
\begin{equation*}
    \lim_{\xi\to\infty}\mathrm{Cov}(X_v^{[\xi]},S^{[\xi]}) = \lim_{\xi\to\infty}\mathrm{Cov}(X_u^{[\xi]}, S^{[\xi]}), \; u\in\mathcal{V}_0.
\end{equation*}
Therefore, we have
\begin{align}
   \lim_{\xi\to\infty}  \mathrm{Var}(W^{[\xi]}) &= \lim_{\xi\to\infty}  \frac{1}{(d^{[\xi]})^2}\mathrm{Var}(S^{[\xi]}) 
    = \lim_{\xi\to\infty}\frac{1}{(d^{[\xi]})^2}\sum_{v\in\mathcal{V}}\mathrm{Cov}(X_v^{[\xi]}, S^{[\xi]}). 
    \label{eq:proofWLLN-2}
\end{align}
Given \eqref{eq:proofWLLN-d}, \eqref{eq:proofWLLN-1}, and since  $\alpha_{\mathrm{sup}}\in[0,1)$, \eqref{eq:proofWLLN-2} becomes
\begin{equation*}
  \lim_{\xi\to\infty}   \mathrm{Var}(W^{[\xi]}) \leq 
    \lim_{\xi\to\infty} C \lambda_{\mathrm{sup}} \left({1+\chi \alpha_{\mathrm{sup}} \frac{((\chi-1)\alpha_{\mathrm{sup}})^{\xi-1} - 1}{(\chi-1)\alpha_{\mathrm{sup}} - 1}}\right)\left({1+\chi \frac{(\chi - 1)^{\xi-1} - 1}{\chi-2}}\right)^{-1} = 0. 
\end{equation*}
By Chebyshev's inequality, 
\begin{equation}\label{eq:wl}
\mathrm{Pr}\left(\left|W^{[\xi]} - 
\mathbb{E}[W^{[\xi]}]
\right| > \varepsilon\right) 
\leq \varepsilon^{-2} \mathrm{Var}(W^{[\xi]}).
\end{equation}
We have $\lim_{\xi \to \infty} \mathrm{Var}(W^{[\xi]}) = 0$ since the sequence $\{\mathcal{B}^{(\chi)[\xi]},\xi\in\mathbb{N}\}$ converges to a Bethe lattice as $\xi\to\infty$. The right-hand side of \eqref{eq:wl} vanishes. This implies $W^{[\xi]} \to \mathbb{E}[W^{[\xi]}]$ in probability. Since $\mathbb{E}[W^{[\xi]}] \leq \sup_{v \in \mathcal{V}}\mathbb{E}[B_v] \lambda_{\mathrm{sup}}$, the result holds.

\subsection{Proof of Theorem~\ref{th:convlinRS}}
We have $h_{v,d^{[\xi]}}^{\text{lin}}(S^{[\xi]})  =  \mu_v + d^{[\xi]} a_{v,d^{[\xi]}}\frac{S^{[\xi]} - \mathbb{E}[S^{[\xi]}]}{d^{[\xi]}}.$ Using $a_{v,d^{[\xi]}} =\mathcal{O}({1}/{d^{[\xi]}})$ and Proposition~\ref{th:LawLargeNumbers}, the result follows.

\section{Factorization of the joint pmf}
\label{sect:factorization}

In the proof of Theorem~\ref{th:flexible}, we argued from the thinning-based stochastic representation of any member of the family $\mathbb{MPMRF}$ satisfies the conditional independencies required to make it a tree-structured MRF as per Definition~\ref{def:mrf}.
From these conditional independencies, we have directly, by successive conditioning along the tree,
\begin{equation}
p_{\boldsymbol{N}}(\boldsymbol{x}) = \prod_{v\in\mathcal{V}} p_{N_v|\bigcap_{j=1}^{v-1}N_j=x_j}(x_v) = p_{N_r}(x_r) \prod_{v\in\mathcal{V}\backslash\{r\}} p_{N_v|N_{\mathrm{pa}(v)} = x_{\mathrm{pa}(v)}}(x_v),\quad\quad \boldsymbol{x} \in \mathbb{N}^d,
    \label{eq:facto-1}
\end{equation}
which we recognize as the joint pmf of a Gibbs distribution. 
One can find the following definition in \cite{koller2009probabilistic}, adapted for discrete random variables.

\begin{definition}[Gibbs distribution]
Let $\mathcal{V}_1,\ldots,\mathcal{V}_m$ be subsets of $\mathcal{V}$, with $m \leq |\mathcal{V}|$, and define $\phi_1,\ldots,\phi_m$ as some functions $\phi_i : \mathbb{R}^{|\mathcal{V}_i|} \to \mathbb{R}$, $i \in \{1,\ldots,m\}$. The joint pmf of a vector of discrete random variables $X = (X_v,\; v \in \mathcal{V})$ following a Gibbs distribution admits the representation
\mbox{$p_{\boldsymbol{X}}(\boldsymbol{x}) = \frac{1}{Z} \prod_{i=1}^{m} \phi_i\bigl( (x_v,\; v \in \mathcal{V}_i) \bigr),\; \boldsymbol{x}\in\mathbb{N}^d$}, where $Z$ is a normalizing constant. A Gibbs distribution factorizes on $\mathcal{G} = (\mathcal{V},\mathcal{E})$ if $\mathcal{V}_1,\ldots,\mathcal{V}_m$ are all cliques of $\mathcal{G}$.
\end{definition}

On a tree $\mathcal{T}$, cliques are any vertex or pair of vertices connected by an edge. The joint pmf of $\boldsymbol{N}$ in \eqref{eq:jointpmf-vecteurN} is that of a Gibbs distribution factorizing on $\mathcal{T}$.
The Hammersley--Clifford Theorem states that a random vector $\boldsymbol{X} = (X_v)_{v \in \mathcal{V}}$ with positive probabilities on $\prod_{v\in\mathcal{V}}\mathrm{supp}(X_v)$ is a MRF defined on $\mathcal{G}$ if and only if it follows a Gibbs distribution factorizing on $\mathcal{G}$. 
Therefore, to design the family $\mathbb{MPMRF}$, satisfying the conditional independencies of a MRF and Poisson marginals, one could have proceeded directly using the factorized joint pmf in \eqref{eq:facto-1}, and, to ensure undirectedness, revert the conditional probabilities to joint probabilities. We have
\begin{align}
p_{\boldsymbol{N}}(\boldsymbol{x}) = p_{N_r}(x_r)  \prod_{v\in\mathcal{V}\backslash\{r\}} \frac{p_{N_{\mathrm{pa}(v)},N_v}(x_{\mathrm{pa}(v)}, x_v)}{p_{N_{\mathrm{pa}(v)}}(x_{\mathrm{pa}(v)})} 
&= \frac{p_{N_r}(x_r)}{p_{N_r}(x_r)^{\mathrm{deg}(r)}}  \frac{\prod_{(u,v)\in\mathcal{E}}p_{N_u,N_v}(x_u, x_v)}{\prod_{v\in\mathcal{V}\backslash\{r\}}p_{N_v}(x_v)^{\mathrm{deg}(v)-1}}\notag\\
&= \frac{\prod_{(u,v)\in\mathcal{E}}p_{N_u,N_v}(x_u, x_v)}{\prod_{v\in\mathcal{V}}p_{N_v}(x_v)^{\mathrm{deg}(v)-1}}, \quad \boldsymbol{x}\in\mathbb{N}^d,
\label{eq:facto-2}
\end{align}
where the second equality comes from the fact that every vertex is the parent of $\mathrm{deg}(v)-1$ other vertices, except the root,  for which it is $\mathrm{deg}(r)$. The factorization as in \eqref{eq:facto-2} is referred to as the node-edge factorization of a MRF. The model will be undirected and invariant under any rooting of the tree if all bivariate joint pmfs in the multiplicands in \eqref{eq:facto-2} are symmetric in that respect. It is well known to be the case for Poisson random variables joined through binomial thinning: letting $\xi_{u,v} = \alpha_{(u,v)}\sqrt{\lambda_u\lambda_v}$,
\begin{equation*}
    p_{N_u,N_v}(x_u,x_v) = \mathrm{e}^{-(\lambda_u+\lambda_v -\xi_{u,v})}\sum_{i=1}^{\min(x_u,x_v)} \frac{(\lambda_u - \xi_{u,v})^{m-k}(\lambda_v - \xi_{u,v})^{n-k}\xi_{u,v}^{k}}{(x_u - k)!(x_v-k)!k!}  ,\quad x_u,x_v\in\mathbb{N},
\end{equation*}
see \cite{mckenzie1988some}, which we incidently recognize as the joint pmf of the bivariate Poisson based on common shocks. Substituting the bivariate and marginal pmfs in \eqref{eq:facto-2}, one recovers \eqref{eq:jointpmf-vecteurN}.

\newpage
\setcounter{page}{1}
\setcounter{section}{0}
\setcounter{table}{0}
\setcounter{figure}{0}
\section{Supplementary material}
\subsection{Numerical illustration of MPMRF and MPCS equivalence}\label{sup:equivalence}
This section presents a numerical example illustrating the equivalence between the MPCS and MPMRF representations.  For the tree structure depicted in Figure~2 of the main text, with parameters 
$\boldsymbol{\lambda} = (4, 3, 2.5, 1.5, 1.2)$ and 
$\boldsymbol{\alpha} = (0.7, 0.5, 0.6, 0.4)$,
Table~\ref{tab:check} reports some values of the pgf obtained under both formulations. The MPCS parameters implied by this specification are reported in Table~\ref{tab:paramgamma}.

\begin{minipage}[t]{0.55\textwidth}
\begin{center}
\captionof{table}{MPCS and MPMRF pgf values}
\label{tab:check}
\includegraphics[width=\textwidth]{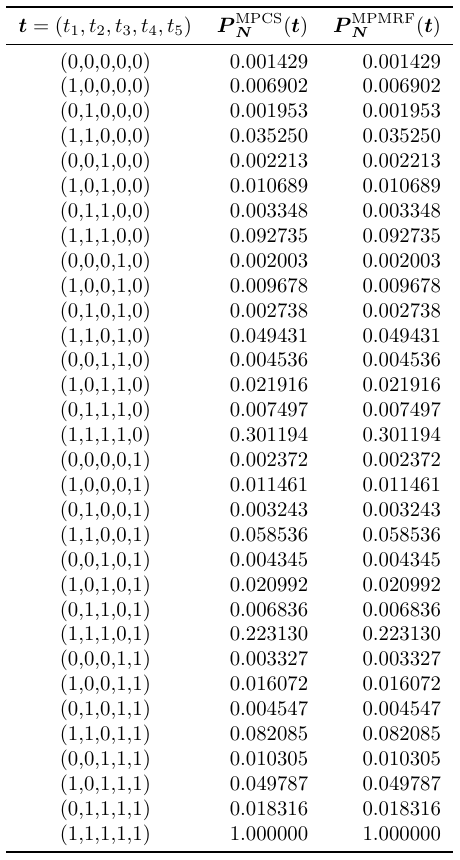}
\end{center}
\end{minipage}
\hfill
\begin{minipage}[t]{0.38\textwidth}
\centering
\captionof{table}{Parameters $\gamma_{\mathcal{W}}$ for each set $\mathcal{W}$}
\label{tab:paramgamma}
\begin{tabular}{cr}
\toprule
Set $\mathcal{W}$ & Parameter $\gamma_{\mathcal{W}}$ \\
\midrule
\{1\} & 1.57513 \\ 
\{2\} & 0.31262 \\ 
\{3\} & 1.31807 \\ 
\{4\} & 0.43747 \\ 
\{5\} & 0.10157 \\ 
\{1,2\} & 0.42823 \\ 
\{2,3\} & 0.33810 \\ 
\{3,4\} & 0.37987 \\ 
\{3,5\} & 0.08819 \\ 
\{1,2,3\} & 0.37184 \\ 
\{2,3,4\} & 0.50718 \\ 
\{2,3,5\} & 0.16772 \\ 
\{3,4,5\} & 0.03894 \\ 
\{1,2,3,4\} & 0.16417 \\ 
\{1,2,3,5\} & 0.14563 \\ 
\{2,3,4,5\} & 0.03381 \\ 
\{1,2,3,4,5\} & 0.14255 \\ 
\bottomrule
\end{tabular}
\end{minipage}

\newpage
\subsection{Algorithms to compute the distribution of $S$}\label{sup:algo-s}
This section details the algorithms used to compute the distribution of $S$ in a discrete and continuous context. In Algorithms~\ref{algo:pmfJ-discret} and~\ref{algo:pmfJ-erlang}, $\boldsymbol{A}_{d \times d}$ is ordered in topological order, with $r = 1$. 
\begin{algorithm}[h]
\caption{Computing the pmf of $S$ : discrete claim amount distributions.}
\label{algo:pmfJ-discret}
\KwData{$\boldsymbol{\alpha}$-weighted adjacency matrix $\boldsymbol{A}_{d \times d}$; parameters $\boldsymbol{\lambda}$; discrete severity pmfs $\boldsymbol{p}_{B_1}, \dots, \boldsymbol{p}_{B_d}$.}
\KwResult{Pmf of $S$, $\boldsymbol{p}_{S} = (p_S(0), \dots,  p_S({n_\mathrm{fft} - 1})).$} \label{alg:ligne-0}
 Set $n_\mathrm{fft}$ to be a large power of 2 \;
\For{\label{alg:ligne-1}
each vertex $v = 1, 2, \dots, d$}{
   Extend $\boldsymbol{p}_{B_v}$ to length $n_{\mathrm{fft}}$ with zeros\;
Compute the discrete Fourier transform (DFT) $\widehat{\boldsymbol{p}}_{B_v} = \mathrm{DFT}({\boldsymbol{p}}_{B_v})$\;}\label{alg:ligne-2}  
\For{each severity index $\ell = 1, \dots, n_\mathrm{fft}$}{
Initialize $\boldsymbol{H} = (H_{ij})_{i\times j \in \mathcal{V}\times\mathcal{V}}$, a matrix of ones\;
\label{alg:ligne1}
\For{each vertex $w = d, (d-1), \dots, 2$}{
   Find the parent of $w$,  $\pi_w = \inf\{j : A_{w, j} > 0\}$\;
   \label{alg:ligne2}
   Set the thinning coefficient $\theta_w = A_{\pi_w, w} \times \sqrt{\lambda_w / \lambda_{\pi_w}}$\;
   \label{alg:ligne3}
   Compute $h_{\ell,w} = \widehat{p}_{B_w}(\ell) \times \prod_{j} H_{w,j}$\;
   \label{alg:ligne4}
   Update $H_{\pi_w,w}$ to be $(1 - \theta_w) + \theta_w \times h_{\ell, w}$\;
   \label{alg:ligne5}
}
Compute $h_{\ell, 1} = \widehat{p}_{B_1}(\ell) \times \prod_{j} H_{1,j}$ and let $\theta_1 = 0$\;
\label{alg:ligne6}
Compute $\widehat{p}_S(\ell) = \prod_v \exp\{(\lambda_v -  A_{\pi_v, v} \times \sqrt{\lambda_v \lambda_{\pi_v}})(h_{\ell,v} - 1)\}$\;
\label{alg:ligne7}}
Obtain $\boldsymbol{p}_S$ by inverse DFT of $\widehat{\boldsymbol{p}}_S$\;
\label{alg:ligne8}
\Return $\boldsymbol{p}_{S}$. 
\end{algorithm}

\begin{algorithm}[h]
\caption{Computation of the cdf of $S$ : mixed Erlang claim distributions.}
\label{algo:pmfJ-erlang}
\KwData{$\boldsymbol{\alpha}$-weighted adjacency matrix $\boldsymbol{A}_{d \times d}$; parameters $\boldsymbol{\lambda}$ and $\boldsymbol{\beta}$; Erlang weights matrix $\boldsymbol{\zeta}_{d \times n_\text{fft}}$.}
\KwResult{Cdf of $S$, denoted as $F_{S}(x).$}
 Set $n_{\mathrm{fft}}$ to be a large power of 2\;
Compute $\beta_{\max} = \max(\beta_v, {v\in\mathcal{V}})$ and $q_{v} = \beta_{v}/\beta_{\max}$ for all $v\in\mathcal{V}$\;
\For{each vertex $v = 1, 2, \dots, d$}{
Construct $\boldsymbol{p}_{\widetilde{K}_v} = \bigl(0, 
\bigl(\,\textstyle\sum_{k=1}^{n_{\mathrm{fft}}} \zeta_{v,k}\, p_{K_{v,k}}(\ell)\,\bigr)_{\ell=1}^{\,n_{\mathrm{fft}}-1}\bigr)$\;
Compute the DFT 
$\widehat{\boldsymbol{p}}_{\widetilde{K}_v} = \mathrm{DFT}(\boldsymbol{p}_{\widetilde{K}_v})$\;}
\For{each index $\ell = 1, \dots,n_\mathrm{fft}$}{
Apply steps~\ref{alg:ligne1} to~\ref{alg:ligne6} of Algorithm~\ref{algo:pmfJ-discret}, 
using $\widehat{p}_{\widetilde{K}_v}(\ell)$ instead of $\widehat{p}_{B_v}(\ell)$ 
for every $v\in\mathcal{V}$\;
Compute $\widehat{p}_W(\ell) = \prod_v \exp\{(\lambda_v -  A_{\pi_v, v} \times \sqrt{\lambda_v \lambda_{\pi_v}})(h_{\ell,v} - 1)\}$\;}
Compute $\boldsymbol{p}_W$ by taking the inverse DFT of $\widehat{\boldsymbol{p}}_W$\;
\Return $F_{S}(x)  = p_W(0) + \sum_{k = 1}^{n_{\mathrm{fft}}} p_W(k) H(x; k, {\beta_{\max}}),\quad x \geq 0$.\label{alg2:ligne17}
\end{algorithm}

\newpage

\subsection{Algorithm for the computation of the expected allocations}\label{sup:algo-alloc}
This section presents the details of the algorithm used to compute the expected allocations. 
\begin{algorithm}[h]
\label{algo:ExpectedAllocations}
\caption{Computing the expected allocations of $X_v$ to $S$.} 
\KwIn{$\boldsymbol{\alpha}$-weighted adjacency matrix $\boldsymbol{A}_{d\times d}$; parameters  $\boldsymbol{\lambda}$; discrete severity pmfs $\boldsymbol{p}_{B_1}, \dots, \boldsymbol{p}_{B_d}$; vertex studied $v$.}
\KwOut{Vector $\boldsymbol{a}=(a_{\ell})_{\ell=1,\ldots,n_{\mathrm{fft}}}$ such that $a_{\ell}=\mathbb{E}[X_v\,\mathbbm{1}_{\{S=\ell-1\}}]$.}

Modify $\boldsymbol{A}$ to be topologically ordered according to root $v$; adjust the vector $\boldsymbol{\lambda}$ and $\{\mathbf{p}_{B_w},\, w\in\mathcal{V}\}$ accordingly. This can be done using Algorithm~5 of \cite{cote2025tree}\;
Set $n_{\mathrm{fft}}$ to be a large power of $2$\;
Compute the DFT $\widehat{\boldsymbol{f}}_S=\mathrm{DFT}(\boldsymbol{f}_S)$ using Algorithm~\ref{algo:pmfJ-discret}\;
Compute the DFT $\widehat{\boldsymbol{\phi}}_{B_1}$ of the vector $\left(k  p_{B_1}(k)/\mathbb{E}[B_1]\right)_{k=1}^{n_\mathrm{fft}}$, with $\mathbb{E}[B_1] = \sum_{\ell} \ell \times p_{B_1}(\ell)$\;
\For{each severity index $\ell = 1,\ldots, n_{\mathrm{fft}}$}{
Apply steps~\ref{alg:ligne1} to~\ref{alg:ligne5} of Algorithm~\ref{algo:pmfJ-discret}\;
Compute $h_{\ell,1}=\widehat{\boldsymbol{\phi}}_{B_1}(\ell)\times\prod_{j} H_{1,j}$ and let $\theta_1=0$\;
Compute ${\widehat{p}}_{K^{(1)} + S}(\ell) = \exp(-{2\pi i(\ell -1)}/{n_{\mathrm{fft}}}) h_{\ell,1} \widehat{f}_S(\ell)$\;
}
Compute ${\boldsymbol{p}}_{K^{(1)} + S}$ by taking the inverse DFT
of $\widehat{\boldsymbol{p}}_{K^{(1)} + S}$\;
Return $\boldsymbol{a} = \lambda_1 \mathbb{E}[B_1] {\boldsymbol{p}}_{K^{(1)} + S}$.
\end{algorithm}

\subsection{Tree sensitivity analysis}\label{sup:tree-sensitivity}

We conduct a sensitivity analysis evaluating the convergence of the estimated tree $\widetilde{\mathcal{T}}$ to the true tree 
$\mathcal{T}$. To isolate the effect of sample size on this convergence, we consider settings with homogeneous mean parameters and homogeneous dependence parameters. For the tree structures $\mathcal{T}_1$, $\mathcal{T}_2$ and $\mathcal{T}_3$ in Figure~\ref{fig:trees-analysis}, we set $\boldsymbol{\lambda}_1 = (2)_{v \in \mathcal{V}_1}$, $\boldsymbol{\lambda}_2 = (3)_{v \in \mathcal{V}_2}$, $\boldsymbol{\lambda}_3 = (4)_{v \in \mathcal{V}_3}$, and assign the same dependence parameter $\alpha = 0.3$ to each edge. For each tree of size $d \in \{3,5,10\}$, by adapting Algorithm 2 from \cite{cote2025tree} to accommodate flexible mean parameter, we generate $m = 1000$ datasets  with size  $n \in \{50, 100, 200, 350, 500,1000\}$.

For every simulated dataset, we compute the estimated tree $\widetilde{\mathcal{T}}$ and measure its proximity to the true $\mathcal{T}$ using the normalized Hamming distance
\begin{equation*}
 d_H(\mathcal{T},\widetilde{\mathcal{T}}) = \frac{1}{2(d - 1)} \left(\sum_{e \in \mathcal{E}} 
 \mathbbm{1}\{e \notin \widetilde{\mathcal{E}}\} + \sum_{e \in \widetilde{\mathcal{E}}} 
 \mathbbm{1}\{e \notin {\mathcal{E}}\}\right).
\end{equation*}
The Hamming distance measures the number of edge deletions and insertions necessary to transform one graph into another \citep{donnat2018tracking}. 

For each pair $(d,n)$, we compute the average Hamming distance between the estimated tree $\widetilde{\mathcal{T}}$ and the true tree $\mathcal{T}$ across the $1000$ samples. We also evaluate the probability of exact recovery, $\pi_{\mathrm{rec}} = \Pr(\widetilde{\mathcal{T}} = \mathcal{T}).$ We report the estimated values for all pairs in Table~\ref{tab:sensitivity}. The results indicate that the use of empirical correlations yields greater reliability for recovering the dependence structure of the MPMRF model.
\begin{figure}[ht]
    \centering
    \begin{subfigure}[b]{0.3\textwidth}
        \centering
        \begin{tikzpicture}[
  every node/.style={circle, draw, inner sep=2pt},
]
  \node (n2) at (0,0) {};
  \node[left= 0.2cm of n2]  (n1) {};
  \node[right= 0.2cm of n2] (n3) {};

  \draw (n1) -- (n2) -- (n3);
\end{tikzpicture}
        \caption{Tree with 3 vertices, $\mathcal{T}_1$}
        \label{fig:tree1}
    \end{subfigure}\hfill
    \begin{subfigure}[b]{0.3\textwidth}
        \centering
        \begin{tikzpicture}[
  every node/.style={circle, draw, inner sep=2pt},
]
  \node (n1) at (0,0) {};
  \node[left= 0.2cm of n1] (n2) {};
  \node[right= 0.2cm of n1] (n3) {};
  \node[above= 0.2cm of n3] (n4) {};
  \node[right= 0.2cm of n3] (n5) {};

  \draw (n1) -- (n2);
  \draw (n1) -- (n3);
  \draw (n3) -- (n4);
  \draw (n3) -- (n5);
\end{tikzpicture}
        \caption{Tree with 5 vertices}
        \label{fig:tree2}
    \end{subfigure}\hfill
    \begin{subfigure}[b]{0.3\textwidth}
        \centering
        \begin{tikzpicture}[
  every node/.style={circle, draw, inner sep=2pt}
]
  \node (n3) at (0,0) {};

  \node[right= 0.2cm of n3]           (n2) {};
  \node[above= 0.2cm of n3]  (n4) {};
  \node[left= 0.2cm of n3]  (n5) {};
  \node[above right = 0.36cm of n3] (n6) {};
  \node[above left= 0.36cm of n3] (n7) {};

  \node[right= 0.2cm of n2] (n1) {};   
  \node[left= 0.2cm of n5] (n8) {};   
  \node[right= 0.2cm of n6] (n9) {};   

  \node[left= 0.2cm of n8] (n10) {};  

  \draw (n1) -- (n2);
  \draw (n2) -- (n3);
  \draw (n3) -- (n4);
  \draw (n3) -- (n5);
  \draw (n3) -- (n6);
  \draw (n3) -- (n7);
  \draw (n5) -- (n8);
  \draw (n8) -- (n10);
  \draw (n6) -- (n9);
\end{tikzpicture}
        \caption{Tree with 10 vertices}
        \label{fig:tree3}
    \end{subfigure}
    \caption{Three undirected tree structures for the sensitivity analysis.}
    \label{fig:trees-analysis}
\end{figure}

\begin{table}[h]
\centering
\caption{Monte Carlo estimates of the mean normalized Hamming distance 
$\overline{d}_H$ and the recovery probability $\hat{\pi}_{\mathrm{rec}}$ 
for tree reconstruction under two MST-building criteria, based on 1000 samples.}
\label{tab:sensitivity}
\begin{subtable}{0.48\textwidth}
\centering
\caption{Empirical mutual information.}
\begin{tabular}{ccrr}
\toprule
$d$ & $n$ & $\overline{d}_H$ & $\hat{\pi}_{\mathrm{rec}}$ \\
\midrule
\multirow{6}{*}{3}  & 50   & 0.262    & 0.476 \\
  & 100  & 0.200    & 0.601 \\
  & 200  & 0.088    & 0.824 \\
  & 350  & 0.027    & 0.946 \\
  & 500  & 0.0055   & 0.989 \\
  & 1000 & 0        & 1     \\
\midrule
\multirow{6}{*}{5}  & 50   & 0.524    & 0.023 \\
  & 100  & 0.436    & 0.071 \\
  & 200  & 0.274    & 0.243 \\
  & 350  & 0.120    & 0.596 \\
  & 500  & 0.0448   & 0.826 \\
  & 1000 & 0.00025  & 0.999 \\
  \midrule
\multirow{6}{*}{10} & 50   & 0.734    & 0     \\
 & 100  & 0.657    & 0     \\
 & 200  & 0.494    & 0.001 \\
 & 350  & 0.276    & 0.046 \\
 & 500  & 0.141    & 0.215 \\
 & 1000 & 0.00533  & 0.953 \\
\bottomrule
\end{tabular}
\end{subtable}
\hfill
\begin{subtable}{0.48\textwidth}
\centering
\caption{Empirical correlation.}
\begin{tabular}{ccrr}
\toprule
$d$ & $n$ & $\overline{d}_H$ & $\hat{\pi}_{\mathrm{rec}}$ \\
\midrule
\multirow{6}{*}{3}  & 50   & 0.089   & 0.809 \\
 & 100  & 0.0398  & 0.919 \\
 & 200  & 0.005   & 0.990 \\
 & 350  & 0       & 1     \\
 & 500  & 0       & 1     \\
 & 1000 & 0       & 1     \\
\midrule
\multirow{6}{*}{5}  & 50   & 0.194   & 0.381 \\
 & 100  & 0.0802  & 0.709 \\
 & 200  & 0.0152  & 0.939 \\
 & 350  & 0.001   & 0.996 \\
 & 500  & 0       & 1     \\
 & 1000 & 0       & 1     \\
\midrule
\multirow{6}{*}{10} & 50   & 0.299   & 0.023 \\
 & 100  & 0.134   & 0.253 \\
 & 200  & 0.0226  & 0.816 \\
 & 350  & 0.00133 & 0.988 \\
 & 500  & 0       & 1     \\
 & 1000 & 0       & 1     \\
\bottomrule
\end{tabular}
\end{subtable}
\end{table}
\subsection{Procedure for datasets construction}\label{sup:datasets}
\subsubsection{Peak-over-threshold}\label{subsec:preprocessing}
We define extreme rainfall events as days when precipitation amounts exceed station-specific thresholds. We adopt the peaks-over-threshold (POT) approach to jointly model the frequency and severity of extreme rainfall events across weather stations using the MPMRF risk model. The POT framework decomposes extremes into two components: the counts exceeding a threshold $N$, typically modeled by a Poisson distribution, and the exceedance amounts $X$. For a sufficiently high threshold $u$, the excess random variable  $Y_i = X_i - u \,\mid\, X_i > u$ is asymptotically distributed as a generalized Pareto distribution (GPD), with $i = 1,2,\ldots,n$, where $n$ denotes the total number of POT-independent exceedances \citep{chavez2005generalized}. For $y > 0$ and $\sigma_u > 0$, let $F_Y(y)$ denote the cdf of the GPD. Its stationary form can be written as 
\begin{equation*}
F_Y(y) =
\begin{cases}
1 - \left( 1 + \dfrac{\xi y}{\sigma_u} \right)^{-1/\xi}, & \xi \neq 0, \\
1 - \mathrm{e}^{-{y}/{\sigma_u}}, & \xi = 0,
\end{cases}
\end{equation*}
where $y$ is the excess over the threshold, and $\xi$ and $\sigma_u$ denote respectively the shape parameter and the scale parameter conditional on the threshold $u$.

The threshold value controls the size of the POT events, their independence, and the adequacy of the GPD. For threshold selection, we use the mean excess function. This method describes the expectation of the exceedance above a given threshold $u$  and is defined as $e(u) = \mathbb{E}[X - u \mid X > u]$.  For the case of a GPD, the mean excess function is an increasing linear function of $u$ when the shape parameter satisfies $\xi < 1$ \citep{embrechts1997modelling}:
\begin{equation*}
    e(u)
    = \frac{\hat{\sigma} + \hat{\xi} u}{1 - \hat{\xi}}
    \qquad\text{with } \hat{\sigma} + \hat{\xi} u > 0.
\end{equation*}

\begin{figure}[t]
    \centering
    \includegraphics[width=\linewidth]{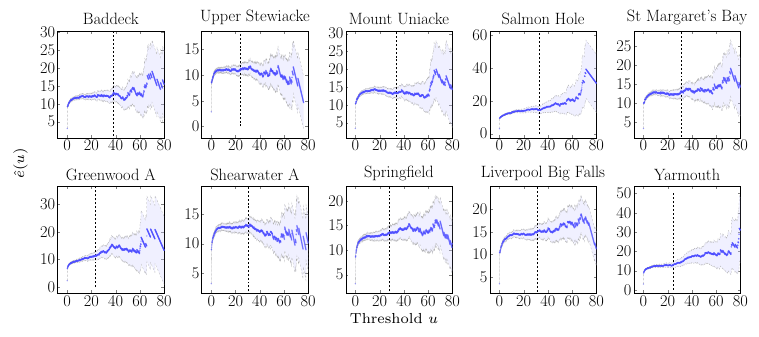}
    \caption{Mean residual life plots of daily rainfall used for threshold selection across stations.}
    \label{fig:mrl}
\end{figure}

Thus, the mean excess function approach consists of plotting the empirical estimator $\hat{e}(u)$ and selecting a threshold $u>0$ such that the curve is approximately linear for $x > u$. We report the dataset's mean residual life plots in Figure~\ref{fig:mrl}. The light blue area around the marks represents the 95\% confidence interval.

\begin{figure}[t]
    \centering
    \includegraphics[width=\linewidth]{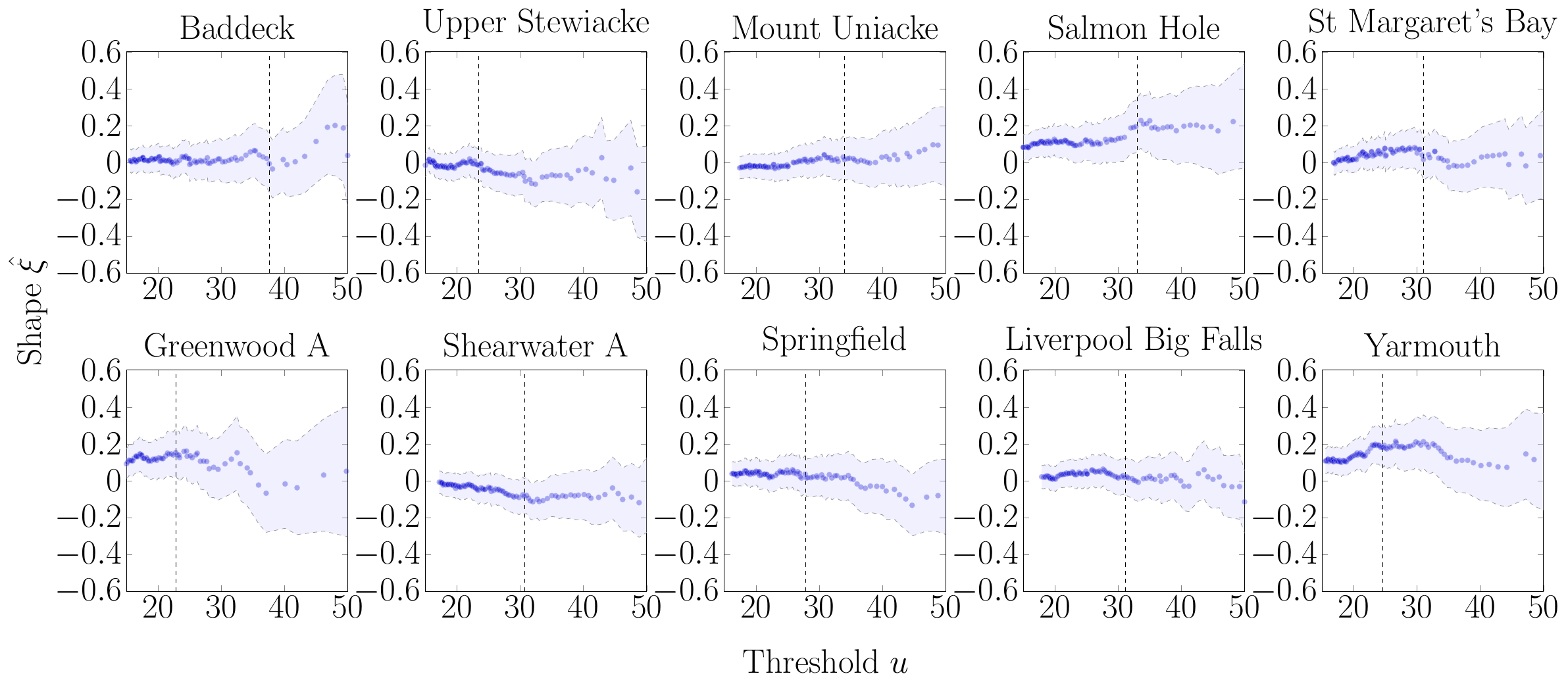}
    \caption{Shape plots of $\hat{\xi}$ for threshold selection across stations.}
    \label{fig:hill}
\end{figure}

To support the choice of an appropriate threshold, we also rely on shape plots (Hill-type plots), which display the estimated shape parameter  $\hat{\xi}$ as a function of the threshold $u$. Since varying the threshold affects only the scale parameter $\sigma$ while the shape parameter $\xi$ should remain stable, an appropriate threshold corresponds to the smallest value above which $\hat{\xi}$ is approximately constant \citep{beirlant2006statistics}. Figure \ref{fig:hill} presents the stability plot for each station. Marks correspond to an estimate of the shape parameter obtained by repeatedly fitting a GPD to the excesses over a decreasing sequence of thresholds. We evaluated 100 models, with the number of threshold exceedances ranging from 700 to 15. The light blue area around the marks represents the 95\% confidence interval.

In Figures \ref{fig:mrl} and \ref{fig:hill}, the vertical dashed lines represent the chosen thresholds at each station. We report the exact values in Table~3; across stations, the threshold ranges from the 97th to the 99th percentiles of daily rainfall amounts.  
The threshold selected for Baddeck (ID~1) lies toward the upper end of the stability region. However, it is the lowest threshold for which the Poisson assumption for exceedance counts, the GPD modelling assumption for excesses, and the independence of exceedances are simultaneously validated.

\subsubsection{Independence of exceedances}
\begin{figure}[t]
    \centering
    \includegraphics[width=\linewidth]{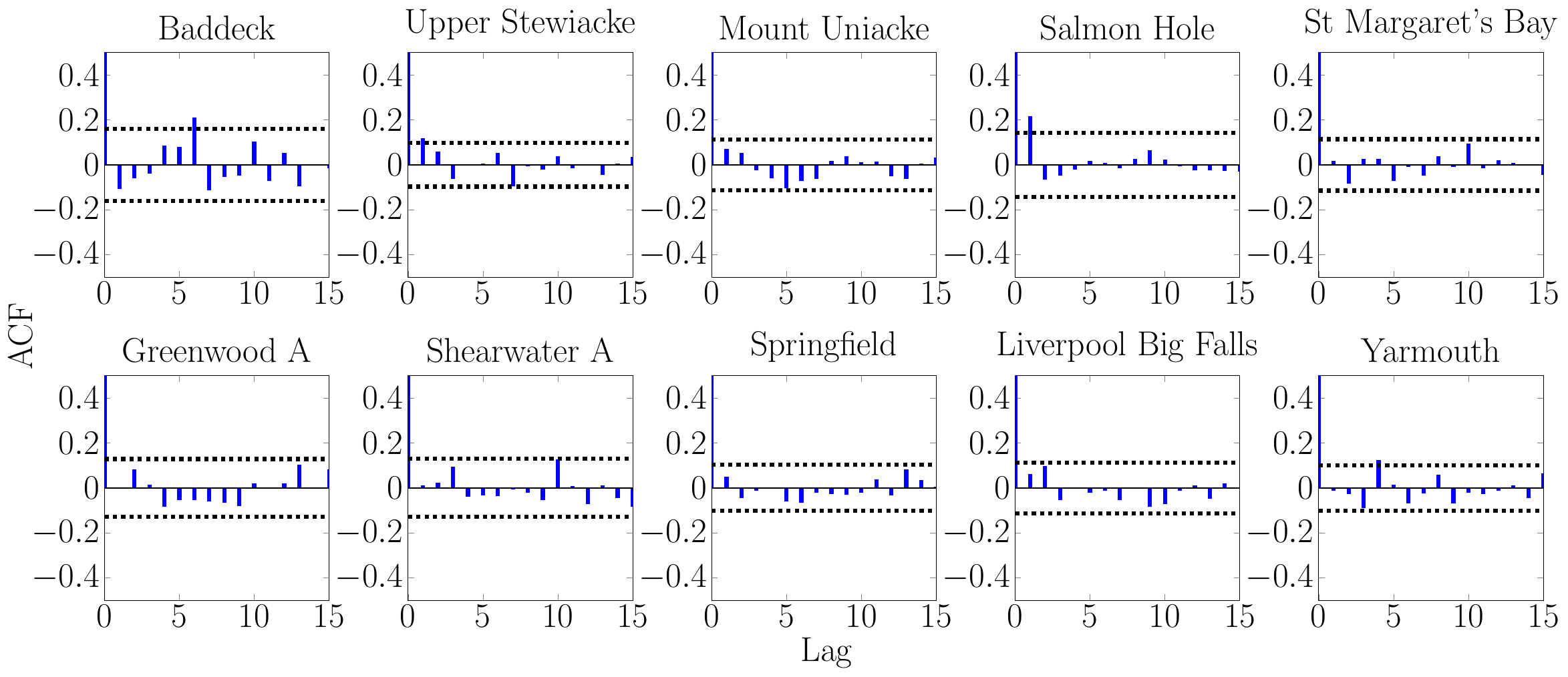}
    \caption{Autocorrelation functions of rainfall exceedances across stations.}
    \label{fig:acfprecip}
\end{figure}

The POT framework relies on the independence assumption of exceedances. In this study, in accordance with  \cite{thiombiano2017nonstationary}, consecutive days with threshold exceedances are grouped into clusters to overcome temporal dependence. Table~3 of the main text displays the number of clustered events per station; most of the clusters consist of a single day, while a smaller proportion include two or three consecutive days. The severity of each extreme event is defined as the maximum daily rainfall within the cluster. For instance, when exceedances occur on two consecutive days, only the larger of the two is retained.  The Ljung-Box test, computed using the rule-of-thumb lag choice for nonseasonal data $\lfloor\min(10, n/5)\rfloor$, where $n$ is the number of observations \citep{hyndman2018forecasting}, yields non-significant $p$-values ($p>0.05$) for all stations. 
We report the maximum daily rainfall autocorrelation function plots, further validating the independence assumption of exceedances for each station, in Figure \ref{fig:acfprecip}. In all cases, the sample autocorrelations at lags $h \geq 1$ remain majoritarily within the 95\% confidence bounds ($\pm 1.96/\sqrt{n}$), and we observe no systematic decay or oscillatory patterns.

\subsubsection{Independence across years}
For the analysis, we only included years in which the proportion of missing data across all weather stations during the rainy season (May through September) did not exceed 10\%. The Shearwater A station had a proportion of missing data just below 10\% in a particular year. To verify whether any extreme weather events occurred on days with missing data, we consulted nearby stations. Since we found no evidence of such events, we did not exclude any additional years from the analysis. After preprocessing, Datasets~2 and 3 comprised 43 yearly multivariate observations of extreme rainfall events. Dataset 3 also includes the precipitation amount associated with each event.

\begin{figure}[t]
    \centering
    \includegraphics[width=\linewidth]{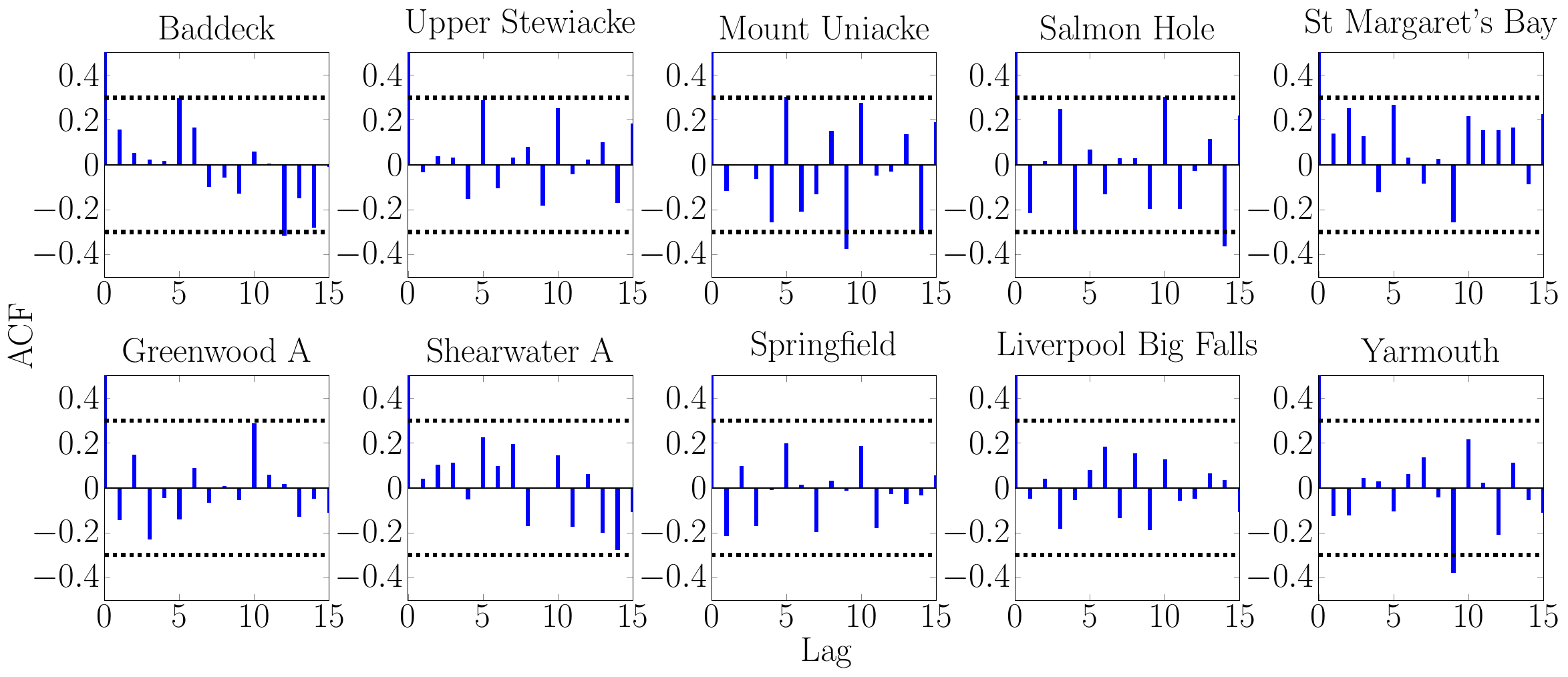}
    \caption{Autocorrelation functions of annual extreme rainfall counts across stations.}
    \label{fig:acf}
\end{figure}

To support the assumption of independence across years, we examined both the Ljung-Box test and the empirical autocorrelation function (ACF). The Ljung-Box test, computed using the rule-of-thumb lag choice for nonseasonal data $\lfloor\min(10, 43/5)\rfloor=8$ yields non-significant $p$-values ($p>0.05$) for all stations. Figure~\ref{fig:acf} presents the ACFs for each station.  Taken together, the ACFs and Ljung-Box test indicate the absence of meaningful temporal autocorrelation, supporting the assumption that annual counts may be treated as independent.

\subsubsection{Poisson-GP modeling assumptions}
According to extreme value theory, the annual occurrences of exceedances above specified high thresholds are expected to follow a Poisson distribution \citep{coles2001classical}. To validate this assumption for each station, we conducted a chi-squared goodness-of-fit test on the yearly counts of exceedances. 
We grouped exceedance-count classes so that all expected frequencies were at least 5. The chi-squared statistic was then computed with $K- 2$ degrees of freedom, where $K$ represents the total number of bins after grouping. We present in Table~\ref{tab:chi2} the chi-squared statistic, the associated degrees of freedom, the corresponding $p$-value. The resulting $p$-values for all stations were greater than 0.3, indicating that there is no substantial evidence to reject the hypothesis that the marginal event counts are Poisson-distributed. 

We complement this test with the empirical dispersion index $D = {\widehat{\mathrm{Var}}(N_v)}/{\widehat{\mathbb{E}}[N_v]}$, for each station. We present the values in Table~\ref{tab:chi2}. The values lie between $0.95$ and $1.4$, which is consistent with the Poisson assumption considering the few observations. 
\begin{figure}[t]
\begin{minipage}[t]{0.4\textwidth}
\centering
\captionof{table}{Goodness-of-fit statistics for the Poisson marginals.}
\begin{tabular}{lrrr|r}
\toprule
ID & {$\chi^2$} stat. & {df} & {$p$-value} & $D$ \\
\midrule
1 & 2.63 & 4 & 0.62 & 1.24 \\ 
  2 & 2.24 & 3 & 0.52 & 1.02 \\ 
  3 & 3.04 & 4 & 0.55 & 1.31 \\ 
  4 & 0.60 & 4 & 0.96 & 1.39 \\ 
  5 & 2.02 & 4 & 0.73 & 1.25 \\ 
  6 & 0.95 & 4 & 0.92 & 1.21 \\
  7 & 4.61 & 4 & 0.33 & 1.24 \\ 
  8 & 1.46 & 3 & 0.69 & 1.33 \\ 
  9 & 2.57 & 3 & 0.46 & 1.02 \\ 
  10 & 1.71 & 3 & 0.63 & 0.96 \\ 
\bottomrule
\end{tabular}
\label{tab:chi2}
\end{minipage}
\hfill
\begin{minipage}[t]{0.5\textwidth}
\centering
\captionof{table}{Goodness-of-fit statistics for the GPD marginals.} 
\begin{tabular}{rrrr|r}
  \toprule
 ID & $\chi^2$ stat. & df & $p$-value & AD stat. \\ 
  \midrule
1 & 11.90 & 10 & 0.29 & 0.59 \\ 
  2 & 9.56 & 16 & 0.89 & 0.29 \\ 
  3 & 16.32 & 14 & 0.29 & 0.44 \\ 
  4 & 13.03 & 12 & 0.37 & 0.48 \\ 
  5 & 10.87 & 13 & 0.62 & 0.53 \\ 
  6 & 11.91 & 12 & 0.45 & 0.47 \\ 
  7 & 19.29 & 16 & 0.25 & 0.58 \\ 
  8 & 10.77 & 17 & 0.87 & 0.35 \\ 
  9 & 13.86 & 15 & 0.54 & 0.36 \\ 
  10 & 11.58 & 16 & 0.77 & 0.36 \\ 
\bottomrule
\end{tabular}
\label{tab:gpdtest}
\end{minipage}
\end{figure}
To visually assess the adequacy of the Poisson assumption for the yearly event counts, we provide in Figure~\ref{fig:poissoncdf} the empirical cdfs alongside their fitted counterparts (obtained via maximum likelihood estimation). Most stations exhibit a fitted Poisson model that closely match the empirical cdfs. However, Shearwater (ID~7) and Springfield (ID~8) show cdfs slightly understated at lower counts and overstated in the upper tail. These discrepancies indicate a mild deviation from the Poisson assumption.
\begin{figure}[t]
    \centering
    \includegraphics[width=\linewidth]{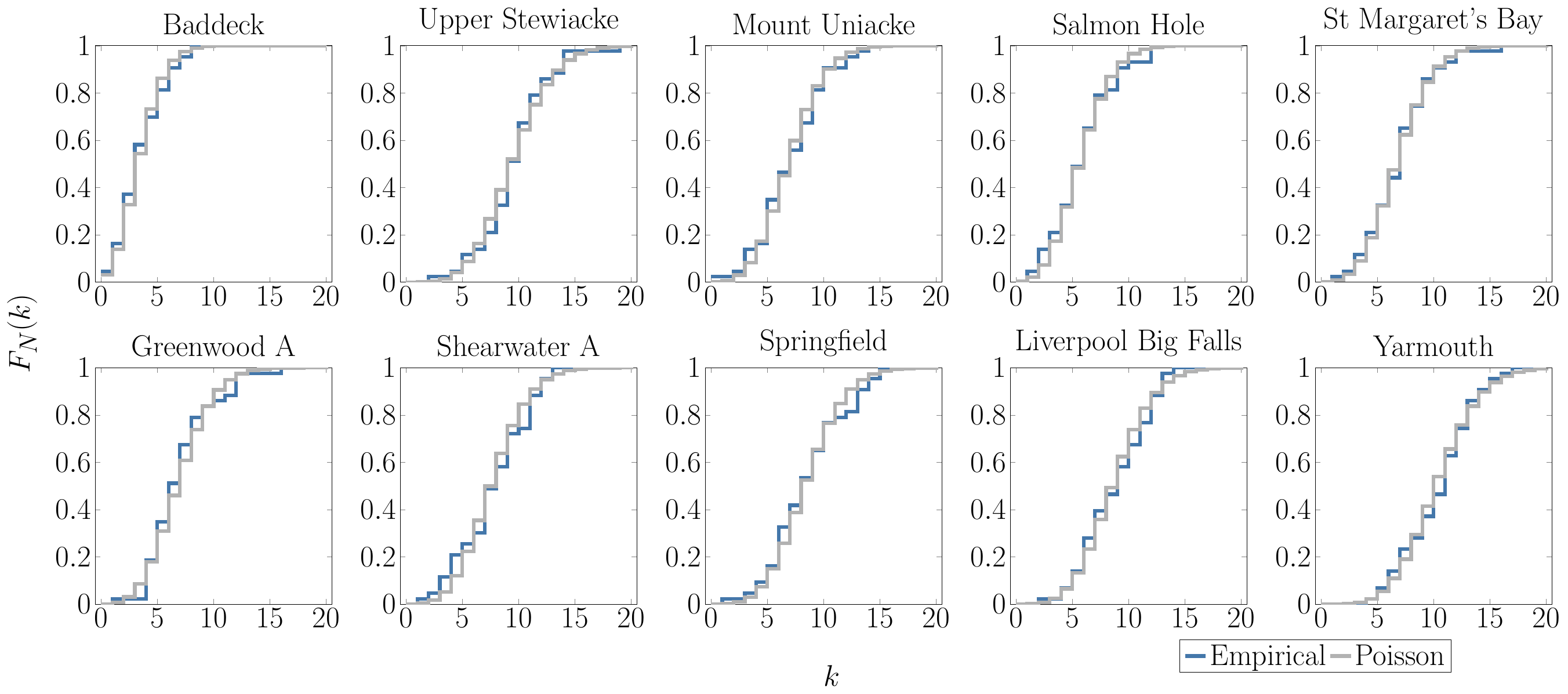}
    \caption{Empirical and theoretical Poisson cdfs at each station.}
    \label{fig:poissoncdf}
\end{figure}

In light of the limited sample size and the predominantly near-Poisson dispersion levels across stations, we retain the Poisson distribution as the marginal specification for the remainder of the analysis. 

To evaluate the adequacy of the GPD for modeling severities, we applied a chi-squared goodness-of-fit test to the exceedances at each station. At the $10\%$ significance level, the test did not reject the null hypothesis that the excesses follow a GPD. Table~\ref{tab:gpdtest} reports the chi-squared statistics together with their associated degrees of freedom and $p$-values. For additional emphasis on tail behavior, we also present the Anderson-Darling (AD) statistic, which places greater weight on discrepancies in the upper tail of the distribution. All AD statistics are lower than the 10\% confidence AD statistic, 1.933.

\begin{figure}[t]
    \centering
    \includegraphics[width=0.99\linewidth]{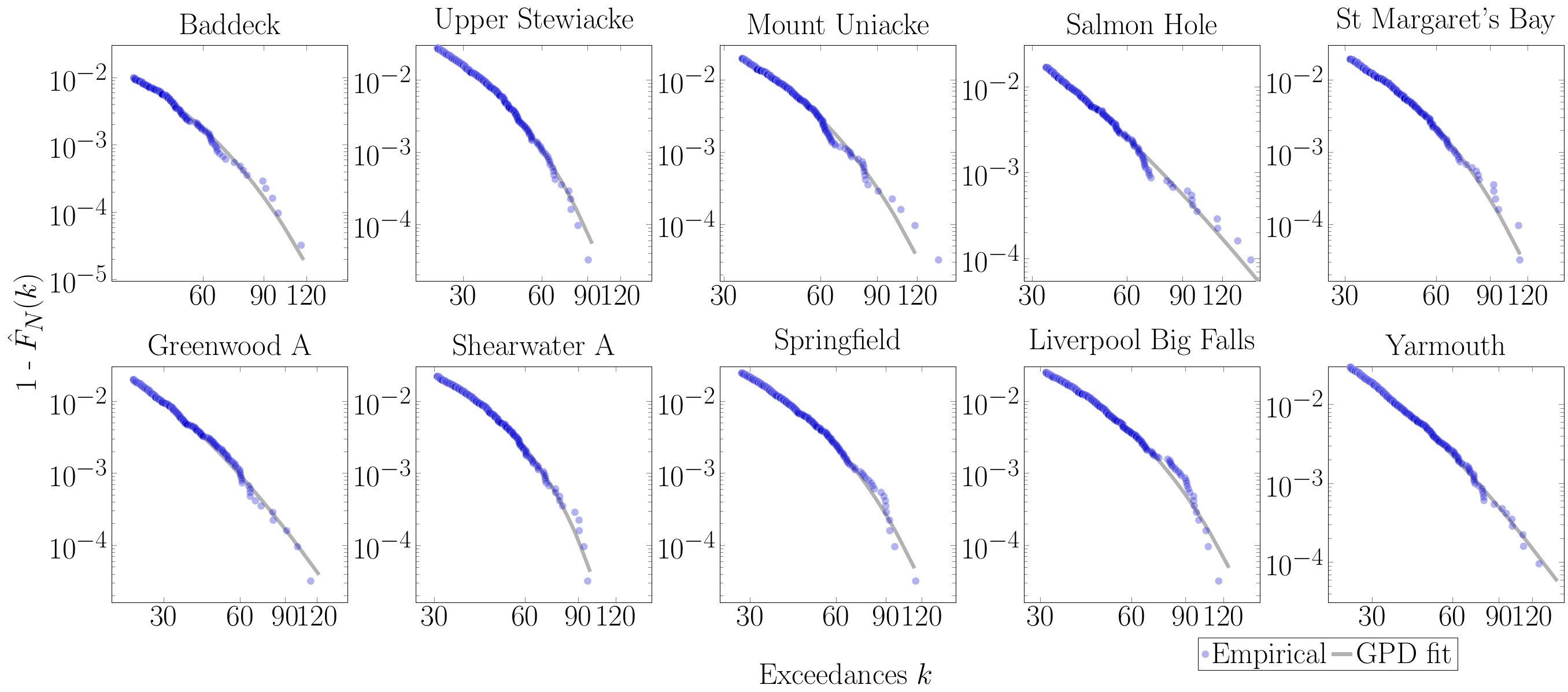}
    \caption{Log-log tail plots of severity excesses for all stations.}
    \label{fig:tailplots}
\end{figure}

We also provide tail plots on a log-log scale in Figure \ref{fig:tailplots}. Salmon Hole (ID~3) exhibits the highest $\chi^2$ statistic; as shown in Figure~\ref{fig:tailplots}, the empirical tail deviates from the fitted GPD in the upper range, indicating reduced tail fidelity for this station.

Overall, based on the combined evidence from goodness-of-fit statistics and tail-plot diagnostics, the GPD provides an appropriate marginal model for the exceedance severities across all stations.
 
\subsubsection{Comparison of risk allocation rules}\label{sup:comp}
Table~\ref{tab:station_ccov_ctvar_diff} reports the exact relative contributions corresponding to Figure~9 in the main text, which compares the allocation of each risk under covariance-based and TVaR-based rules.

\begin{table}[bh]
\centering
\caption{Relative $C^{\mathrm{Cov}}_\kappa(X_v, S)$, Relative $C^{\mathrm{TVaR}}_\kappa(\widetilde{X}_v, \widetilde{S})$, and their difference for each station, with $\kappa = 0.99$.}
\begin{tabular}{crrr}
\toprule
ID & $C^{\mathrm{Cov}}_\kappa(X_v, S)$ (\%) & $C^{\mathrm{TVaR}}_\kappa(\widetilde{X}_v, \widetilde{S})$ (\%), & $C^{\mathrm{Cov}}_\kappa(X_v, S)$ $-$ $C^{\mathrm{TVaR}}_\kappa(\widetilde{X}_v, \widetilde{S})$ (\%) \\
\midrule
1 & 5.65 & 5.63 & 0.02 \\ 
  2 & 10.23 & 10.17 & 0.06 \\ 
  3 & 11.71 & 11.77 & -0.06 \\ 
  4 & 9.82 & 10.09 & -0.27 \\ 
  5 & 9.80 & 9.80 & 0.00 \\ 
  6 & 8.69 & 8.78 & -0.09 \\ 
  7 & 9.52 & 9.40 & 0.12 \\ 
  8 & 11.29 & 11.24 & 0.05 \\ 
  9 & 12.48 & 12.36 & 0.12 \\ 
  10 & 10.81 & 10.76 & 0.05 \\ 
\bottomrule
\end{tabular}
\label{tab:station_ccov_ctvar_diff}
\end{table}


\end{document}